\documentclass{aa}      
\usepackage{amsmath}     
\usepackage{graphicx}  
\usepackage{epstopdf}    
\usepackage{txfonts}
\usepackage{natbib}
\usepackage{mathrsfs} 
\usepackage{fontenc}
\usepackage{bm} 
\usepackage{pdflscape} 
\usepackage{color}  
\usepackage{longtable}   
\usepackage{multirow}    
\usepackage{xcolor}   

\begin{document}

\title{Study of a sample of faint Be stars in the exofield of CoRoT}
\subtitle{Part III. Global spectroscopic characterization and astrophysical parameters of the central stars 
\thanks {The complete version of Tables~\ref{corr_par_prom_t} and \ref{corr_par_mode_t} are available in electronic form at the CDS via anonymous ftp to cdsarc.u-strasbg.fr (130.79.128.5) or via http://cdsweb.u-strasbg.fr/cgi-bin/qcat?J/A+A/}} 

\author{ J. Zorec \inst{1} 
\and     A.M. Hubert \inst{2}
\and     C. Martayan \inst{3}
\and     Y. Fr\'emat \inst{4}                        
}
\institute{Sorbonne Universit\'e, CNRS, UPMC, UMR7095 Institut d'Astrophysique de Paris, 98bis Bd. Arago, F-75014 Paris, France; \email{zorec@iap.fr} 
\and GEPI, Observatoire de Paris, PSL Research University, CNRS UMR 8111, 5 place Jules Janssen, 92190 Meudon, France; \email{Anne-Marie.Hubert@obspm.fr} 
\and European Organization for Astronomical Research in the Southern  Hemisphere, Alonso de Cordova 3107, Vitacura, Santiago de Chile, Chile
\and Royal Observatory of Belgium, 3 Av. Circulaire, B-1180 Bruxelles, Belgium 
}
 
\offprints{J. Zorec: \email{zorec@iap.fr}}     
\date{Received ..., ; Accepted ...,} 

\abstract
{The search and interpretation of non-radial pulsations from Be star light curves observed with the CoRoT satellite requires high-quality stellar astrophysical parameters.} 
{The present work is devoted to the spectroscopic study of a sample of faint Be stars observed by CoRoT in the fourth long run (LRA02).}
{The astrophysical parameters were determined from the spectra in the $\lambda\lambda 4000-4500$ \AA\ wavelength domain observed with the VLT/FLAMES instruments at ESO. Spectra were fitted with models of stellar atmospheres using our GIRFIT package. Spectra obtained in the 
$\lambda\lambda 6400-7200$ \AA\ wavelength domain enabled the confirmation or, otherwise, a first identification of Be star candidates.} 
{The apparent parameters ($T_{\rm eff},\log g,V\!\sin i$) for a set of 19 B and Be stars were corrected for the effects induced by the rapid rotation. These allowed us to determine: 1) stellar masses that are in agreement with those measured for detached binary systems; 2) spectroscopic distances that agree with the GAIA parallaxes; and 3) centrifugal/gravity equatorial
force ratios of $\sim0.6-0.7,$  which indicate that our Be stars are subcritical rotators. A study of the Balmer 
H$\alpha$, H$\gamma$ and H$\delta$ emission lines produced: 1) extents of the circumstellar disk  
(CD) emitting regions that agree with the interferometric inferences in other Be stars; 2) $R-$ dependent exponents $n(R)=\ln[\rho(R)/\rho_o]/\ln(R_o/R)$ of the CD radial density distributions; and 3) CD base densities $\rho_o$ similar to those inferred in other recent works.}     
{The H$\gamma$ and H$\delta$ emission lines are formed in CD layers close to the central star. These lines produced a different value of the exponent $n(R)$ than assumed for H$\alpha$. Further detailed studies of H$\gamma$ and H$\delta$ emission lines could reveal the physical properties of regions where the viscous transport of angular momentum to the remaining CD regions is likely to originate from. The subcritical rotation of Be stars suggests that their huge discrete mass-ejections and concomitant non-radial pulsations might have a common origin in stellar envelope regions that become unstable to convection due to rotation. If it is proven that the studied Be stars are products of binary mass transfer phases, the errors induced on the estimated $T_{\rm eff}$ by the presence of stripped sub-dwarf O/B companions are not likely to exceed their present uncertainties.} 

\titlerunning{Study of a sample of faint Be stars in the exofield of CoRoT. III}
\authorrunning{J. Zorec et al.}   
\maketitle

\section{Introduction}\label{intro}

  The photometric data obtained by the CoRoT satellite \citep{weiss_2004,baglin_2006} concern a  wide range of variable objects, including Be stars, which are rapidly, albeit subcritical, rotating objects \citep{cranmer_2005,zorec_2016}. During their main sequence lifespan, these stars build up a circumstellar disk (hereafter CD) \citep{jaschek1981,collins1987}, whose presence is revealed by line emissions and continuum flux excess, mainly in the visible and infrared spectral domains \citep{rivinius_2013}. \par
 The CoRoT photometric data of Be stars are particularly relevant because they carry signatures of  their non-radial pulsation modes. As pulsating objects, Be stars lie in the instability region of the Hertzsprung-Russell  (HR) diagram between (or even partially shearing) the instability regions of $\beta$~Cephei stars and the  slowly pulsating-B stars (SPBs, \citet{miglio_2007,
walczak2015,saio2017,szewczuk2017,burssens2020}. Unlike Be stars, SPBs have often been considered slow rotators. However, long ground-based photometric surveys \citep[e.g.,][]{salmon2014} and recent four-year data provided by the Kepler mission \citep[][and references therein]{pedersen2021} have enabled the identification of rapid rotators among B-star pulsators (previously considered SPBs), where properties of their internal structure and pulsations might perhaps become an approximation for Be stars \citep{moravveji2016}.\par

  The analysis of the CoRoT photometric data of Be stars carried out by \citet{huat_2009} and \citet{sema2018} revealed complex frequency spectra exhibiting multiple-periodicity and groups of closely separated frequencies, as well as isolated frequencies. This same behavior was also observed in data of Be stars obtained in other space missions: MOST \citep{walker2003,
walker2005a,walker2005b}, SMEI \citep{jackson2004,goss2011}, KEPLER \citep{borucki2010,kurtz2015,papics2017}, BRITE-Constellation \citep{weiss2014,baade2016}, and TESS \citep{ricker2016,labadie2022}. \par

 The groups of closely separated frequencies can be considered as combinations of g-mode pulsations, which cannot be produced by spots in the stellar surface \citep{kurtz2015}. One of the most important findings in the CoRoT data of HD 49330 \citep{huat_2009}, later also observed in another 7 Be stars analyzed by \citep{sema2018}, is the presence of light outbursts accompanied by changes mainly of the photometric power spectrum. They are supposed to be manifestations of discontinuous mass-ejections, similar to those reported by \citet{hub1998} using  HIPPARCOS photometry, as well as in stars of the LMC and SMC observed by the MACHO \citet{cook_1995,keller_2002} and the OGLE \citet{mennickent_2002, dewit_06} surveys. Although these ejections are supposed to provide the required mass to build up a CD as described by the viscous decretion model
(VDD) \citep{lee1991,ghore2021,haubois2012}, the mechanisms causing the ejections proper are still heavily debated. On the one hand, from theoretical grounds the combination of prograde g-modes of stellar pulsation could lead to sporadic mass ejections at circumstances when stellar rotation is near critical \citep{osaki_1986,kee2016,saio2017}. On the other hand, based on spectroscopic studies, \citet{baade2016} suggested a possible relation between a stellar oscillation called ``difference frequency," which somehow should result from the combination of two other observed close pulsation frequencies, with sporadic mass ejection events. A series of correlations \citep[e.g.,][]{sema2018}, theoretical predictions of pulsating modes \citep[e.g.,][]{saio2013}, and recent results \citep[e.g.,][]{papics2017} confirm that Be stars should be considered as genuine non-radial pulsators.\par

  Two preceding papers by \cite{sema2013} and \citet{sema2018} were dedicated to spectroscopic studies and analysis of non-radial pulsation modes in faint Be stars observed in the initial run (IRA01) and the first two long runs (LRC01 and LRA01) of the CoRoT mission. The present work deals with a sample of faint Be stars belonging to the fourth long run (LRA02), whose main objective is to achieve a global spectroscopic characterization for them. The specific aim of the present paper is to determine reliable stellar astrophysical parameters of a new set of faint Be stars. In this work we consider Be stars to behave as single objects. \par 
   
  Although the astrophysical parameters are here determined using spectra obtained in the $4000-4500$ \AA\ wavelength domain, spectra in a wavelength region near the H$\alpha$ line were also obtained in order to confirm the Be star nature of the studied stars. To take full advantage of the obtained spectroscopic data, we also inquire the physical structure of the CD in the studied Be stars. Our approach is based on first principles of radiation transfer that control the formation of Balmer lines. It enables to probe the extents, and the density distribution in the CDs of the studies Be stars. \par
    
   The present work is organized as follows. Section~2 briefly identifies the sky region of the studied Be stars and specifies the source where the CoRoT data were extracted. Section~3 presents the spectroscopic observations of the program objects. Section 4 gives a short description of the spectral characteristics of the selected Be stars. Section 5 describes the methods used to determine the primary and secondary apparent stellar astrophysical parameters and their treatment for rapid rotation effects. Some correlations between the Balmer line emission intensities and the near-infrared photometric color indices are shown in Sect. 6. A method to describe the physical characteristic of CD of the program stars is detailed in Sect. 7. A short summary of the results obtained is given in Sect. 8. \par  
 
\section{CoRoT Observations}

 The telescope of the CoRoT satellite was pointed to the ``anticentre" of the Milky Way to observe the LRA02 field whose coordinates are 
$\alpha=103^{\circ}52$, $\delta=-04^{\circ}38$ and $Roll=06^{\circ}00$. Half of the field was dedicated to observation of bright stars devoted to programs of seismology and  the other half to observation of faint stars with magnitudes $11<V<17$ mag to additional programs, such as the search for exoplanets. The upper part of the LRA02 field was very close to the lower part of the IRA01 field, as shown in \citet[][Fig. III.1.1.]{deleuil_2016}\par
  The LRA02 exofield contains stars within coordinates $06^h47^m30^s<\alpha<06^h53^m50^s$ and $-05^{\circ}51^{'}30{"}<\delta<-03^{\circ}06^{'}30^{"}$. In this field, 11\,448 stars were observed over 112 days from 16 November 2008 to 8 March 2009.CoRoT data, namely, the light curves with several levels of corrections and headers, of all faint stars are available at CDS\footnote[1]{http://vizier.u-strasbg.fr/viz-bin/VizieR?-source=B/corot}. \par 

\section{GIRAFFE observations}
\subsection{Observations and reduction of spectra}\label{oaros}
 
A spectroscopic campaign  with the VLT/FLAMES instrumentation at ESO\footnote[2]{Program ID : 086.D-0212(A), (PI, C. Neiner)} was undertaken to obtain spectra of the faint Be stars identified in the LRA02 exofield of CoRoT. Observations were conducted in 2010 (November and December) and in 2011 (January). At least three spectra were obtained for each target; one spectrum with 3600s of exposure time in the MEDUSA MODE at medium resolution, R=6400, in the LR2 setup (396.4 – 456.7 nm), hereafter referred to as the ``blue spectrum," and two or three successive spectra with R=8600 in the LR6 setup (643.8 – 718.4 nm), hereafter referred to as the 
``red spectra," having 780s of exposure time each. A total of 2209 stars have been observed in 18 fields, of which 2189 ones in two blue and red setups, and 19 stars in only one. We thus dispose spectra  for about 19 \% of the stars that were photometrically observed by CoRoT in the LRA02 exofield. GIRAFFE observations have been conducted at a nearly full moon. Consequently, the stellar spectra are most often polluted by telluric absorption lines and the solar spectrum due to the scattered moonlight. We made the reduction of spectra with the use of the ESO GIRAFFE  pipeline in the same way as in \cite{sema2013} (Paper I) for Be stars studied in the first exofields of CoRoT. Then, we used the IRAF package to separate all the spectra contained in each exposure made in 18 GIRAFFE fields (labeled from $a$ to $r$). The reduced spectra were calibrated in wavelength, corrected from heliocentric velocity and cleaned for cosmics. For each star, the red spectra were summed up to increase their S/N .\par 
  
\subsection{Setting up the faint Be star sample  in the LRA02 exofield of CoRoT} 

 In a first step, we made the selection of emission-line stars by inspecting the H$\alpha$ line in the red spectrum. We identified about 40 emission-line stars and confirmed two of them: CoRoT 110751872 (TYC 4808-1005-1) and CoRoT 110751876 (EM*AS 136). Moreover, we observed that the H$\alpha$ line profiles of  many fainter B/A stars are often polluted by night sky emission lines \cite[e.g.,][]{hanuschik_2003} which could bring on confusion with regard to the true nature of these objects. We finally selected only 20 Be stars. Other stars are cool or much fainter stars that have noisy spectra polluted by OH lines \citep{osterbrock_1996} in their H$\alpha$ line profile. \par
  The selected CoRoT Be star candidates are listed in Table~\ref{tbl1}. They have been identified in UCAC4 catalog (Fourth U.S. Naval Observatory CCD Astrograph Catalog) and GAIA DR3 catalog \citep{vallenari2022}. Information on GIRAFFE spectra and conditions of their exposure times are given in Table~\ref{tabr}. The mean signal-to-noise ratios (S/Ns) range from 40 to 400. The fraction of moonlight that contaminates the spectra is also mentioned.\par
  
 In a second step, we investigated the blue GIRAFFE spectra of all Be star candidates to identify the photospheric lines in order to determine their spectral type and the corresponding photospheric parameters. Most spectra were obtained for a full moon (see Table~\ref{tabr}). The exposure times in the blue wavelength range are longer than those in the red one. Therefore, the spectra of the faintest stars ($15<V<17$ mag) were more heavily polluted by sky background in the LR2 setup than in the LR6 one. The contamination due to the moonlight was estimated using the processed data of our objects and of their sky environments, which are available in the ESO Science Spectrum archive. We subtracted the moonlight contribution to the flux of each faint star. Thanks to these corrections, we could confirm the B spectral type of all stars  in Table~\ref{tbl1} (19 stars) and the Be character for 18 of them. \par 
 
\begin{table}
\caption{Be star candidates in the LRA02 exofield of CoRoT} 
\label{tbl1}
\centering
\tabcolsep 2.5pt
{\scriptsize
\begin{tabular}{ccccccl}
\hline\hline
CoRoT ID & N$^o$ & V & R & GAIA DR3 ID & G & SpT\\
         & & mag & mag & & mag  & \\
\hline
{\scriptsize\object{CoRoT 103000272}} & 1 & 13.244 & 13.124 & {\scriptsize 3105803586041696768} & 13.138  & {\scriptsize B8\,V} \\
{\scriptsize\object{ CoRoT 103032255}} & 2 & 13.551 & 13.352 & {\scriptsize 3105810496648942720} & 13.400  & {\scriptsize B7\,II} \\
{\scriptsize\object{ CoRoT 110655185}} & 3 & 15.670 & 15.300 & {\scriptsize 3102172212671872256} & 15.401  & \\
{\scriptsize\object{ CoRoT 110655384}} & 4 & 16.073 & 15.750 & {\scriptsize 3101951348274296704} & 15.862  & \\
{\scriptsize\object{ CoRoT 110655437}} & 5 & 15.590 & 15.640 & {\scriptsize 3102171250599205248} & 15.813  & \\
{\scriptsize\object{ CoRoT 110662847}} & 6 & 14.114 & 13.780 & {\scriptsize 3105813279788040320} & 13.905  &{ \scriptsize B3\,V} \\
{\scriptsize\object{ CoRoT 110663174}} & 7 & 16.464 & 15.900 & {\scriptsize 3102281133042324096} & 15.983  & \\
{\scriptsize\object{ CoRoT 110663880}} & 8 & 14.513 & 13.985 & {\scriptsize 3105801253879634944} & 14.023  & {\scriptsize B8\,I} \\
{\scriptsize\object{ CoRoT 110672515}} & 9 & 16.062 & 15.563 & {\scriptsize 3101912109453608576} & 15.567 &  \\
{\scriptsize\object{ CoRoT 110681176}} & 10 & 12.058 & 11.936 & {\scriptsize 3102777420808111104} & 11.985  & {\scriptsize B8\,V} \\
{\scriptsize\object{ CoRoT 110688151}} & 11 & 11.311 & 11.150 & {\scriptsize 3102105206881278848} & 11.242  & {\scriptsize B6\,III} \\
{\scriptsize\object{ CoRoT 110747131}} & 12 & 14.180 & 14.470 & {\scriptsize 3101888744831685888} & 14.586 & \\
{\scriptsize\object{ CoRoT 110751872}} & 13 & 11.402 & 11.330 & {\scriptsize 3102285324930474624} & 11.374  & {\scriptsize B5\,III} \\
{\scriptsize\object{ CoRoT 110751876}} & 14 & 11.443 & 11.310 & {\scriptsize 3105805651925926784} & 11.324  & \\
{\scriptsize\object{ CoRoT 110752156}} & 15 & 15.875 & 15.341 & {\scriptsize 3102336314781346816} & 15.246  & \\
{\scriptsize\object{ CoRoT 110827583}} & 16 & 15.331 & 14.960 & {\scriptsize 3101956326137090560} & 15.087  & \\
{\scriptsize\object{ CoRoT 300002611}} & 17 & 16.395 & 15.916 & {\scriptsize 3105402234235943424} & 15.643  & \\
{\scriptsize\object{ CoRoT 300002834}} & 18 & 16.004 & 15.445 & {\scriptsize 3105481124194678656} & 15.441  & \\
{\scriptsize\object{ CoRoT 300003290}} & 19 & 15.471 & 15.058 & {\scriptsize 3105385054366798336} & 15.175  & \\ 
\hline
\multicolumn{7}{l}{Notes. Column 2: Star number; Column 3: V magnitude from the AAVSO photometric}\\
\multicolumn{7}{l}{all-sky survey (APAAS) reported  in the UCAC4 Catalog;\ Column 4: magnitude in the R} \\ 
\multicolumn{7}{l}{band from the CoRoT faint stars database (Vizier) On-line Data Catalog: B/CoRoT; Co-}\\ 
\multicolumn{7}{l}{lumn 5: GAIA DR3 ID; Column 6:  magnitude G from the GAIA EDR3 Catalog; Colu-}\\  
\multicolumn{6}{l}{mn 7: Spectral types determined by spectroscopy in \citet{sebastian_2012}.}\\
\hline 
\end{tabular} 
}
\end{table}

\section{Short description of the spectral characteristics of the program Be stars}\label{sdotpbs}

\begin{figure*}[]
\centerline{\includegraphics[scale=1.0]{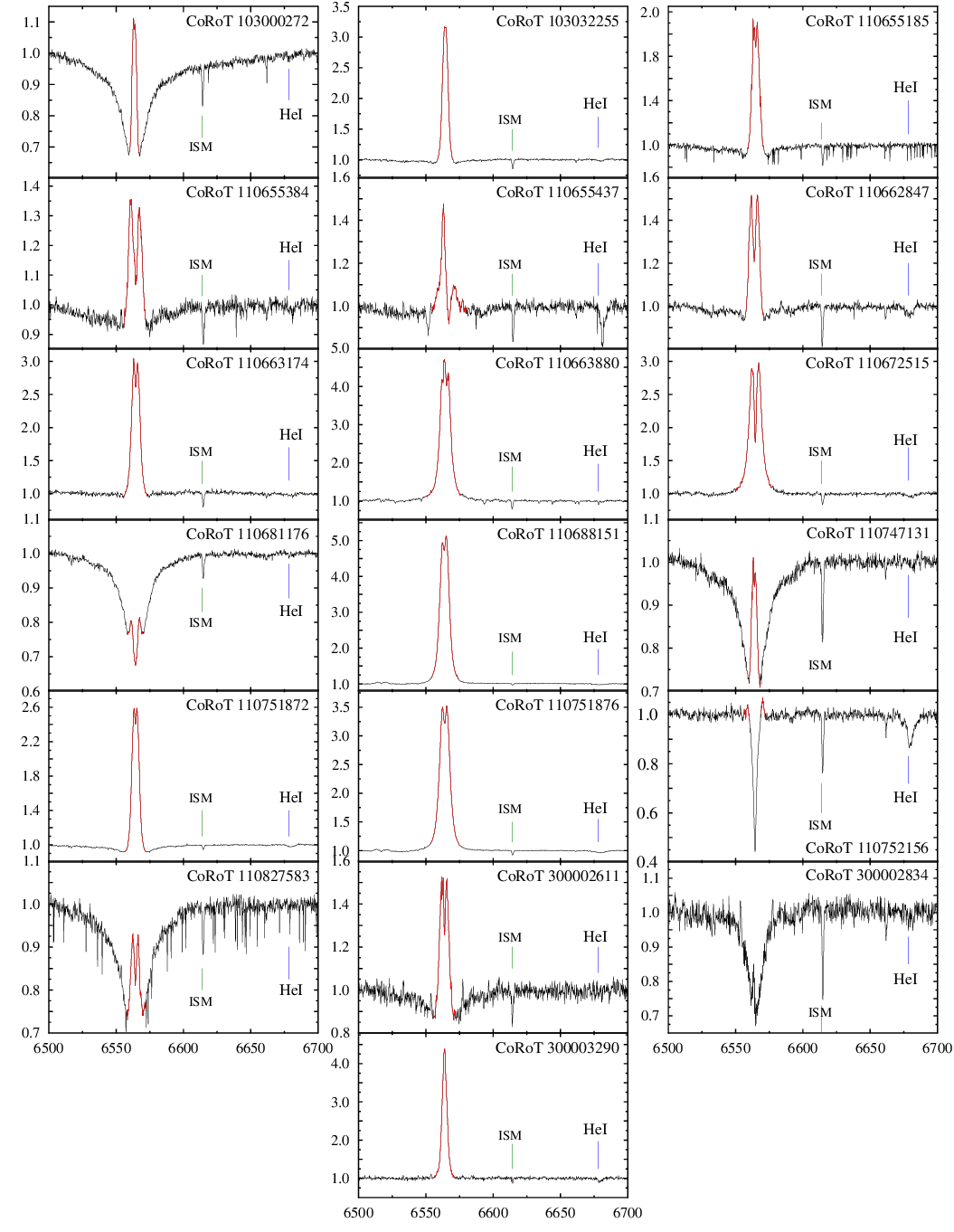}} 
\caption{\label{fig_ha} Spectra in the H$\alpha$ ($\lambda$6562 \AA) line region of the studied Be stars. This figure also indicates the positions of the \ion{He}{i}\,6678 line and the diffuse interstellar absorption band at $\lambda$6614 \AA. The H$\alpha$ emission component
of circumstellar origin is marked in red. The small infilling emission in CoRoT 300002834 is probably not of circumstellar origin.}
\end{figure*}

 We present a short description of the spectral characteristics of each Be star candidate in the LRA02 exofield of CoRoT. Several quantities related with Balmer emission lines and shell features are reported. We note that diffuse interstellar bands (DIBs) are often detected in spectra. Most of them are identified at $\lambda\lambda$ 4232, 4430, 4501, 6614, 6660, 6993, 7060, and 7120 \AA\ (Jenniskens et Desert 1994; Galazutdinov et al. 2000). The spectral region around the H$\alpha$ line from $\lambda$ 6500 to 6700 \AA\ of the program stars is shown in Fig.~\ref{fig_ha}. \par 

{\it CoRoT 103000272}: H$\alpha$ shows a faint, double-peaked emission, which is nearly symmetrical ($V\ge R$, 
$\Delta_{\rm peaks}=50$ km\,s$^{-1}$) and centrally superposed on the broad photospheric line profile. No emission is detected in the blue spectrum. We note the presence of very weak metallic lines (\ion{Fe}{ii}) in the absorption as well as several DIBs. \par

{\it CoRoT 103032255}: H$\alpha$ shows a strong, single and symmetrical emission line profile. The photospheric red lines \ion{He}{i} $\lambda\lambda$ 6678 and 7065 $\AA$ are rather weak. In the blue domain, a very weak emission is present at each side of the core of the photospheric H$\gamma$ line profile. \par 

{\it CoRoT 110655185}:  H$\alpha$ shows a symmetrical and  double-peaked emission line profile ($\Delta_{\rm peaks} = 110$ km\,s$^{-1}$) of moderate intensity superimposed on the photospheric line profile; other wavelength ranges in the red spectrum are polluted by telluric lines. A weak emission is present at each side of the core of the photospheric H$\gamma$ and H$\delta$ line profiles. \par

{\it CoRoT 110655384}: H$\alpha$ depicts a very weak, double-peaked emission with $V\ge R$ and $\Delta_{\rm peaks}=290$ km\,s$^{-1}$ superimposed on the broad photospheric line profile. Weak nebular lines (sky lines) are visible on each H$\alpha$ side. No emission is detected in blue Balmer lines. \par 

{\it CoRoT 110655437}: from its CoRoT light curve, this faint object is known as an eclipsing binary whose period is 3.96685 d \citep{klagyivik2017}. In the red spectrum, H$\alpha$ displays a complex line profile with an asymmetric double-peaked  emission ($V\gg R$, $\Delta_{\rm peaks}=367$ km\,s$^{-1}$) disturbed by two narrow absorptions or shell features, the first one at $RV=+155$ km\,s$^{-1}$, the second at $RV=-540$ km\,s$^{-1}$. \ion{He}{i} $\lambda$ 6678 and 7065 show a double-like structure with a sharp absorption component located at the center of a broad one. Weak nebular lines are present on each side of H$\alpha$. It is difficult to search signatures of a binary component in the blue spectrum, which is noisy and has a lower spectral resolution than the red one. Other comments on this object are presented in Sect.~\ref{tbsnf}.\par

{\it CoRoT 110662847}: H$\alpha$ displays a symmetrical, double-peaked emission ($\Delta_{\rm peaks}=210$ km\,s$^{-1}$) superimposed on the photospheric line profile. Red \ion{He}{i} lines $\lambda\lambda$ 6678 and 7065 are conspicuous. No emission is detected in the blue lines. \par

{\it CoRoT 110663174}: H$\alpha$ shows a nearly symmetrical, double-peaked emission line profile (V$\ge$R, $\Delta_{\rm peaks}=120$ km\,s$^{-1}$) of moderate intensity superposed on the photospheric line profile. The blue spectrum of this faint star is very noisy and  disturbed by the presence of the full moon in the sky at the epoch of the observation. After correction of this spectrum from sky lines contribution, the core of the photospheric H$\gamma$ line profile is disturbed by emission. \par

{\it CoRoT 110663880}: H$\alpha$ displays a strong emission line profile with a complex, possibly triple structure in its core. The blue spectrum of this faint star is very noisy and strongly affected by the presence of the full moon in the sky at the epoch of the observation. After correction of this spectrum from sky lines contribution, the core of the photospheric H$\gamma$ and H$\delta$ line profiles appeared to be disturbed by a double weak emission, itself disturbed by a central shell feature. \par

{\it CoRoT 110672515}: H$\alpha$ displays a symmetrical, double-peaked emission ($\Delta_{\rm peaks}=225$ km\,s$^{-1}$) with a deep H shell feature (RV=+34 kms$^{-1}$). Red He I lines are conspicuous; we also note the presence of a metallic shell, mainly of \ion{Fe}{ii} lines, which depict a weak, double emission on each side of  an absorption feature. H$\gamma$ and H$\delta$ line profiles are affected by emission on each side of their photospheric core. A shell is also present (H and probably metals) in the blue domain. \par

{\it CoRoT 110681176}: H$\alpha$ shows a very weak emission, which is nearly symmetrical ($V\le R$, $\Delta_{\rm peaks}=295$ km\,s$^{-1}$) in the core of the photometric line profile. Blue and red \ion{He}{i} lines are weak. No emission is observed in the blue Balmer lines. \par

{\it CoRoT 110688151}: Among the Be stars of the sample, this one displays a significant contamination of its spectra due to lines of circumstellar origin. H$\alpha$ depicts a strong, closely double-peaked emission which is nearly symmetrical ($V\le R$, $\Delta_{\rm peaks}=120$ km\,s$^{-1}$). Red \ion{He}{i} lines are conspicuous. In the blue domain, the photospheric H$\gamma$ and H$\delta$ line profiles are disturbed in their core by a double-peaked emission. Moreover, the strongest \ion{Fe}{ii} lines of multiplets 27, 28, 37, 38 in the blue spectrum, and of multiplets 40 and 74 in the red one are present in emission with a double structure. Metallic shell lines highly pollute the blue \ion{He}{i} and \ion{Mg}{ii} lines commonly used to determine stellar astrophysical parameters. \par

{\it CoRoT 110747131}: H$\alpha$ depicts a weak, closely doubled-peaked emission, which is nearly symmetrical (V $\geq$ R; $\Delta_{\rm peaks}=70$ km\,s$^{-1}$), in the core of the photospheric line profile. No emission is detected in blue lines. \par

{\it CoRoT 110751872}: H$\alpha$ shows a symmetrical, closely double-peaked emission of moderate intensity ($\Delta_{\rm peaks}=100$ km\,s$^{-1}$) superimposed on the photospheric profile. Red \ion{He}{i} lines are conspicuous. No emission is detected in blue lines. \par

{\it CoRoT 110751876}: This star displays lines formed in the circumstellar medium that contaminate its spectra. H$\alpha$ shows a rather strong, closely double-peaked emission line ($\Delta_{\rm peaks}=140$  km\,s$^{-1}$), which is symmetrical. In the blue domain, Balmer lines are affected by emission on each side of their photospheric core. Blue \ion{He}{i} absorption lines are deep and strong. As in CoRoT 110688151, the stronger lines of the main \ion{Fe}{ii} multiplets are present with a double emission profile in spectra, which makes particularly difficult the fitting of blue \ion{He}{i} and \ion{Mg}{ii} lines. \par 

{\it CoRoT 110752156}: H$\alpha$ shows a very weak photospheric line profile  disturbed on each side by a double-peaked emission (V$\leq$R; $\Delta_{\rm peaks}=525$ km\,s$^{-1}$). Moreover, this complex line presents a deep and sharp core in absorption (H shell, RV = +35 kms$^{-1}$). Red \ion{He}{i} lines are conspicuous, \ion{He}{i} 6678 depicts a double structure that could be explained as due to binarity, DIP or other facts... The blue spectrum of this faint star is strongly affected by the presence of the full moon in the sky at the epoch of the observation. After correction from sky lines contribution, this spectrum remains noisy, however \ion{He}{i} lines are conspicuous as well as a shell H feature in H$\gamma$. \par

{\it CoRoT 110827583}: H$\alpha$ displays a weak, symmetrical and double-peaked emission ($\Delta_{\rm peaks}=170$ km\,s$^{-1}$) in the core of the broad photospheric line profile. A lot of deep and narrow absorption lines, probably of telluric origin, pollute the red spectrum of this faint object. Moreover, weak nebular lines (OH) are also present on each side of the core of H$\alpha$. No emission is detected in the blue domain. After the subtraction of moonlight, the blue spectrum corresponds to a B type star.\par 

{\it CoRoT 300002611}: H$\alpha$ is a weak, symmetrical and double-peaked emission ($\Delta_{\rm peaks}=150$ km\,s$^{-1}$) superimposed to the broad and weak photospheric line. Nebular lines are conspicuous in the vicinity of H$\alpha$. Spectra are noisy, the S/N is poor in the bluer part of the blue domain, however weak blue \ion{He}{i} lines are detected in this star. Moreover, a weak emission seems to be  present at each side of the core of the photospheric H$\gamma$ line profile, as well as in H$\delta$, but with lower intensity. \par

{\it CoRoT 300002834}: Very weak and narrow emission lines (OH) are observed over the wings of the H$\alpha$ photospheric line; nebular lines [\ion{S}{ii}] 6717 and 6731 seem also to be present. The emission component in the H$\alpha$ core could not have a circumstellar origin. Thus, the object is not considered a Be star. We note that fainter early-type stars in the same GIRAFFE field as well as in some others have similar characteristics in their red spectrum.\par 

{\it CoRoT 300003290}: H$\alpha$ shows a strong, single-peaked emission. Very weak sky lines are observed on each side of the line. The photospheric H$\gamma$ line is disturbed by a weak, narrow and single-peaked emission in its core. The same is observed in H$\delta$, but to a lesser degree. As the star is faint, and consequently the  blue spectrum is rather noisy, it is difficult to identify whether circumstellar \ion{Fe}{ii} lines are present or not. \par 

\section{Astrophysical parameters}\label{fp}

  The astrophysical parameters derived here are meant to help to estimate  the masses, ages, and radii of the program stars. They will enable calculations of rotational frequencies to test the pulsation-instability properties of stars and compare them to   theoretical predictions. The determination of the astrophysical parameters implies several steps.\par 
First, a correction from the ``veiling" effect of spectra displaying emission lines is needed, which is induced by the  disk. Then, the veiling-corrected spectral energy distribution (SED) allows us to access to the ``apparent" stellar astrophysical parameters. The apparent parameters are those of classical plane-parallel models of stellar atmospheres that correctly represent the spectral characteristics of the studied spectral wavelength region. In our case, this corresponds to the $\lambda\lambda\,4000-4500$ \AA \ interval, emitted by the observed stellar hemisphere, which is deformed by the rapid stellar rotation and has non-uniform surface distributions of gravity and effective temperature. Finally, using models of stellar atmospheres that account for the effects due to rapid rotation, mostly stellar geometrical deformation and gravity darkening (GD), the apparent astrophysical parameters are translated into parent non-rotating counterpart astrophysical parameters (pnrc), which are ultimately used to determine the stellar mass and its age. In this step, we also derive the $V\!\sin i$ parameter corrected for the GD effect, known as Stoeckley's underestimation. \par 

\subsection{The veiling effect}\label{ve}

  The observed continuum energy distribution of Be stars corresponds to the stellar energy emitted by the central stars, modified by the presence of the CD. According to these cases, the positive or negative flux excess produced by the circumstellar environment can represent a significant fraction of the total flux in the visible spectral domain. The apparent residual intensities of spectral lines are meant to be genuine photospheric components that are usually analyzed to infer the astrophysical stellar parameters; thus, they may not only be affected by emission and absorption components produced in the CD, but also can be over- or under-normalized by the continuum energy distribution modified by the flux excess of circumstellar origin. These perturbations have been known from the Fifties, and are called the ``veiling effect" \citep{jane1959,haw1961}. Since then, this effect has been widely studied in the framework of T-Tauri stars \citep{baba1990}, but also for Be stars \citep[see, e.g.,  ][]{hube1982,ball1995,sema2013}. In order to avoid complex calculations of radiation transfer effects in circumstellar media, which in Be stars are still poorly mastered, we propose taking into account the veiling effect using heuristic expressions that describe the relation between line spectrum $F^{\rm obs}(\lambda)$ and the continuum spectrum $F_{\rm c}^{\rm obs}(\lambda)$ emitted by the star+disk system, with $F^*(\lambda)$ and $F^*_{\rm c}(\lambda)$ attributed to the stellar photosphere. The expression used in this work to carry out the correction of spectra by the veiling effect was discussed in \citet{sema2013}. Some details regarding this correction and the values of the corresponding veiling factors are given in Appendix~\ref{cooavcs}.\par  
 
\subsection{Apparent astrophysical parameters}\label{app_par}

  To derive the apparent astrophysical parameters, plane-parallel model atmospheres were computed for effective temperatures ranging from $T_{\rm eff}=8000$ K to 55\,000 K and gravities $\log g=2.5$ to 4.5 dex.  We computed the temperature structure of the atmospheres with the ATLAS9 FORTRAN program \citep{kur1993,cast1997}. For effective temperatures of 15\,000 K $\leq T_{\rm eff} \leq$ 27\,000 K, the non-LTE level populations were calculated for the considered atoms using TLUSTY \citep{hublan1995} by keeping the temperature and density distributions fixed. We adopted the solar chemical abundances given by \citet{gresau1998}. For model atmospheres with effective temperatures lower than $T_{\rm eff}=15\,000$ K, full LTE was assumed, while those having $T_{\rm eff} > 27\,500$ K are models from the OSTAR2002 non-LTE grid \citep{lanhub2003}. The grid of fluxes we use during the fitting procedure was built with SYNSPEC. Other computing details of models are given in \citet{frem2005,frem2006}.\par  
  The apparent effective temperature and surface gravity are based on the fit of spectral regions not affected by the CE emission/absorption, which encompass wings of hydrogen lines as well as those of helium, carbon, oxygen, silicon, and \ion{Mg}{ii} lines in the spectral domain from 4000 to 4500 \AA. The normalized spectra of the studied stars, 
$\Phi^*(\lambda)$, namely, those corrected for the veiling effect with Eq.~(\ref{veil3}), are Doppler-shifted according to the radial velocity. The fit of the veiling-corrected spectra is carried out by comparing them with a grid of synthetic spectra using the robust least squares method employed in the MINUIT minimization package developed at CERN. This package was transcribed into a FORTRAN language and is named GIRFIT \citep{frem2006}. The synthetic spectra were convoluted with a Gaussian function for the required instrumental resolution.\par 
 
   In the most general form, the fitting procedure of spectra with MINUIT considers  as an unknown the parametric set $(T_{\rm eff},\log g, V\!\sin i,V_{\rm Z})$, where $V_{\rm Z}$ is the radial velocity not corrected for the Earth movement around the Sun. The fitting procedure is controlled with the $\chi2$ test computed over zones in the 4000 to 4500 Å wavelength range where the line circumstellar emissions and shell absorptions and the diffuse interstellar absorption bands are carefully avoided. For three studied stars (stars Nos. 4, 13, and 14) we determined the $V\!\sin i$ using the Fourier transform
(FT) method, which was then considered as a fixed  quantity. The FT method was applied to the lines \ion{He}{i} 4026 and  4472 and \ion{Mg}{ii} 4481\ \AA. These objects are the only ones in our sample for which the FT $V\!\sin i$ parameters issued from the three spectral lines are: on the one hand, mutually consistent within the average GIRFIT uncertainty and, on the other hand, consistent with the value obtained with the GIRFIT fitting method. \par
   We note that in spite of the robust character of the MINUIT algorithm, fits of equal quality according to the $\chi^2$ test with slightly different values of the set ($T_{\rm eff},\log g,V\!\sin i$) can be obtained; that is to say: the noise introduces degeneracies in the combination of different parameters, in particular, between $\log g$ and $V\!\sin i$. This phenomenon is worth to be considered in some detail since $M_{\rm V}$ and $\log L/L_{\odot}$ rely sensitively on the value of $\log g$, mainly in late B spectral types. We then introduced a refining protocol to determine the astrophysical parameters that is detailed in the following section.\par 

\subsection{Refined astrophysical parameters}\label{refined_par} 

 In the refining operation of astrophysical parameters, we used the following BCD (Barbier-Chalonge-Divan) calibrations: $M_{\rm V}=M_{\rm V}(\lambda_1,D)$ \citep{zobri1991}, $T_{\rm eff}=T_{\rm eff}(\lambda_1,D)$ \citep{zorcid2009}, and $\log g=\log g(\lambda_1,D)$ \citep{zorec1986}. To this panoply of relations, we added the expression for the color excesses $E(B-V)=[(B-V)_{\rm obs}-(B-V)_*]$, where the intrinsic color index 
$(B-V)_*$ is a function of $(T_{\rm eff},\log g)$. These come from the synthetic UBV colors updated by Castelli F. in 2011 of the model atmospheres published in \citet{caskur2003}. We employed the photometric quantities $V_{\rm obs}$ and $(B-V)_{\rm obs}$ derived from the $(G,BP,RP)$ magnitudes of the GAIA photometric system \citep{riello2021}, which represent a uniform photometric data base for our program stars. This uniformity is less obvious among the UBV photometric data proper.\par
  We assumed that the corrections $\Delta T_{\rm eff}$ and $\Delta\log g$ meant to be applied to the first determination of ($T_{\rm eff},\log g$) can be obtained from the deviations $\Delta M_{\rm V}$ and $\Delta E(B-V)$ that best approach the condition $\lim (\Delta d/d)\to 0$ derived from Pogson's formula \citep{pogson1856}: 

\begin{equation}
\Delta d/d=0.46[\Delta M_{\rm V}+R_{\rm V}\Delta E(B-V)]
\label{dmde}
.\end{equation}
 
\noindent In Eq.~(\ref{dmde}), $d$ is the stellar distance and $R_{\rm V}=A_{\rm V}/E(B-V)=3.1$ because no other values of $R_{\rm V}$ toward the direction of our program stars are known. In this treatment, the color excess $E(B-V)_{\rm T}=(B-V)_{\rm obs}-(B-V)(T_{\rm eff},\log g)$  has two components:

\begin{equation}
\displaystyle E(B-V)_{\rm T} = E(B-V)_{\rm ISM} + E(B-V)_{\rm CD},
\label{ebv_t}
\end{equation}
 
\noindent where $E(B-V)_{\rm ISM}$ is the interstellar extinction proper and $E(B-V)_{\rm CD}$ corresponds to the reddening produced on the Paschen continuum by the CD.  We estimated $E(B-V)_{\rm CD}$ with the empirical relation given as a function of the equivalent width $W_{\alpha}$ of the H$\alpha$ line emission component used in \citet{gko2016}, which is based on previous determinations by \citet{radd2013} and \citet{dach1988}. Because the magnitude excess $\Delta V_{\rm CD}$ in the $V$ magnitude due to the CD, the reddening $E(B-V)_{\rm CD}$ does not conform to the law $R_{\rm V}=A_{\rm V}/E(B-V)=3.1$ that is otherwise valid for the interstellar medium, we must dissociate the circumstellar reddening from the interstellar one and apply it directly to $V_{\rm obs}$. While $E(B-V)_{\rm ISM}$ was estimated using Eq.~(\ref{ebv_t}) and the noted relation between $E(B-V)_{\rm CD}$ and  $W_{\alpha}$, the magnitude excess $\Delta V_{\rm CD}$ was obtained from Eq.~(8) in \citet{zobri1991}.\par
  When dealing with Eq.~(\ref{dmde}), knowledge of $d$ is never required. We simply parameterized the ratio $\Delta d/d$ from $-0.5$ to $+0.5$, and for each $\Delta d/d$ we tested a series of values $\Delta M_{\rm V}$ and $\Delta E(B-V), $  which (via the above BCD calibrations) enable us to calculate the corrections $\Delta T_{\rm eff}$ and $\Delta \log g$. The chosen fit with GIRFIT then corresponds  to the smallest possible  value found of $|\Delta d/d|$ that leads to the best $\chi^2$ test. From this fit, we get not only the apparent astrophysical parameters ($T_{\rm eff},\log g,V\!\sin i)$, but also $M_{\rm V}$ and $E(B-V)_{\rm ISM}$. In most cases, it is $|(\Delta d/d)|\lesssim0.05$. \par
  Specifically, BCD calibrations of astrophysical parameters are displayed in tabular forms and can be reverted to: $M_{\rm V}=M_{\rm V}(T_{\rm eff},\log g)$ and $E=E(B-V)(T_{\rm eff},\log g)$, from which  the required relations readily follow:

\begin{eqnarray}
\begin{array}{rcl}
\displaystyle \Delta M_{\rm V} &  = & \displaystyle (\partial M_{\rm V}/\partial T_{\rm eff})\Delta T_{\rm eff}+ (\partial M_{\rm V}/\partial \log g)\Delta\log g,\\
\displaystyle \Delta E &  = & \displaystyle (\partial E/\partial T_{\rm eff})\Delta T_{\rm eff}+ (\partial E/\partial \log g)\Delta\log g, \\
\end{array}
\label{corr-tg}
\end{eqnarray}   

\noindent needed to derive $\Delta T_{\rm eff}$ and $\Delta\log g$ from 
$\Delta M_{\rm V}$ and $\Delta E(B-V)$. The best fits of the observed spectra previously corrected for the veiling effect obtained with GIRFIT,  subject to the noted iteration procedure, are shown in Appendices as Figs.~\ref{fig_fit1}, \ref{fig_fit2}, \ref{fig_fit3}, and \ref{fig_fit4}. The apparent astrophysical parameters derived are given in Table~\ref{par_app}. We refer to them as the ``primary apparent astrophysical parameters." \par 
  Due to the noted subtleties that have to be taken into account to calculate the spectroscopic distances of Be stars, it is worth specifying Pogson's formula, as it is used in the present work: 
 
 \begin{eqnarray}
\left. \begin{array}{rcl}
\displaystyle \log d_{\rm spect.} &  = & \displaystyle 0.2\,(V_*-M_{\rm V})+1.0 \\
\displaystyle V_* &  = & \displaystyle (V_{\rm obs}-\Delta V_{\rm CD})-3.1\,E(B-V)_{\rm ISM}
\end{array}
\right\rbrace.
\label{pog}
\end{eqnarray}  

  Once the final couple ($T_{\rm eff},\log g$) was adopted, we calculated the uncertainties associated to the estimated distances marked in tables of results, which depend only on the  measurement errors of $V_{\rm obs}$ and 
$(B-V)_{\rm obs}$, $E(B-B)_{\rm ISM}$ and $I_{\alpha}$. Each of these parameters is represented as $X+\epsilon_{\rm X}$, where the error 
$\epsilon_{\rm X}$ is simulated through a Monte Carlo procedure. The errors $\epsilon_{\rm X}$ are assumed having a normal distribution characterized by the respective standard dispersion $\sigma_{\rm X}$ reported in our tables. The reported values of $d_{\rm spect.}$ in  Table~\ref{l1d_dist} are the averages $\langle d_{\rm spect.}\rangle$ of $10^4$ estimates, and $\sigma_{\rm d}$ represents their standard deviation. The same procedure was adopted also for $\langle d_{\rm GAIA}\rangle=\langle 1/(\pi+\epsilon_{\pi})\rangle$, where $\epsilon_{\pi}$ is considered to have a normal distribution with the standard deviation published for each GAIA DR3 parallax $\pi$ \citep{gaia_a,gaia_b,gaia_c}. The errors associated to $M_{\rm V}$ in table Table~\ref{l1d_dist} are also due to $V_{\rm obs}$, $(B-V)_{\rm obs}$, $E(B-B)_{\rm ISM}$ and $I_{\alpha}$. \par

\begin{table*}[]
\centering
\caption[]{\label{par_app} Primary and secondary apparent astrophysical parameters.}
\tabcolsep 5.0pt
\begin{tabular}{ccccc|crcrc}
\hline\hline
\noalign{\smallskip}
Star & $T_{\rm eff}$ & $\log g$ & $V\!\sin i$&  $V_{\rm Z}$ & $\log L/L_{\odot}$ & $M/M_{\odot}$ 
\ \ \ \ & $t/t_{\rm MS}$ & $R/R_{\odot}$ \ \ \ \ & $V_{\rm c}$\\
 & $\pm\sigma_{T_{\rm eff}}$ & $\pm\sigma_{\log g}$ & $\pm\sigma_{V\!\sin i}$ & $\pm\sigma_{V_{\rm Z}}$ &
 $\pm\sigma_{\log L/L_{\odot}}$ & $\pm\sigma_{M/M_{\odot}}$ \ \ \ \ & $\pm\sigma_{t/t_{\rm MS}}$ & $\pm\sigma_{R/R_{\odot}} \ \ \ $ & $\pm\sigma_{V_{\rm c}}$ \\
 & K &   dex  &    km\,s$^{-1}$ &  km\,s$^{-1}$ & & & & & km\,s$^{-1}$\\ 
\noalign{\smallskip}\hline
\noalign{\smallskip}
 1 & 11450 $\pm$ \ \ 570  & 3.84 $\pm$ 0.12 &  120 $\pm$ 13 & \ \ 38.2 $\pm$ \ \ 2.9 
 & 2.318 $\pm$ 0.101 &  3.44 $\pm$ 0.17 & 0.88 $\pm$ 0.06 &  3.66 $\pm$ 0.35 & 348 $\pm$ 11  \\ 
  
 2 & 13670 $\pm$ \ \ 690  & 3.56 $\pm$ 0.12 &  230 $\pm$ 25 & \ \  75.8 $\pm$ \ \ 3.2
 & 3.076 $\pm$ 0.105 &  5.07 $\pm$ 0.30 & 1.03 $\pm$ 0.01 &  6.15 $\pm$ 0.62 & 324 $\pm$ 09  \\
  
 3 & 11180 $\pm$ \ \ 600  & 3.58 $\pm$ 0.13 &  220 $\pm$ 25 & 106.0 $\pm$ \ \ 3.2
 & 2.584 $\pm$ 0.110 &  3.77 $\pm$ 0.22 & 1.02 $\pm$ 0.02 &  5.22 $\pm$ 0.55 & 304 $\pm$ 10  \\
  
 4 & 13610 $\pm$ \ \ 659  & 4.22 $\pm$ 0.11 &  370 $\pm$ 40 & \ \  69.1 $\pm$ 11.9
 & 2.264 $\pm$ 0.094 &  3.60 $\pm$ 0.17 & 0.30 $\pm$ 0.16 &  2.43 $\pm$ 0.21 & 442 $\pm$ 13  \\
  
 5 & 16030 $\pm$ \ \ 740  & 3.70 $\pm$ 0.11 &  230 $\pm$ 22 & 117.0 $\pm$ \ \ 8.5
 & 3.288 $\pm$ 0.092 &  6.01 $\pm$ 0.29 & 0.97 $\pm$ 0.03 &  5.71 $\pm$ 0.49 & 368 $\pm$ 11  \\
 
 6 & 15590 $\pm$ \ \ 779 & 3.47 $\pm$ 0.11 &  250 $\pm$ 25 & \ \  47.1 $\pm$ \ \ 6.6
 & 3.513 $\pm$ 0.102 & 6.62 $\pm$ 0.42 & 1.02 $\pm$ 0.00 & 7.83 $\pm$ 0.75 & 330 $\pm$ 08  \\
  
 7 & 14030 $\pm$ \ \ 730  & 3.88 $\pm$ 0.13 &  270 $\pm$ 30 & \ \ 56.6 $\pm$ \ \ 2.7
 & 2.752 $\pm$ 0.105 &  4.50 $\pm$ 0.24 & 0.85 $\pm$ 0.07 &  4.02 $\pm$ 0.41 & 381 $\pm$ 13  \\
  
 8 & 22830 $\pm$ 1187  & 3.76 $\pm$ 0.15 &  200 $\pm$ 39 & \ \  45.9 $\pm$ \ \ 3.5
 & 4.095 $\pm$ 0.127 & 10.70 $\pm$ 0.84 & 0.89 $\pm$ 0.06 &  7.13 $\pm$ 0.89 & 442 $\pm$ 16  \\
 
 9 & 14210 $\pm$ \ \ 720  & 3.37 $\pm$ 0.10 &  250 $\pm$ 27 &  \ \ 57.2 $\pm$ \ \ 0.7
 & 3.423 $\pm$ 0.097 &  6.21 $\pm$ 0.38 & 1.03 $\pm$ 0.00 & 8.50 $\pm$ 0.75 & 307 $\pm$ 07  \\
  
10 & 10430 $\pm$ \ \ 550  & 3.59 $\pm$ 0.12 &  220 $\pm$ 24 & \ \  76.0 $\pm$ \ \ 2.1
 & 2.399 $\pm$ 0.103 &  3.38 $\pm$ 0.18 & 1.02 $\pm$ 0.02 &  4.85 $\pm$ 0.47 & 299 $\pm$ 09  \\
  
11 & 14300 $\pm$ \ \ 700 & 4.00 $\pm$ 0.13 &  300 $\pm$ 33 & \ \  61.5 $\pm$ \ \ 2.0
 & 2.652 $\pm$ 0.105 &  4.33 $\pm$ 0.23 & 0.72 $\pm$ 0.11 &  3.45 $\pm$ 0.35 & 405 $\pm$ 14  \\
 
12 & 11650 $\pm$ \ \ 580  & 3.58 $\pm$ 0.14 &  200 $\pm$ 20 & \ \  46.8 $\pm$ \ \ 4.6
 & 2.673 $\pm$ 0.113 &  3.98 $\pm$ 0.24 & 1.02 $\pm$ 0.02 &  5.33 $\pm$ 0.61 & 309 $\pm$ 11  \\
 
13 & 13750 $\pm$ \ \ 690  & 3.82 $\pm$ 0.13 &  \ 260 $\pm$ 29 & \ \ 78.8 $\pm$ \ \ 0.3
 & 2.777 $\pm$ 0.107 &  4.50 $\pm$ 0.24 & 0.90 $\pm$ 0.06 &  4.31 $\pm$ 0.44 & 368 $\pm$ 12  \\
  
14 & 22160 $\pm$ 1140  & 4.00 $\pm$ 0.15 &  325 $\pm$ 36 & \ \ 60.5 $\pm$  \ \ 4.3
 & 3.714 $\pm$ 0.121 &  8.75 $\pm$ 0.58 & 0.65 $\pm$ 0.13 &  4.88 $\pm$ 0.59 & 484 $\pm$ 19  \\
 
15 & 21830 $\pm$ 1500  & 3.58 $\pm$ 0.17 &  285 $\pm$ 31 & \ \ 57.2 $\pm$ \ \ 3.8
 & 4.206 $\pm$ 0.156 & 11.11 $\pm$ 1.17 & 0.99 $\pm$ 0.04 & 8.87 $\pm$ 1.34 & 401 $\pm$ 17  \\
 
16 & 10790 $\pm$ \ \ 550 & 3.97 $\pm$ 0.15 &  220 $\pm$ 24 & \ \ 68.9 $\pm$  \ \ 3.4
 & 2.021 $\pm$ 0.116 &  2.95 $\pm$ 0.17 & 0.76 $\pm$ 0.11 &  2.93 $\pm$ 0.34 & 360 $\pm$ 14  \\
 
17 & 12000 $\pm$ \ \ 600 & 4.04 $\pm$ 0.14 &  350 $\pm$ 38 & \ \ 57.4 $\pm$  \ \ 6.4
& 2.192 $\pm$ 0.112 & 3.30 $\pm$ 0.19 & 0.66 $\pm$ 0.13 & 2.88 $\pm$ 0.33 & 386 $\pm$ 14  \\

18 & 13490 $\pm$ \ \ 670 & 3.57 $\pm$ 0.11 &  225 $\pm$ 24 & \ \ 86.8 $\pm$  \ \ 6.4
& 3.034 $\pm$ 0.097 & 4.94 $\pm$ 0.27 & 1.03 $\pm$ 0.01 & 6.03 $\pm$ 0.54 & 325 $\pm$ 09  \\
 
19 & 17200 $\pm$ 1270 & 3.40 $\pm$ 0.16 & 100 $\pm$ 12 & \ \ 72.1 $\pm$  \ \ 6.9
& 3.851 $\pm$ 0.135 & 8.27 $\pm$ 0.76 & 1.02 $\pm$ 0.00 & 9.48 $\pm$ 1.38 & 336 $\pm$ 11  \\
\hline
\end{tabular} 
\end{table*}

\subsection{Secondary apparent astrophysical parameters}\label{der_par}

  We use the term ``secondary apparent astrophysical parameters" to apply to all those that tightly depend on the ``primary apparent astrophysical parameters" and were not obtained from the fitting operation of spectra. Some of them also rely on models of stellar evolution. These are: the BCD parameters $(\lambda_1,D)$, the bolometric luminosity, $\log L/L_{\odot}$, the stellar mass, $M/M_{\odot}$, equatorial radius, $R/R_{\odot}$, the equatorial linear critical velocity, $V_{\rm c}$, and the fractional stellar age, $t/t_{\rm MS}$ ($t$ is the age and $t_{\rm MS}$ is the time that a non rotating star spends in the main sequence evolutionary phase). \par 
  The BCD quantities $(\lambda_1,D)$ are determined from the adopted ($T_{\rm eff},\log g$) parameters, which in turn lead to the BCD spectral classification of the studied Be stars \citep{zobri1991,zorcid2009}. This set of secondary apparent parameters is given in Table~\ref{l1d_dist}.\par
  Adopting the bolometric correction $BC(T_{\rm eff})$ by \citet{flow1996}, to avoid possible conflicts with the zero point of bolometric corrections, we determined the bolometric ``apparent" absolute magnitude $M_{\rm bol}$ following the prescription from \citet{torr2010}:  
  
\begin{equation}
\displaystyle M_{\rm bol} = M^{\odot}_{\rm bol}+[M_{\rm V}-M^{\odot}_{\rm V}]+[BC(T_{\rm eff})-BC(\odot)].
\label{mv_bol}
\end{equation} 
 
\noindent From this, we can derive the bolometric luminosity, $\log L/L_{\odot}$. Once $\log L/L_{\odot}$ and $T_{\rm eff}/T^{\odot}_{\rm eff}$ are determined, we estimate the stellar radius $R/R_{\odot}$. Finally,
by interpolating in the grids of stellar evolutionary models without rotation given by \citet{geor2013} for solar metallicity Z = 0.014, we obtain the stellar masses $M/M_{\odot}$ and its fractional ages $t/t_{\rm MS}$. All these secondary apparent astrophysical parameters are reported in Table~\ref{par_app} with their uncertainties.\par
\begin{figure}[]
\centerline{\includegraphics[scale=0.9]{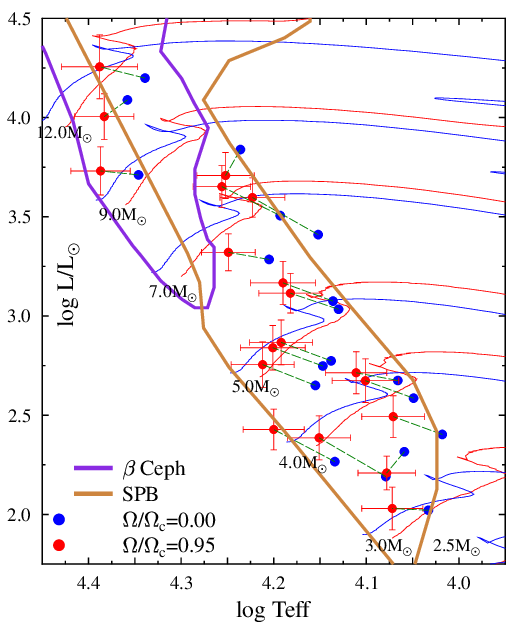}}  
\caption{\label{hr_diag} HR diagram of the studied stars. Evolutionary tracks are from \citet{geor2013} for $\Omega/\Omega_{\rm c}=0.0$ (blue) and for rotation rates at the ZAMS $\Omega/\Omega_{\rm c}=0.95$ (red). Blue points: Stars with their apparent parameters; red points: Stars with their pnrc parameters obtained for $\Omega/\Omega_{\rm c}=0.95$. Instability strips of $\beta$~Ceph- ({\color{violet}\rule{0.8cm}{0.1cm}}) and SPB-type ({\color{brown}\rule{0.8cm}{0.1cm}}) pulsators adapted from \citet{walczak2015} for OPLIB opacities and metallicity of $Z=0.015$.}
\end{figure}

 
 To complete the presentation of the program objects, we show their HR diagram in Fig.~\ref{hr_diag}, where we have included the instability zones calculated by \citet{walczak2015} for non rotating objects using the OPLIB new Los Alamos opacities. Similar calculations can be found in \citet{miglio_2007,burssens2020} for other opacities, and rotating  stars in \citet{saio2017} and \citet{szewczuk2017}, where  the widening of the instability regions due to stellar rotation can be seen. \par

\begin{table*}[]
\centering
\caption[]{\label{l1d_dist} Indirect apparent quantities, interstellar reddening, and distances.}
\begin{tabular}{rcclccccrr} 
\hline\hline
\noalign{\smallskip}
Star & $D$ & $\lambda_1$ & MK & $V_{\rm obs}$ & $ (B-V)_{\rm obs}$ & $M_{\rm V}\pm\sigma_{\rm V}$ & 
$E(B-V)_{\rm ISM}\pm\sigma_{\rm E_{\rm ISM}}$ & $d_{\rm spect.}\pm\sigma_{\rm d}$  & $d_{\rm GAIA}\pm\sigma_{\rm d}$\\
     & dex &   \AA       &          &  mag       &  mag & mag  &  mag & pc & pc \\
\noalign{\smallskip}\hline 
   1 & 0.415 & 45.8 & B9\,IV      & 13.259 &  0.442 & -0.40 $\pm$ 0.12 &  0.525 $\pm$ 0.049 &  2556 $\pm$ 179  &  2540 $\pm$ \ \ \ \ 86  \\
   2 & 0.308 & 34.6 & B7\,III     & 13.526 &  0.452 & -1.90 $\pm$ 0.12 &  0.554 $\pm$ 0.047 &  5425 $\pm$ 365  &  5175 $\pm$ \ \ 391  \\
   3 & 0.425 & 36.1 & B9\,III     & 15.698 &  0.782 & -1.13 $\pm$ 0.13 &  0.841 $\pm$ 0.052 &  6858 $\pm$ 469  &  6572 $\pm$ 1936 \\
   4 & 0.310 & 61.6 & B6-7\,III   & 16.048 &  0.582 & +0.17 $\pm$ 0.12 &  0.692 $\pm$ 0.049 &  5584 $\pm$ 377  &  5433 $\pm$  1241 \\
   5 & 0.252 & 39.9 & B5\,III     & 16.233 &  0.966 & -1.97 $\pm$ 0.14 &  1.119 $\pm$ 0.051 &  8768 $\pm$ 576  &  8774 $\pm$  2741 \\
   6 & 0.242 & 34.9 & B5\,III     & 14.105 &  0.612 & -2.65 $\pm$ 0.13 &  0.760 $\pm$ 0.049 &  7509 $\pm$ 498  &  7393 $\pm$  1114 \\
   7 & 0.313 & 45.0 & B6\,IV      & 16.391 &  0.950 & -0.97 $\pm$ 0.14 &  1.046 $\pm$ 0.052 &  6574 $\pm$ 439  &  6464 $\pm$  1757 \\
   8 & 0.144 & 50.0 & B2\,III-IV  & 14.479 &  1.015 & -3.09 $\pm$ 0.18 &  1.167 $\pm$ 0.052 &  6092 $\pm$ 403  &  5972 $\pm$ \ \ 698  \\
   9 & 0.268 & 32.5 & B6\,III     & 15.872 &  0.791 & -2.61 $\pm$ 0.12 &  0.886 $\pm$ 0.051 & 13862 $\pm$ 925  & 13603 $\pm$  4457 \\
  10 & 0.472 & 37.7 & B9.5\,III   & 12.059 &  0.301 & -0.80 $\pm$ 0.12 &  0.363 $\pm$ 0.050 &  2227 $\pm$ 162  &  2184 $\pm$ \ \ \ \ 84  \\
  11 & 0.302 & 50.5 & B6\,III-IV  & 11.284 &  0.167 & -0.66 $\pm$ 0.12 &  0.221 $\pm$ 0.046 &  1758 $\pm$ 124  &  1706 $\pm$ \ \ \ \ 60  \\
  12 & 0.399 & 35.7 & B9\,III     & 14.834 &  0.698 & -1.26 $\pm$ 0.12 &  0.791 $\pm$ 0.051 &  5344 $\pm$ 366  &  5326 $\pm$ \ \  751  \\
  13 & 0.322 & 43.0 & B6-7\,IV    & 11.418 &  0.174 & -1.11 $\pm$ 0.14 &  0.274 $\pm$ 0.045 &  2135 $\pm$ 148  &  2073 $\pm$ \ \ \ \ 80  \\
  14 & 0.164 & 57.4 & B2.5\,V     & 11.411 &  0.344 & -2.17 $\pm$ 0.14 &  0.510 $\pm$ 0.047 &  2489 $\pm$ 169  &  2510 $\pm$ \ \  129  \\
  15 & 0.143 & 44.6 & B2\,III     & 15.855 &  1.211 & -3.59 $\pm$ 0.19 &  1.434 $\pm$ 0.053 & 10000 $\pm$ 648  &  9476 $\pm$  2950 \\
  16 & 0.445 & 55.7 & B9\,IV      & 15.304 &  0.641 & +0.21 $\pm$ 0.13 &  0.698 $\pm$ 0.052 &  3857 $\pm$ 276  &  3829 $\pm$ \ \  400  \\
  17 & 0.379 & 54.5 & B7-8\,V     & 16.454 &  1.444 & +0.02 $\pm$ 0.12 &  1.521 $\pm$ 0.055 &  2185 $\pm$ 142  &  2260 $\pm$ \ \  186  \\
  18 & 0.316 & 34.8 & B7\,III     & 15.929 &  1.057 & -1.81 $\pm$ 0.12 &  1.189 $\pm$ 0.052 &  6482 $\pm$ 424  &  6255 $\pm$ 1463 \\
  19 & 0.202 & 36.7 & B4\,III     & 15.557 &  0.910 & -3.21 $\pm$ 0.14 &  1.054 $\pm$ 0.051 & 12481 $\pm$ 819  & 13504 $\pm$ 4243 \\
\hline
\end{tabular}
\end{table*}

\subsection{Astrophysical parameters corrected for effects carried by the rapid rotation}\label{rot_eff}

  We consider that the stars are rigid rotators characterized by a uniform surface rate of angular velocity ratio $\Omega/\Omega_{\rm c}$, where $\Omega_{\rm c}$ is the surface angular velocity at critical rotation. The apparent astrophysical parameters obtained in Sect.~\ref{app_par} can then be given according to the following formal expression:

\begin{equation}
\displaystyle P_{\rm app} = P_{\rm pnrc}\,F_{\rm P}(M,t,\Omega/\Omega_{\rm c},i),
\label{pnrc}
\end{equation}

\noindent where $P$ stands for parameters as effective temperature, $T_{\rm eff}$, surface gravity, $g$, bolometric luminosity, $L$, and projected rotational velocity, $V\!\sin i$. On the left-hand side of Eq.~(\ref{pnrc}), these quantities are considered ``apparent," that is, they are issued from models of stellar atmospheres and evolution without rotation that fit the observed spectral domain. On the right side of Eq.~(\ref{pnrc}), the parameters $P$ correspond to the parent-non-rotating-counterparts \citep[$pnrc$;][]{frem2005}, which are multiplied by functions that represent the effects carried by a rapid rotation in an object of actual mass, $M,$ and age, $t$, whose surface angular velocity is $\Omega$ and its rotational axis is seen according to an aspect angle, $i$. \par 
 The functions $F_{\rm P}(M,t,\Omega,i)$ in Eq.~(\ref{pnrc}) were calculated with FASTROT \citep{frem2005} as follows. For a series of stellar masses, ages, and rotational rates, we calculated 2D models of stellar structure assuming rigid rotation over the whole star \citep{zor2011,zor2012}. The geometrical deformations of stars induced by the rapid rotation produce a surface gravity that depends on the latitude and a non uniform distribution of the effective temperature, which is calculated by the gravity darkening formula given by \citet{rieu2011}. Using local plane-parallel model atmospheres, we recomposed the apparent stellar spectra in the $\lambda\lambda 4000-4500$ \AA \ wavelength region for objects of several masses, ages, angular velocity rates $\Omega/\Omega_{\rm c}$, and inclination angles, $i$. The system of equations in Eq.~(\ref{pnrc}) is given in a tabular form whose solutions for the unknown quantities $(M,t,\Omega/\Omega_{\rm c},i)$ are obtained by iteration. Since only $T_{\rm eff}$, $\log g$ and $V\!\sin i$ as entry apparent parameters can be considered independent, we solved these equations for a series of imposed values of $\Omega/\Omega_{\rm c}$. Due to gravitational darkening, the observed (apparent) $V\!\sin i$ are underestimated \citep{sto1968,tow2004,
frem2005}. The amount of this underestimation is automatically taken into account in the solution of Eq.~(\ref{pnrc}). The evolutionary tracks of rotating stars used were calculated by \citet{geor2013} and are given in terms of effective temperatures and bolometric luminosity averaged over the surface of the geometrically deformed rotating objects. This imposes that at each iteration step, the corresponding averaged quantities over the stellar surface are calculated from the running pnrc parameters. Moreover, at each iteration step we need to determine the rotational velocity of stars in the ZAMS to identify the right evolutionary track needed to interpolate the stellar mass and age at the running stellar age, which is consistent with the imposed rotational rates $\Omega/\Omega_{\rm c}$. Once a cycle of iterations is over, we calculate the rotational frequency, $\nu_{\rm r}$, given by:  

\begin{equation}
\displaystyle \nu_{\rm r} = 0.02\,\left[\frac{V_{\rm eq}(\Omega/\Omega_{\rm c},M,t)}{R_{\rm eq}(\Omega/\Omega_{\rm c},M,t)}\right] \ \ \ \ {\rm cycles/day},
\label{nur}
\end{equation}

\noindent where $V_{\rm eq}$ is the equatorial linear velocity given in km\,s$^{-1}$ and $R_{\rm eq}$ is the equatorial radius in solar units of the star distorted by its rotation.\par 
  Each entry parameter in Eq.~(\ref{pnrc}) is considered with its uncertainty assumed having a normal (Gaussian) distribution. We produced $10^4$ Monte Carlo trials of all entry parameters and each time we sought a new solution for the system using Eq.~(\ref{pnrc}). Sometimes, the set of entry parameters ($T_{\rm eff}\pm\epsilon_{T_{\rm eff}},\log g\pm\epsilon_{\log g},\log L/L_{\odot}\pm\epsilon_{\log L},V\!\sin i\pm\epsilon_{V\!\sin i}$) may not correspond to realistic objects. In such cases, the functions $F_{\rm P}(M,t,\Omega,i)$ cannot produce a reliable solution and the trial is considered as having no solution. Although the uncertainties of the entry parameters are assumed to have normal distributions, the distributions of the solution quantities do not always have normal distributions. So, the averages of parameters with non symmetrical distribution do not coincide with the mode of the distribution (mode = parameter corresponding to the maximum of the distribution). In order to illustrate the obtained differences in the estimated parameters, we produced two tables of results: one of them gives the average values of solutions with their classical standard $1\sigma$ deviations, while the other is for modes, where the ``sigmas" correspond to the uncertainties related to the identification of the modes. Similar solutions to the system  Eq.~(\ref{pnrc}) have been previously done in \citet{huhi2000,hub2003,
frem2005,zor2005,vin2006,mart2006,mart2007,sema2013,zorec_2016}. In Appendix \ref{aparfcfreotobe}, Tables~\ref{corr_par_prom_t} and \ref{corr_par_mode_t} give the averages and modes of distributions of astrophysical parameters and rotational frequencies corrected for rotational effects assuming $\Omega/\Omega_{\rm c}=0.95$, respectively. These tables give: the pnrc effective temperature, $T_{\rm eff}$, $\log g$, true $V\!\sin i$, pnrc bolometric luminosity in solar units, $\log L/L_{\odot}$, mass $M/M_{\odot}$, critical linear equatorial velocity, $V_{\rm c}$, estimated inclination angle, $i$, true fractional age, $t/t_{\rm MS}$, the true stellar age, $t$, and the rotational frequency, $\nu_{\rm r}$, in cycles/day. All these quantities are given with their respective $1\sigma$ dispersion calculated from the obtained distribution of solutions. Similar solutions obtained for imposed values $\Omega/\Omega_{\rm c} = 0.8, 0.9, 0.999,$ and 1.0 are given in Tables~D.1 and D.2, respectively, accessible as online data.\par  

\section{Some empirical relations concerning the CD in the studied Be stars}\label{pccd}

  The close relation between the spectroscopic distances and those measured by GAIA, provide us with some confidence on the reliability of the inferred astrophysical parameters. We can then try to benefit from some information on the properties of CDs of the studied stars, drawn from the emission in spectral lines. Thanks to the derived astrophysical parameters, good estimates of the neat emission line profiles and of the amount of emission filling up the photospheric component of the here observed Balmer lines H$\alpha$, H$\gamma,$ and H$\delta$ can be obtained, at least when it is possible according to the S/N of the spectra.\par
  
\subsection{Total emission in the observed Balmer lines}\label{tebl}
    
\begin{figure*}[] 
\centerline{\includegraphics[scale=0.75]{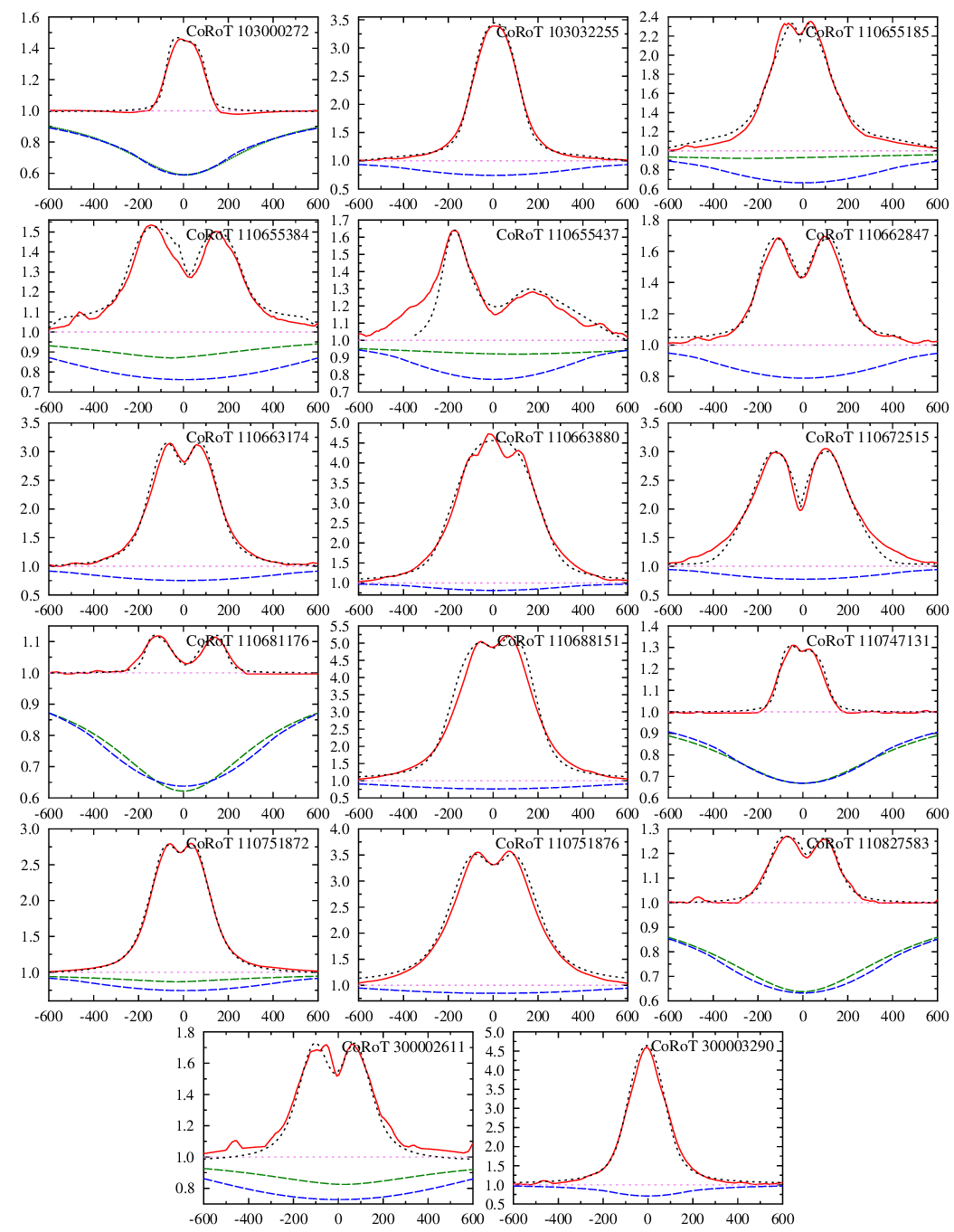}} 
\caption{\label{spct_ha} Smoothed H$\alpha$ line-emission components (red profiles). The emission residual intensities are relative to the photospheric line calculated according to the astrophysical parameters (T$_{\rm eff},\log g, V\!\sin i$) of each object. The figure shows the photospheric component broadened by rotation (blue profiles). It includes\ the empirical pseudo-photospheric component (green profiles, see Sect.~\ref{tebl}) in cases of partial emission feeling up of the absorption photospheric line. For the stars 103000272, 110681176, 110747131, and 110827583, the fit of the observed pseudo-photospheric absorption component closely reproduces the actual photospheric line.}
 \end{figure*}

  In cases when the emission in a line clearly exceeds the continuum over its entire wavelength range, we can simply add to this latter the amount of emission that fills up the photospheric absorption component derived from the stellar astrophysical parameters. If this filling up is partial,
here the apparent photospheric absorption profile underlying the emission is called  the pseudo-photospheric absorption.  We estimated the pseudo-photospheric absorption and add to this last the difference between the pseudo-photospheric and the photospheric proper absorption line profile. The pseudo-photospheric (pph) absorption line profile is easily fitted using the relation $\psi(\lambda) = \exp\left\{-1/[a+b\times (\lambda-\lambda_o)^c]\right\}$ introduced by \citet{ball1995,chau2001}, where the constants $a, b,$ and $c$ are determined using only three points in the line profile. In Fig.~\ref{spct_ha}, the red lines show the total H$\alpha$ line-emission components reduced to the continuum intensity level $I_{\lambda}/I_{\rm c}=1$ (violet dotted line). Thus, the total emission profile (red line) makes up the observed profile proper, to which the residual intensity between the photospheric profile and the pseudo-photospheric absorption profile has been added. The blue dashed lines correspond to the photospheric absorption line profile broadened according to the measured $V\!\sin i$ parameter. The green dashed lines are the pseudo-photospheric absorption line profiles. The equivalent width of the total H$\alpha$ emission corresponds then to the area below the red profile and the $I_{\lambda}/I_{\rm c}=1$ line, that is, $W_{\rm H\alpha}=W^{\rm em}_{\rm H\alpha}+W^{\rm ph}_{\rm H\alpha}-W^{\rm pph}_{\rm H\alpha}$. In Table~\ref{w_int_2}, we  give the equivalent widths $W_{\rm H\alpha}$, $W^{\rm em}_{\rm H\alpha}$, $W^{\rm ph}_{\rm H\alpha}$, and $W^{\rm pph}_{\rm H\alpha}$.

\begin{table}[]
\centering
\caption[]{\label{w_int_2} Equivalent widths and normalized fluxes of the emission components in the H$\alpha$ and H$\delta$ lines.}
\tabcolsep 3.0pt
\begin{tabular}{rrrrr|cc} 
\hline\hline
\noalign{\smallskip}
Star & $W^{\rm em}_{\rm H\alpha}$ & $W^{\rm ph}_{\rm H\alpha}$ & $W^{\rm pph}_{\rm H\alpha}$ & $I_{\rm H\alpha}$ & $W^{\rm em}_{\rm H\delta}$ & $I_{\rm H\delta}$ \\
     &  \AA   & \AA  &  \AA   & \AA   &  \AA   & \AA \\
\noalign{\smallskip}\hline
   1 &  1.8 &  9.0 & 8.9 &  0.6 &  0.0 & 0.0 \\
   2 & 10.8 &  5.7 & 0.0 &  7.7 &  0.0 & 0.0 \\
   3 &  7.3 &  8.0 & 3.7 &  3.8 &  1.2 & 0.4 \\
   4 &  4.3 &  8.2 & 4.6 &  3.6 &  0.9 & 0.3 \\
   5 &  2.0 &  5.1 & 3.7 &  2.0 &  0.7 & 0.3 \\
   6 &  3.5 &  4.8 & 0.0 &  4.7 &  0.0 & 0.0 \\
   7 & 14.1 &  6.6 & 0.0 & 10.0 &  0.5 & 0.2 \\
   8 & 34.8 &  3.4 & 0.0 & 39.0 &  0.7 & 0.8 \\
   9 & 20.8 &  5.2 & 0.0 & 12.8 &  1.2 & 0.5 \\
  10 &  0.8 &  9.4 & 0.0 &  3.0 &  0.0 & 0.0 \\
  11 & 35.8 &  6.9 & 0.0 & 21.2 &  0.7 & 0.3 \\
  12 &  1.4 &  7.5 & 0.0 &  3.2 &  0.0 & 0.0 \\
  13 & 14.3 &  6.6 & 2.9 &  8.5 &  0.1 & 0.0 \\
  14 & 24.5 &  4.0 & 0.0 & 27.8 &  0.7 & 0.70\\
  15 &  0.0: &  3.2 & 0.0 &  0.0 &  0.0 & 0.0 \\
  16 &  1.9 & 10.7 & 0.0 &  3.9 &  0.0 & 0.0 \\
  17 &  5.1 &  9.0 & 5.5 &  3.2 &  1.3 & 0.4 \\
  18 &  0.0: &  5.9 & 0.0 &  0.0 &  0.0 & 0.0 \\
  19 & 15.3 &  4.1 & 0.0 & 12.6 &  1.3 & 1.7 \\
\hline
\end{tabular} 
\end{table} 

\begin{figure*}[]
\centerline{\includegraphics[scale=0.75]{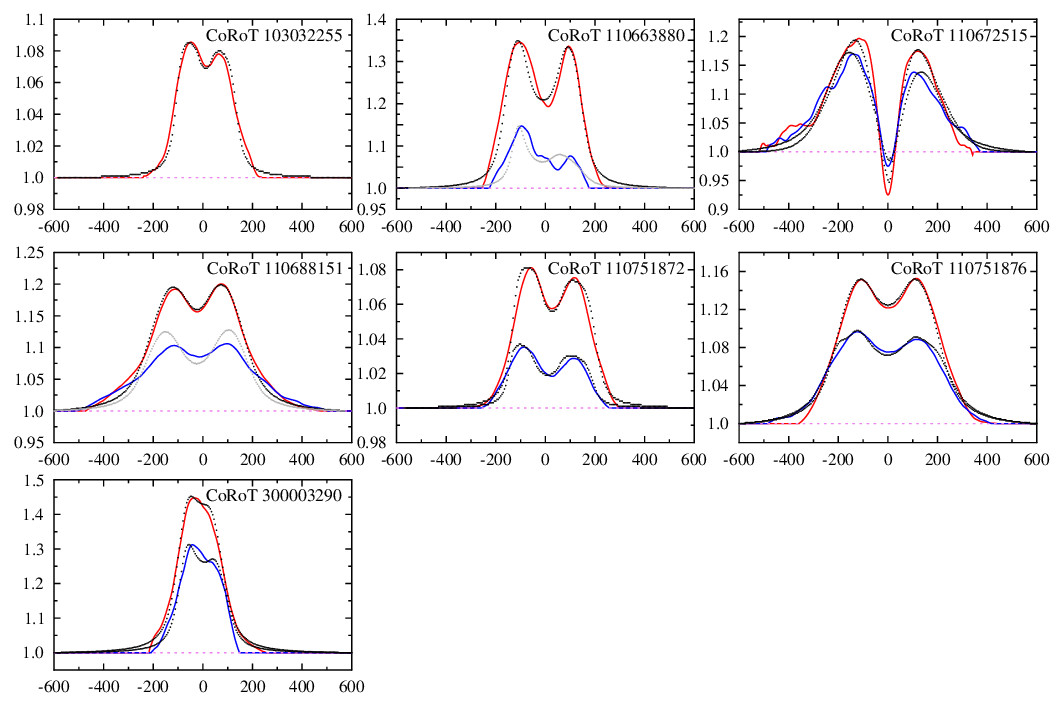}} 
\caption{\label{spct_hghd} Smoothed H$\gamma$ (red) and H$\delta$ (blues) line-emission components. The emission residual intensities are relative to each photospheric line. Emissions in these Balmer lines are seen also in some other stars, but they are too noisy and not exploitable.}
\end{figure*}

  The line emission profiles in the H$\gamma$ and H$\delta$ lines are determined through the residual intensities between the emission proper and the photospheric line profile. We also assume that the emission in these lines is weak enough to take the pseudo-photospheric profile as the genuine photospheric absorption. In Fig.~\ref{spct_hghd}, we reproduce the emission line profiles in H$\gamma$ and H$\delta$ lines. Some emission in these lines can also be present in other stars, but they are too noisy to carried out a clear extraction. We attempted to determine the equivalent width for all detectable emissions; they are given in Tables~\ref{w_int_2} and \ref{w_int_1} and were previously used in Sect.~\ref{ve} to calculate the veiling factor. \par 
  
\subsection{Other measurements on the emission line profiles}\label{omemlp}

\begin{figure*}[]
\centerline{\includegraphics[scale=0.7]{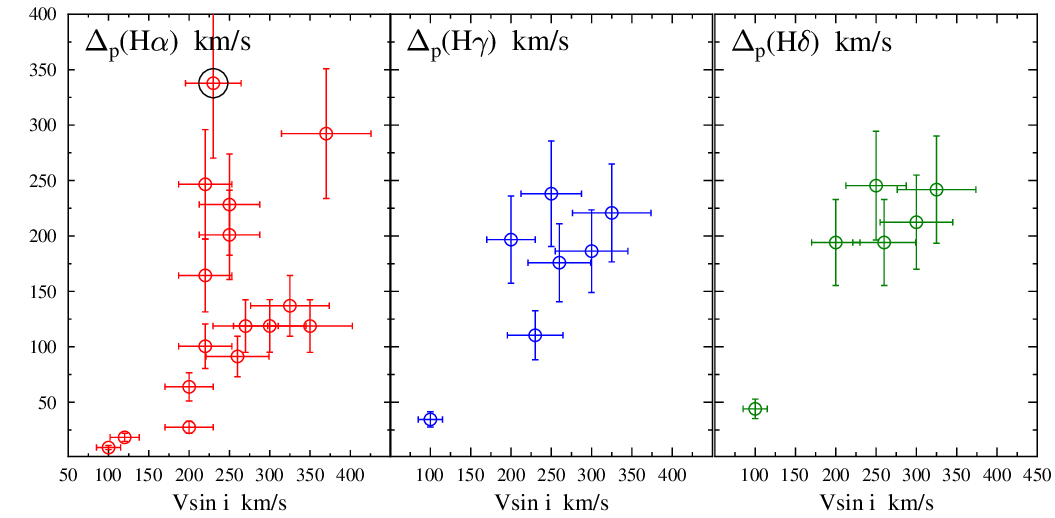}} 
\centerline{\includegraphics[scale=0.7]{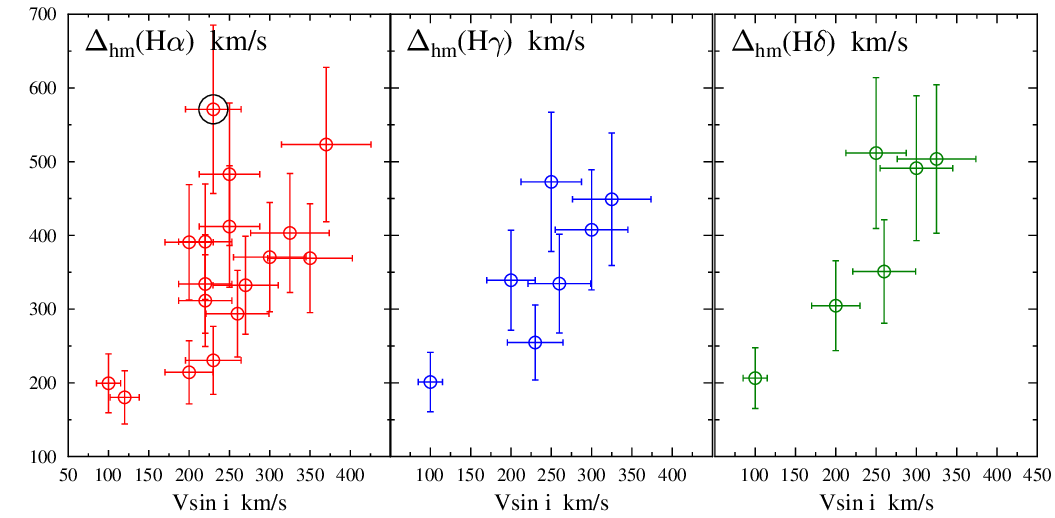}}
\caption{\label{fwhm_1_2} Relations with the $V\!\sin i$ of the peak separations $\Delta_{\rm p}$ and line widths at half intensity $\Delta_{\rm hm}$ of the emission components in the $H\alpha$ (red), $H\gamma$ (blue) and $H\delta$ (green) lines. A black circle surrounds the binary star N$^o$~5.} 
\end{figure*}

  An estimate of the total extent of the CD line emission formation region is frequently done using a relation which assumes that the disk is optically thin and that it has a rotation law that is dependent of the distance $R/R_o$ ($R_o$ is radius of the base of the disk) from the central star \citep{huang1972}. Following the reasoning in \citet{arcos2017} we can write Huang's radius as 
  
\begin{equation}  
(\displaystyle R/R_o)_{\rm H} = (2.7V\!\sin i/\Delta_{\rm p})^{1/j},
\label{eq_huang}
\end{equation}

\noindent where $\Delta_{\rm p}$ is the distance in km/s between the emission peaks, and $j=1/2$ if the rotation is Keplerian or $j=1$ if the disk conserves the angular momentum. The factor of $2.7$ in Eq.~\ref{eq_huang} differs from that used in \citet{arcos2017}, because stellar deformations were calculated following \citet{frem2005} for 
$\Omega/\Omega_{\rm c}=0.88$. In Table~\ref{dpdhm}, we give the measured emission peak separations and the disks extents $R/R_o$ determined assuming Keplerian rotation ($j=0.5$). To avoid confusion, we note that Huang's radius $(R/R_o)_{\rm H}$ in Table~\ref{dpdhm} is calculated with 
$\Delta_{\rm p}$, while $(R/R_o)_{\rm hm}$ represents a kind of Huang's CD extent determined with the full-widths-at-half-maximum (FWHM), $\Delta_{\rm hm}$, of the respective Balmer emission lines. The radius $(R/R_o)_{\rm H}$ is meant to represent the extent of the entire region in the CD where the emission is raised, if the CD is optically thin, while $(R/R_o)_{\rm hm}$ represents just a part of this line-forming region. The width of the base of emission line should in principle correspond to $(R/R_o)_{\rm base}=1$. However, the wings of emission lines are affected by other broadening mechanisms apart from rotation, so that the relation in Eq.~\ref{eq_huang} cannot identify well defined distances in the CD when using the FWHM or the width at the base of emission lines.
Relations with the $V\!\sin i$ of the peak separations $\Delta_{\rm p}$ and line widths at half intensity $\Delta_{\rm hm}$ of the emission components in the $H\alpha$, $H\gamma$ and $H\delta$ lines are shown in Fig.~\ref{fwhm_1_2}.\par 

\begin{table*}[] 
\centering
\caption[]{\label{dpdhm} Separation of emission peaks $\Delta_{\rm p}$, line emission FWHM, $\Delta_{\rm hm}$, and Huang's radius of the CD according to $\Delta_{\rm p}$ and $\Delta_{\rm hm}$.}
\tabcolsep 4.0pt
\begin{tabular}{rrrrr|rrrr|rrrr}
\hline\hline
\noalign{\smallskip}
     & \multicolumn{4}{c}{H$\alpha$} & \multicolumn{4}{c}{H$\gamma$} & \multicolumn{4}{c}{H$\delta$} \\
Star & $\Delta_{\rm p}$ & $\Delta_{\rm hm}$ & $(R/R_o)_{\rm H}$ & $(R/R_o)_{\rm hm}$ &
       $\Delta_{\rm p}$ & $\Delta_{\rm hm}$ & $(R/R_o)_{\rm H}$ & $(R/R_o)_{\rm hm}$ &
       $\Delta_{\rm p}$ & $\Delta_{\rm hm}$ & $(R/R_o)_{\rm H}$ & $(R/R_o)_{\rm hm}$ \\
     & km/s & km/s & & & km/s & km/s &&& km/s & km/s && \\
\noalign{\smallskip}
\hline
\noalign{\smallskip}       
   1 &   --- & 180 &    --- &  4 &     --- &   --- &   --- &  ---  &    ---  &   ---  &  --- &  ---  \\  
   2 &   --- & 230 &    --- &  9 &   103 & 255 &  45 &  7  &    ---  &   ---  &  --- &  ---  \\
   3 & 111 & 312 &   33 &  4 &     --- &   --- &   --- &  ---  &    ---  &   ---  &  --- &  ---  \\ 
   4 & 296 & 523 &   13 &  4 &     --- &   --- &   --- &  ---  &    ---  &   ---  &  --- &  ---  \\
   5 & 355 & 571 &    4 &  2 &     --- &   --- &   --- &  ---  &    ---  &   ---  &  --- &  ---  \\
   6 & 209 & 412 &   13 &  3 &     --- &   --- &   --- &  ---  &    ---  &   ---  &  --- &  ---  \\
   7 & 126 & 332 &   40 &  6 &     --- &   --- &   --- &  ---  &    ---  &   ---  &  --- &  ---  \\
   8 & 158 & 391 &   13 &  2 &   201 & 339 &   8 &  3  &  205  & 305  &  8 &  4  \\
   9 & 225 & 483 &   11 &  2 &   225 & 473 &  11 &  3  &  241  & 512  &  10 &  2  \\
  10 & 253 & 392 &    6 &  3 &     --- &   --- &   --- &  ---  &    ---  &   ---  &  --- &  ---  \\
  11 & 122 & 371 &   52 &  6 &   182 & 408 &  24 &  5  &  225  & 491  & 15 &  3  \\
  12 &  71 & 214 &   68 &  8 &     --- &   --- &   --- &  ---  &    ---  &   ---  &  --- &  ---  \\
  13 &  95 & 294 &   67 &  7 &   182 & 335 &  18 &  5  &  213  & 351  & 13 &  5  \\
  14 & 142 & 403 &   42 &  5 &   221 & 449 &  17 &  4  &  233  & 504  & 16 &  3  \\
  16 & 170 & 334 &   14 &  4 &     --- &   --- &   --- &  ---  &    ---  &   ---  &  --- &  ---  \\
  17 & 154 & 369 &   40 &  7 &     --- &   --- &   --- &  ---  &    ---  &   ---  &  --- &  ---  \\
  19 &   --- & 199 &    --- & 11 &    40 & 201 &  60 &  2  &   71  & 20  & 18 &  2  \\
\noalign{\smallskip}
\hline
\end{tabular}
\end{table*}     
 
\subsection{Near-IR circumstellar color excess and the Balmer line emissions}\label{irdce}

  To complete the presentation of observational characteristics of the program Be stars, we include in this section a short discussion of their near-IR photometric data. These allow us, on the one hand, to situate them in the color-color diagram and show they do not belong to the B[e] class of stars known as Herbig HAe/Be, and on the other hand, to obtain some relations commonly used to present the Be phenomenon \citep{ashok84,dach1988,hanusch1989,kastmaz89}. However, it is not in the scope of the present work to extract information on the CDs from the photometric data.\par
\begin{figure}[]
\centerline{\includegraphics[scale=0.8]{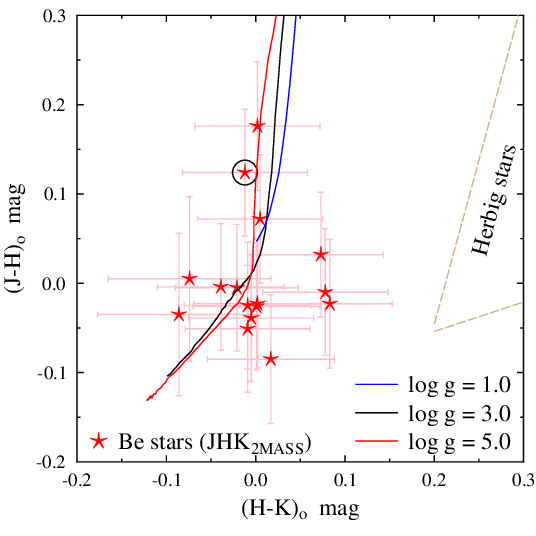}} 
\caption{\label{fig_jhk} Dereddened JHK color-color diagram of the program Be stars. The JHK magnitudes are from 2MASS. In this diagram are also shown the model sequences of the intrinsic colors corresponding to $\log g = 1.0$, $3.0,$ and $5.0$ dex, and the corner occupied by the bluest Herbig HAe/Be stars. A black circle surrounds the binary star N$^o$~5.}
\end{figure}

  
  Be stars present a thermal IR flux excess mainly due to bound-free, free-free transitions, and electron scattered radiation in their CD. To characterize the energy distribution in the visible domain, we could use the $(B-V)$ color index from the UBV Johnson-Cousins photometric system derived from the $(G,G_{\rm BP},G_{\rm RP})$ GAIA photometry using the published transformation relations into the UBV photometric system \citep{riello2021} in the way that has widely been studied by \citet{mouj1998,mouj1999}. However, this color index is heavily marred by the interstellar reddening so that the genuine color excess due to the CD is marred by the uncertainties of the $E(B-V)_{\rm ISM}$ color excess determinations. We then used  the near-IR photometry in the J, H, and K bands to try to characterize the color excess produced by the Be star CD. To this end, the 2MASS photometry \citep{cutri2003} was used. The observed $(J-H)_{\rm obs}$ and $(H-J)_{\rm obs}$ colors were corrected from interstellar extinction using the relations between the color excesses $E(V-J)=2.23\,E(B-V)$,  $E(V-H)=2.55\,E(B-V)$,  and $E(V-K)=2.76\,E(B-V)$ and the reddening relation, $A_{\rm V}=3.1E(B-V),$ established by Mathys J.S (1999) and published in
\citet{cox2000,kinm2002}. The determination of the color excess $E(B-V)$ was detailed in Sect.~\ref{der_par}. The JHK dereddened color-color diagram of the studied stars is shown in Fig.~\ref{fig_jhk} (red stars), where we added the model sequences of intrinsic colors for $\log g=1.0$ to 5.0 dex by \citet{caskur2003} (updated in 2011), and the region occupied by the bluest Herbig HAe/Be stars \citep{hern2005} to show that our program stars were not mistakenly chosen among Herbig HAe/Be objects. The stars numbered 17 and 19 have much larger JHK color uncertainties than the remaining ones, which might explain their extreme blue $(H-K)_o$ colors. \par 
  The color excesses proper in the IR that characterize the flux excess produced by the CD are defined as
 
\begin{eqnarray}
\left. \begin{array}{lcl}
\displaystyle \Delta(J-H)_{\rm CD} & = & \displaystyle (J-H)_o-(J-H)_* \\
\displaystyle \Delta(H-K)_{\rm CD} & = & \displaystyle (H-K)_o-(H-K)_* \\
\end{array}
\right\rbrace,
\label{ir_dev}
\end{eqnarray}

\noindent where $(J-H)_*$ and $(H-K)_*$ are the respective intrinsic colors of stars that underlay the CD. The intrinsic colors are inferred from the synthetic colors indices calculated with LTE models of stellar atmospheres by \citet{caskur2003}, using the apparent $(T_{\rm eff},\log g)$ parameters. Also, $(J-H)_o$ and $(H-K)_o$ are the observed color indices corrected for the ISM extinction.\par 
  Having at our disposal observations in the H$\alpha$ line -- as well as, for some stars, also in the H$\gamma$ and H$\delta$ reliable line emissions -- we can look for a correlation between the flux IR color excess and the intensity of the emission component in the Balmer line emission components. Because the IR flux excess and the Balmer line emissions are produced in rather different regions of the same CD, such correlations can  be of interest in attempting to control the consistency of physical inputs meant to describe the structure of Be star disks and their formation mechanisms. Marginally, we note that the sought correlations are marred by uncertainties due mainly to the non-simultaneity of observations carried of the IR colors and of the Balmer lines, as well as of the physical origins. In fact, since the perturbations produced by the CD on the continuum spectrum and on the line emissions are not produced in the same layers, there may be line emission intensities produced rather far from the star (roughly $2R_o \lesssim R \lesssim 10R_o$) that correspond to positive or negative flux excesses coming from disk layers near the star, generally $R\lesssim2R_o$. In spite of these shortcoming, we can attempt to establish some relations with the data at disposal in the way were previously attempted in \citet{ball1995} between the flux excess $\Delta V$ in the magnitude $V$ and the H$\gamma$ line emission, which revealed a quite reliable and useful correlation, which is described in Sect~\ref{ve} as a way of correcting the observed spectra from the veiling effect in the $\lambda\lambda\,4000-4500$ \AA \ spectral range.\par
\begin{figure*}[]
\centerline{\includegraphics[scale=0.7]{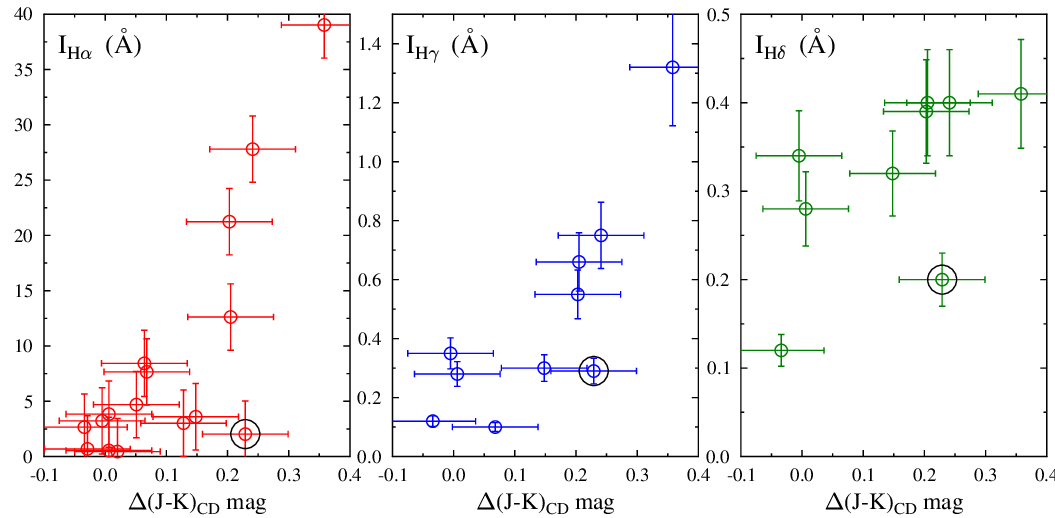}} 
\caption{\label{wf_jhk_1} Relations between the normalized line emission intensities $I_{\rm H\alpha}$ (red), $I_{\rm H\gamma}$ (blue), and $I_{\rm H\delta}$ (green) and the color dereddened color excess $\Delta(J-K)_{\rm CD}$ mag due to the CD. A black circle surrounds the binary star N$^o$ 5.}
\end{figure*}

  The emission intensities at H$\alpha$, H$\gamma$, and H$\delta$ lines are given by: 
  
\begin{equation}
\displaystyle I_{H\alpha,H\gamma,H\delta} = W^{\rm em}_{\rm H\alpha,H\gamma,H\delta}\times\left[\frac{F^{\rm c}_{\rm H\alpha,H\gamma,H\delta}}{F^{\rm c}_{\rm H\alpha,H\gamma,H\delta}(22500,4.0)}\right],
\label{int_em}
\end{equation} 
  
\noindent where $W^{\rm em}_{\rm H\alpha,H\gamma,H\delta}$ are the equivalent widths of the emission components in the $H\alpha$, $H\gamma,$ and $H\delta$ lines, $F^{\rm c}_{\rm H\alpha,H\gamma,H\delta}$ are the fluxes of the continuum spectrum at the center of the lines, and $F^{\rm c}_{\rm H\alpha,H\gamma,H\delta}(22500,4.0)$ are the normalizing fluxes given by a model with $T_{\rm eff}=22\,500$ K and $\log g=4.0$ (roughly a B2V type star). In Table~\ref{w_int_2}, we give the equivalent widths $W^{\rm em}_{\rm H\alpha,H\gamma}$ of the emission component in the H$\alpha$ and H$\gamma$ lines, and the respective intensities $I_{\rm H\alpha}$ and $I_{\rm H\gamma}$. In Fig.~\ref{wf_jhk_1}, we   show the trends determined by $I_{\rm H\alpha}$, $I_{\rm H\gamma}$, and $I_{\rm H\delta}$ against the color index excess $\Delta(J-H)_{\rm CD}$, which are rather well defined, while the correlations of the line emission intensities with $\Delta(H-K)_{\rm CD}$, not shown here, look much more scattered. In Table~\ref{ir_colors}, we give the observed color indices $(J-H)_{\rm obs}$ and $(H-K)_{\rm obs}$, the adopted stellar intrinsic colors $(J-H)_*$ and $(H-K)_*$, and the color index excess $\Delta(J-H)_{\rm CD}$. \par

\begin{table}[] 
\centering
\caption[]{\label{ir_colors} JHK color indices of the studied stars.}
\tabcolsep 2.0pt
\begin{tabular}{rcc|cc|cc}
\hline\hline
\noalign{\smallskip}
Star & $(J\!-\!H)_{\rm obs}$ & $(H\!-\!K)_{\rm obs}$ & $(J\!-\!H)_*$ & 
$(H\!-\!K)_*$ &  $\Delta(J\!-\!H)_{\rm CD}$ & $\Delta(H\!-\!K)_{\rm CD}$ \\ 
\hline\noalign{\smallskip}
   1 &  0.143 & 0.102 &   -0.026 & -0.028  &    0.001 &  0.019 \\   
   2 &  0.154 & 0.119 &   -0.044 & -0.045  &    0.021 &  0.047 \\    
   3 &  0.264 & 0.137 &   -0.022 & -0.027  &    0.018 & -0.012 \\    
   4 &  0.198 & 0.228 &   -0.045 & -0.043  &    0.022 &  0.126 \\    
   5 &  0.482 & 0.223 &   -0.060 & -0.057  &    0.184 &  0.045 \\    
   6 &  0.192 & 0.150 &   -0.056 & -0.055  &    0.005 &  0.046 \\    
   7 &  0.600 & 0.156 &   -0.048 & -0.046  &    0.313 & -0.018 \\    
   8 &  0.549 & 0.247 &   -0.093 & -0.086  &    0.270 &  0.088 \\    
   9 &  0.512 & 0.128 &   -0.047 & -0.048  &    0.275 & -0.010 \\    
  10 &  0.090 & 0.077 &   -0.012 & -0.019  &   -0.014 &  0.020 \\    
  11 &  0.103 & 0.120 &   -0.050 & -0.047  &    0.082 &  0.121 \\    
  12 &  0.454 & 0.112 &   -0.027 & -0.031  &    0.228 & -0.023 \\    
  13 &  0.082 & 0.037 &   -0.046 & -0.045  &    0.040 &  0.024 \\    
  14 &  0.153 & 0.185 &   -0.090 & -0.083  &    0.080 &  0.161 \\    
  15 &  0.420 & 0.296 &   -0.089 & -0.082  &    0.050 &  0.078 \\    
  16 &  0.138 & 0.164 &   -0.018 & -0.021  &   -0.067 &  0.038 \\    
  17 &  0.491 & 0.245 &   -0.032 & -0.032  &    0.037 & -0.042 \\    
  18 &  0.345 & 0.164 &   -0.043 & -0.044  &    0.008 & -0.042 \\    
  19 &  0.409 & 0.226 &   -0.066 & -0.063  &    0.138 &  0.067 \\    
\noalign{\smallskip}
\hline
\multicolumn{7}{l}{Stars numbered 7, 9, and 12 have color indices obtained from the} \\
\multicolumn{7}{l}{(G,BP,RP) magnitudes of the GAIA photometry} \\
\hline
\end{tabular}
\end{table}

\section{Physical characteristics of CDs derived from the interpretation of the Balmer line emission components}\label{pcdilemc} 

  In this section we explore the properties of the CDs derived from the spectroscopic characteristics of the H$\alpha$, H$\gamma,$ and H$\delta$ emission lines. We produce a simple representation of the CD that may contrast in simplicity with more detailed calculations as those by \citet{carbjo2006,carbjo2008,
sigut2007,sigut2009,sigut2013,catan2013,kurf2018}. The simplifications we use are based on first physical principles and are introduced in as physically consistent 
a way as possible.\par
 
\subsection{Assumptions about the geometry and temperature of the CD}
\label{agcd}
 
 To describe the geometrical characteristics of the CD, we use two reference systems centered on the star. In the star system $(X,Y,Z)$, the $Z-$axis contains the rotation axis, while $(X,Y)$ are in the stellar equatorial plane with $X$ on the back sky plane. The system $(x,y,z)$ has the $z-$axis directed towards the observer, and $x$ coincides with $X$. The angle between $Z$ and $z$ is the inclination angle, $i,$ of the star-disk system. The radius vector, $R,$ is measured on the $(X,Y)$ plane. Between $Y$ and $R,$ there is the azimuth angle $\theta$. \par
 We take the CD driven by a quasi-Keplerian speed of rotation $V_{\rm K}(R)=V^o_{\rm K}(R_o/R)^{\gamma/2}$. We include an expansion (or contraction) velocity parameter $V_{\rm exp}$ to account for the emission line profile asymmetry. The integration of the hydrostatic equilibrium equation for the circumstellar environment leads to the axisymmetric density profile of the disk given in \citet{kurf2018}, of which we adopt the simplified expression valid for $Z/R\ll1:$ 

\begin{equation}
\begin{array}{lcl}
\displaystyle \rho(R,z) & = & \rho_oD(R)\exp\left\{-\left[Z/\!\!\sqrt{2}h(R)
\right]^2\right\} \\
\end{array} 
\label{eq2}
,\end{equation}

\noindent which still describes the main geometrical characteristics of disks and enables easier mathematical handling. In Eq.~\ref{eq2}, 
$\rho_o$ is the base density of the disk at the stellar equator, $D(R)$ is a function meant to describe the CD density distribution in the equatorial plane,  and $h(R) = [V_{\rm s}/V_{\rm K}]R$ is the $Z$-scale height of the disk with $V_{\rm s}$ as the sound speed and $V_{\rm K}$ as its quasi-Keplerian rotation. Explicitly

\begin{equation}
\left. \begin{array}{lcl}
h(R) & = & k[(R_o/R_{\odot})^{\gamma}(T_{\rm eff}/T_{\odot})/(M/M_{\odot})]^{1/2}(R/R_o)^{\beta} \\
k & = & \left[{\cal R}/(\sqrt{2}\mu G)\,(R_{\odot}T_{\rm eff}^{\odot}/M_{\odot})\right]^{1/2} \\
\end{array}
\right\}
\label{h}
,\end{equation}   

\noindent with $\cal{R}$ the constant of gases; $\mu$ the average molecular weight; gravitational constant $G$; $R_{\odot}$, $T_{\rm eff}^{\odot}$, and $M_{\odot}$ are the solar radius, effective temperature, and mass, respectively. In the following, we use $\gamma=1$ to otherwise describe deviations from a strict Keplerian rotation. Whether the temperature of CD follows the geometrical dilution of the stellar bolometric radiation or whether it is isothermal, we take either $\beta=(2+\gamma-1/2)/2=1.25$ or $\beta=(2+\gamma)/2=1.5$, respectively.\par 
 The function $D(R)$ can be specified only if a theory is used to describe the formation and further evolution of the CD, as done in VDD models \citep{narita1994,okazaki01,haubois2012,
ghore2021,marr2021}. In this work, we assume that the $R-$dependent density profile of the disk is parameterized as $D(R)=(R_o/R)^n$, where the exponent $n$ is a function of $R$. The density exponent, $n,$ that is reported currently in the literature, actually represents the global slope of the disk density variation with $R$. However, from \citet{haubois2012}, it is clear that depending on the evolution phase of the circumstellar disk, $n(R)$ is a strong function of $R$, mainly in proximity to the star, where emission component in lines such as H$\gamma$ and H$\delta$ are formed. Anticipating the possibility that the regions where the emission of H$\alpha$ and those of H$\gamma$ and H$\delta$ do not respond to the same value of the density exponent $n$, as in \citet{arias2007}, we adopted the following expression for $n(R):$  

\begin{equation} 
\begin{array}{lcl}
\displaystyle n(R) & = & [(n_2-n_1)/\pi]\arctan\left\{Q\times\left[(R/R_o)-(R_D/R_o)\right]\right\}+ \\ 
 & & (1/2)(n_1+n_2),
\end{array}
\label{eq3}
\end{equation} 

\noindent whose value changes from $n(R)=n_1$ within $R_o\lesssim R\lesssim R_D$ to $n(R)=n_2>n_1$ for $R \gtrsim R_D$. In Eq.~(\ref{eq3}), $Q\lesssim10$ is a constant chosen freely. However, no significant changes are carried on the determination of parameters if we adopt a simple step relation $n(R)=n_1$ at $R_o\leq R\leq R_D$ and $n(R)=n_2$ for $R\gtrsim R_D$. \par  
  From the theory of VDD \citep{lee1991,
okazaki01,okazaki07}, we know that in isothermal disks at the steady state, it is $n=3.5$ over the entire disk radial extent. The results obtained by \citet{arias2007} based on the study of \ion{Fe}{ii} emission lines in the blue spectral range of Be stars, suggest that $n_1\lesssim 1.0$ for $R\lesssim R_D\sim 3R_o$, while studies on the SED in the near- and far-IR of Be stars \citep{wat1986,vieira2015} suggest $n_2 \simeq 2.5-3.9$ for $R\gtrsim R_D$. The modeling of the SED in the 3500-10500 \AA\ wavelength interval of a set of Be stars by \citet{mouj1998PhDT,mouj2000,mouj2000b} and \citet{chau2001} suggested an average exponent $<n>=1.6\pm0.2$.  \citet{granada_2010} used $n_1=0.5$ in a study of emissions in the hydrogen Humphry's series, which form in CD regions near the central star. Studies of the temporal evolution of VDD density distributions under different formation scenarios \citep{haubois2012} end up concluding that $n$ varies from 0.5 to 4.5, but sometimes $n<0$ in the base of the disk. The function $n(R)$ on the distance from the star, it is sensitive to the kinematic viscous coefficient and depends on the disk decretion history \citep{haubois2012,vieira2017,ghore2018,ghore2021,marr2021}. \par   
  In a large series of papers, disk models for Be stars were assumed to be isothermal \citep{marlb1969,marlb1978a,marlb1978b,humm1994,catan2013}. More refined discussions on the distribution of the temperature in the CD of Be stars obtained high non-uniform temperature distributions, whose global behavior depends on the physical inputs and assumptions made in the models. \citet{millar1,millar2,carbjo2006,carbjo2008,sigut2007,sigut2013} noted that not far from the star, the disk temperature decreases in the equatorial region and increases somewhat at larger distances and higher $Z-$coordinates. On the contrary, by including viscous heating, \citet{kurf2018} have found that the temperature increases in the equatorial regions and decreases towards larger distances in both $R$ and $Z$. \par
  However, it has long been known that the source function of Balmer lines  is strongly dominated by photoionization processes \citep{thomas1957,
thomas1965,jefferies1968,mihalas1978,hub_mih_2014}. This means that the production and destruction of line photons is dominated by the stellar radiation field. The source function is thus dissociated from local  temperature and density in the line formation region. To formalize a marginal dependencies of the source function with the CD temperature, we assume that the local temperature is determined by the geometrical dilution of the stellar bolometric flux 
$[T(R)/T_{\rm eff}]^4=(1/2)\{1-[1-(R_o/R)^2]^{1/2}]\}$, which is close to the relation found by 
\citet{mouj1998PhDT} $T/T_{\rm eff}=0.19+0.61/R$ in a study of the observed visible energy distribution in 21 Be stars. Moreover, as the line emission efficiency depends on the electron density squared, the emission rapidly decreases with $Z$, reducing thus the extent of effective zone perpendicular to the equatorial plane for the emission of photons. We assume then a uniform temperature in the $Z-$direction.\par 
   Furthermore, we reduce the CD line formation region into an equivalent cylinder having an average half-height $\overline{H}$ and a radial extent $\Delta R$ that extends from a given radius, $R$. Here, $\Delta R$ is the distance over which it is recovered $99\%$ of the disk radial opacity integrated from $R$ to $R+\Delta R$. Using Eq.~\ref{eq2}, it follows that the opacity of the disk in the $Z-$direction averaged over the $\Delta R$ distance, is the same as for a cylinder with height $\overline{H}=3\sqrt{2}\,\overline{h(R)}$, where $\overline{h(R)}$ is the weighted average of the disk density scale-height, $h(R),$ over the distance, 
$\Delta R$  

\begin{equation} 
\overline{h(R)}/R_o \simeq \left[1+\beta(\Delta R/R)\right]\,h(R) \ , \ \ \ {\rm for} \ \ \Delta R/R \lesssim1, 
\label{HM}
\end{equation}  

\noindent where $h(R)$ is given by Eq.~\ref{h}.\par 

 Although in our discussion, $\overline{H}$ is considered a free fitting parameter, the value obtained from the analytical expression is adequate enough in almost all cases to obtain the best fit of a given observed line emission. In Tables~\ref{ha_fit}, \ref{hg_fit}, and \ref{hd_fit}, we refer the model $\overline{H}$ to the value given by $3\sqrt{2}\,\,\overline{h(R)}$.\par 
 
\subsection{The source function}\label{sf}
 
  In the frame of an equivalent two-level atom with continuum, the main characteristic of the source functions for the first Balmer lines is that they are strongly dominated by photionizations and radiative recombinations, where collisional ionization and excitation rates can be neglected \citep{thomas1957,thomas1965,jefferies1968,mihalas1978,
hub_mih_2014}. The non-LTE source function of these lines are then determined only by the radiation field of the central star. The source function of the first Balmer lines with continuum in an equivalent emitting ring with uniform physical properties can then be approached with the following expressions \citep{mihalas1978,cidale1989,hub_mih_2014}
 
\begin{equation}
\displaystyle S_{\lambda}(\tau_{\rm o}) = \left\{
\begin{array}{ll}    
[\eta/(1+\eta)]^{1/2}B^*_{\lambda} & {\rm for}\ \tau_o < 1 \\

[\eta/(1+\eta)]^{1/2}B^*_{\lambda}\tau_{\rm o}^{1/2} & {\rm for} \ \tau_o \geq 1, \\
\end{array}\right.,
\label{eq8}
\end{equation}

\noindent where we do not consider the thermalization of the source function that should be produced for $\tau_{\rm o}\gtrsim10-10^2$. In Eq.~(\ref{eq8}), $\tau_o$ is the optical depth in the central wavelength of the Balmer $\lambda$ line; $\eta$ is the ``sink" term in the line source function, $S_{\lambda}$; $B^*$ is the ``source" factor of $S_{\lambda}$. This approximation was already used in \citet{vin2006} and \citet{azf2007}. The expressions for the radiation sink and the source factors $\eta$ and $B_*$ are given in  \citep{thomas1965,
hub_mih_2014}

   In Tables~\ref{ha_fit}, \ref{hg_fit}, and \ref{hd_fit}, the source function ratios $S^o_{\lambda}/F_*=[\eta/(1+\eta)]^{1/2}B^*/F_*$ are given for the 
H$\alpha$, H$\gamma,$ and H$\delta$ lines of the studied stars. Within the same framework of radiation-dominated transitions  the non-LTE departure coefficients were calculated. \par

\subsection{Formal solution of the radiation transfer for the CD}

  From the simplifications adopted on the geometrical shape of the CD, as a radiation emitting region characterize by a uniform opacity 
$\tau_{\rm Z}$ in the direction perpendicular to the equator, can be considered to behave as an equivalent thin ring, having a total height of $2\overline{H}$ and radius $R$. \par 

\begin{figure}[]
\centerline{\includegraphics[scale=0.8]{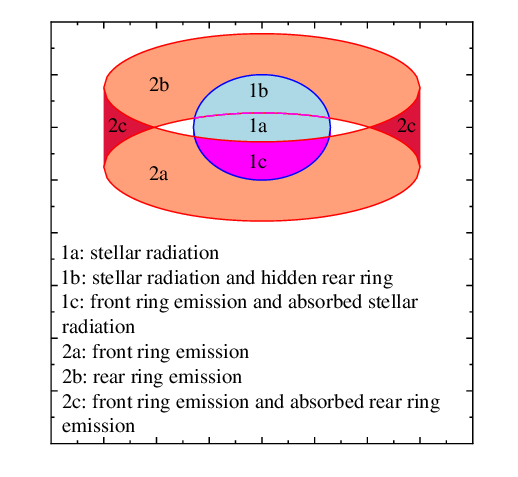}} 
\caption{\label{fanillo} Regions in the disk that contribute in different ways to the observed line emission, according to our simplified representation of the star-disk system.} 
\end{figure}

  Although the simulation of the formation of emission lines from a flat disk of uniform height does not offer particular difficulties \citep[see][]{homa1986}, calculations as a function of the inclination angle, $i,$ in the frame of an equivalent ring are a little more delicate. The shape of the emission ring-region projected towards the observer is sketched in Fig.~\ref{fanillo}. The radiation field coming from the ring-star system can be represented using the formal solution of the equation of radiation transfer at each point $(x,y)$ projected on the background plane of the sky. For optical depths $0\leq\tau_o<1$, according to Eq.\ref{eq8}, the source function $S_{\lambda}=S^o_{\lambda}$ of the studied Balmer lines is constant. The formal solution of the radiation transfer equation applied to the star-disk system applied to the geometrical configuration given in Fig.~\ref{fanillo} is as follows:\ 
  
\begin{equation}
  \left. \begin{array}{lcl}
  \displaystyle I^a_{\lambda}(x,y,V-V_r) & = & \displaystyle I^*_{\lambda}(x,y)\exp[-\tau^f_{\lambda}(x,y,V-V_r)/\mu(x,z)]+\\
  & & S^o_{\lambda}\{1-\exp[-\tau^f_{\lambda}(x,y,V-V_r)/\mu(x,y)]\} \\
   & & ({\rm stellar\ emission\ absorbed\ by\ the\ front}\\
                     &   &{\rm shell)+(emission\ from\ the\ front\ shell})\\
  \displaystyle I^b_{\lambda}(x,y,V-V_r) & = & \displaystyle S^o_{\lambda}\{1-\exp[-\tau^r_{\lambda}(x,y,V-V_r)/\mu(x,y)]\}\times \\
  & &\exp[-\tau^f_{\lambda}(x,y)/\mu(x,y)]+ \\
  & & 
  S^o_{\lambda}\{1-\exp[-\tau^f_{\lambda}(x,y,V-V_r)/\mu(x,y)]\}\\
                     & = & ({\rm emission\ from\ rear\ shell\ 
                        }{\rm absorbed\ by} \\
                     &   &{\rm \ front\ shell})+({\rm emission\ by\ front\ shell})\\
  \end{array}
  \right\}
  \label{eq20}
  ,\end{equation}  

\noindent where $I^*_{\lambda}$ is the contribution to the field of radiation by the central star. In these expressions, $\mu(x,y)$ is the cosine between the normal to the ring at a test point $(x,y)$ and the sight direction, $V$ is the wavelength displacement in the spectral line in velocity units, and $V_r$ is the radial velocity at the point $(x,y),$ defined below in Sect.~\ref{lo} (Eq.~\ref{eq30}).\par     
 For optical depths $\tau_o\geq1$, the formal solution for the radiation field introduces such integrals as:  
 
\begin{equation}
\begin{array}{rcl}
I_{\lambda} &=& S^o_{\lambda}\int_1^{\tau_o}\tau_o^{1/2}\exp(-\tau_{\lambda})\,d\tau_{\lambda},\\
\tau_{\lambda} &=&\tau_o\Phi_{\lambda},\\
I_{\lambda} &=& \Phi_{\lambda}[\tau_o^{1/2}f(\Phi_{\lambda}\tau_o)-f(\tau_o)],\\
\end{array}
\label{ist}
\end{equation}

\noindent where $\Phi_{\lambda}$ is the intrinsic line absorption profile and $f(t)$ is determined by the incomplete gamma function, 
$\gamma(t,3/2)$. \par 

  These relations can be used to compose the field of radiation due to overlapping absorbing and emitting layers of the front and rear parts of the equivalent ring. In the optical depths, a distinction is made between 
$\tau^r_{\lambda}(x,y,V)$ and $\tau^f_{\lambda}(x,y,V)$ of the "rear"  and "front" layers, respectively, to caution that the radiation field is affected by the Doppler effect. This is due to the possibility of several apparent radial velocity compositions produced by the rotation and expansion velocities of the projected shells relative to the observer. \par 
 Finally, the radiation flux in the lines emitted by the star-disk system is given by:

\begin{equation}
\displaystyle F(x,y) = \int_{\cal S}I_{\lambda}(x,y){\rm d}x{\rm d}y
\label{eq19}
,\end{equation}

\noindent where ${\cal S}$ is the surface of the star-disk system projected onto the sky background, which in addition to the ring surface shown in Fig.~\ref{fanillo}, encompasses the projected emitting surface of effective cylinder from $R$ to $R+\Delta R$.\par  
 The expressions in Eq.~\ref{eq20} can be adapted to several configurations that depend on the value of the height $\overline{H}$: 
$\overline{H}\leq R_o$ or $\overline{H}> R_o$ and on the value of the inclination angle $i_{RH}=\arctan(\Delta R/2\overline{H})$. We previously used this procedure in \citet{huhi2000,aml2004,vin2006} and \citet{azf2007}.\par  

\subsection{Line opacity}\label{lo}

  We considered the opacity of the studied Balmer lines parameterized according to their value in the $Z-$direction, which at the respective central wavelength, $\lambda_o$, and in the non-LTE approximation is: 

\begin{equation}
\begin{array}{lcl}
\tau_{\lambda_o} & = & \alpha_{\lambda_o}\overline{N_2(R)_Z}\left\{1-(b_n/b_2)\exp\left[-(hc/\lambda_okT)\right]\right\},\\
\alpha_{\lambda_o} & = & (\sqrt{\pi}e^2)/(m_{\rm e}c)(\lambda_of)/V_D), 
\end{array}
\label{eq88}
\end{equation} 

\noindent where $\overline{N_2(R)}$ represents the population of the atomic level $n=2$ in the $Z-$direction, averaged over the radial distance from $R$ to $R+\Delta R$, $f$ is the oscillator strength of the given transition, $V_D$ is the total Doppler velocity of its intrinsic absorption line profile that encompasses not only thermal but  other possible macroscopic velocity components of statistical nature, and $b_n$ represents the non-LTE departure coefficients for atomic levels $n$. The term $(b_n/b_2)\exp[-(hc/\lambda_okT)] \ll1$ can be neglected in most cases.\par 

 Writing the line opacity as:
 
\begin{equation} 
\tau_Z = \tau_{\lambda_o}\Phi(\Delta\lambda/\Delta\lambda_{\rm D}) ,
\label{tauline}  
\end{equation} 
 
\noindent the intrinsic Balmer line profile $\Phi(\lambda)$ formally corresponds to the composite expression for a Stark profile given in \citet{gray_2008} or \citet{hub_mih_2014}. We calculated it using the tables of Stark profiles given by \citet{stehle1999}. However, in most cases, the gas density in the disk is low, so that the Stark line broadening is 
small.\par 
  The opacity in the $Z-$direction, $\tau_Z$, is adopted in this work as a free parameter. For aspect angles, $i\lesssim i_{RH}=\arctan(\Delta R/\overline{2H})$, the optical depth of the shell as a function of the inclination is then parameterized as $\tau(i)=\tau_Z/\cos i$ and then as 
$\tau(i)=\tau_Z\tan i_{RH}/\sin i$ when $i \gtrsim i_{RH}$. Apart from these $i-$angle dependencies of the calculated line emission characteristics, stronger  inclination angle dependencies are carried by the apparent shape of the entire emission ring-region projected towards the observer, as sketched out in Fig.~\ref{fanillo}. \par 
  The emission intensity at an observed wavelength $\lambda$ corresponds to the sum of the emissions at wavelength offsets $\delta\lambda$ with respect to the central absorption of the atom, which, in turn, is displaced due to the macroscopic movement of the medium with respect to the observer. The absorption (or emission) intensity with which the atom contributes at 
$\lambda$ is finally determined by the profile
$\Phi(\delta\lambda)=\Phi(\lambda-\lambda_o-\Delta\lambda)$, where 
$\Delta\lambda=\lambda_o\frac{V_r}{c}$ and $V_r$ is the total macroscopic radial velocity of the gas at a given point ($R,\theta$) in the ring:

\begin{equation}
\displaystyle V_r = [V_{\rm rot}\sin\theta+V_{\rm exp}\cos\theta]\sin i,
\label{eq30}
\end{equation}  
  
\noindent where $V_{\rm rot}$ is the rotation velocity of the ring and  $V_{\rm exp}=V_{\rm exp}^o(R_o/R)^{\alpha}$ is its ``expansion" velocity that is used only to account for the line profile asymmetries. The quantity $V_{\rm rot}$ is also only to fit at best the width of emission lines, which encompass the actual line broadening by the disk rotation integrated over its formation region and a host of other line broadening mechanisms. Hence, we do not expect to have well defined correlations between $V_{\rm rot}$ and $V^*_{\rm rot}/\sqrt{R}$. \par 
  For the line profile calculations we use velocities instead of wavelengths, so that the argument in the profile $\Phi(\lambda-\lambda_o-\Delta\lambda)$ is readily transformed into velocity displacements   
$[(\lambda-\lambda_o-\Delta\lambda)]/\Delta\lambda_D=$ $(V-V_r)/V_D$, where $V-V_r$ measures in velocity units the offset $\lambda-\lambda_o$ in the observed profile.\par
  The use of equivalent emitting (absorbing) shell with radius $R$, also requires a consideration of the Doppler displacements due to the changes of the macroscopic velocity fields in the $Z-$direction. These contributions are called ``shear" components. Following \citet{homa1986}, it can be shown that the integrated effect on the line profile of the shear velocity components carries an enlarged line Doppler width $\Delta\lambda_{\rm D} \to \Delta_{\rm D}$,
  
\begin{equation}
\Delta_{\rm D} = \Delta\lambda_{\rm D}\left[1+\left(\frac{\lambda_o}{\Delta_{\lambda_{\rm D}}}\frac{V_{sh}}{c}\right)^2\right]^{\frac{1}{2}}
\label{eq65}
,\end{equation}  
  
\noindent where the factor in braces acts as a supplementary broadening agent of the line profile. In Eq. \ref{eq65}, $V_{sh}$ is the contribution of the shear velocity component given by   

\begin{equation}
V_{sh} = [\frac{1}{2}V_{rot}\sin\theta+\alpha.V_{exp}\cos\theta]\frac{\overline{H}}{R}\cos\theta\tan i\sin i
\label{eq66}
.\end{equation}

 Shear effects are anisotropic with strong dependence on the inclination angle, $i,$ and azimuth, $\theta$. These effects has almost no incidence on the optically thin lines, but it saturates somewhat the emission peaks as soon as $\tau_{\lambda_o}\gtrsim1$. In what follows, we assume $\alpha=0$. Disks where $\alpha\neq0$, namely, those with gradients of expansion velocities, can be studied using separate equivalent rings which may then reproduce multi-peak emission line profiles \citep{azf2007}. \par   

\begin{table}[] 
\centering
\caption[]{\label{ha_fit} Parameters obtained from the fit of the observed 
H$\alpha$ emission lines for their CD formation region.}
\tabcolsep 3.0pt
\begin{tabular}{rcc|ccc|cc|rr} 
\hline\hline
\noalign{\smallskip}
     & \multicolumn{2}{c|}{$S^o_{\lambda}/F_c$} & & && \multicolumn{2}{c|}{$\overline{H}/R_o$} & $V_{\rm rot}$ & $V_{\rm exp}$ \\
Star & model & fit & $\tau_{\lambda_o}$ & $R/R_o$ & $\Delta R/R_o$ & model & fit & km/s & km/s \\ 
\noalign{\smallskip}  
 1 & 0.053 &       & 0.30 & 5.5 & 2.3 & 1.06 &      & 270 & $-35$ \\ 
 2 & 0.064 & 0.100 & 0.80 & 4.5 & 2.0 & 0.98 & 1.00 & 150 & 0 \\ 
 3 & 0.052 &       & 1.10 & 7.5 & 3.6 & 1.78 &      & 200 & 0 \\ 
 4 & 0.063 & 0.100 & 1.05 & 4.0 & 1.6 & 0.59 &      & 300 & $-25$ \\
 5 & 0.072 & 0.135 & 1.40 & 2.2 & 1.7 & 0.42 &      & 400 & $-250$ \\ 
 6 & 0.072 &       & 1.05 & 4.3 & 2.2 & 0.98 &      & 250 & 0 \\ 
 7 & 0.065 & 0.160 & 0.90 & 4.0 & 1.7 & 0.81 & 0.90 & 170 & $-5$ \\ 
 8 & 0.094 &       & 1.70 & 4.0 & 6.1 & 1.23 &      & 400 & $-55$ \\ 
 9 & 0.066 & 0.200 & 0.76 & 4.9 & 2.4 & 1.18 &      & 250 & 5 \\ 
10 & 0.047 & 0.100 & 0.40 & 2.0 & 0.8 & 0.34 &      & 230 & $-10$ \\
11 & 0.067 &       & 1.60 & 8.1 & 5.5 & 1.82 &      & 250 & 30 \\
12 & 0.054 &       & 2.00 & 3.3 & 1.0 & 0.59 &      & 150 & $-18$ \\
13 & 0.064 &       & 1.40 & 6.9 & 3.9 & 1.53 &      & 200 & 0 \\
14 & 0.091 &       & 3.00 & 3.5 & 3.6 & 0.76 &      & 270 & 0 \\
16 & 0.050 & 0.070 & 0.35 & 5.0 & 1.8 & 0.85 &      & 230 & $-5$ \\
17 & 0.056 & 0.120 & 0.43 & 5.0 & 2.1 & 0.85 &      & 200 & $-2$ \\ 
19 & 0.076 &       & 2.10 & 5.0 & 3.6 & 1.44 &       & 220 & 0 \\ 
\noalign{\smallskip}
\hline
\end{tabular}

\centering
\caption[]{\label{hg_fit} Parameters obtained from the fit of the observed
H$\gamma$ emission lines for their CD formation region.}

\tabcolsep 3.0pt
\begin{tabular}{rcc|ccc|cc|rr}
\hline\hline
\noalign{\smallskip}
     & \multicolumn{2}{c|}{$S^o_{\lambda}/F_c$} & & && \multicolumn{2}{c|}{$\overline{H}/R_o$} & $V_{\rm rot}$ & $V_{\rm exp}$ \\
Star & model & fit & $\tau_{\lambda_o}$ & $R/R_o$ & $\Delta R/R_o$ & model & fit & km/s & km/s \\ 
\noalign{\smallskip}
 2 & 0.182 &       & 0.06 & 2.1 & 1.4 & 0.42 &      & 230 & $-10$ \\ 
 8 & 0.227 & 0.500 & 0.09 & 2.2 & 2.0 & 0.47 &      & 240 & $-10$ \\
 9 & 0.182 & 0.500 & 0.01 & 1.6 & 1.8 & 0.38 &      & 150 & $-17$ \\
11 & 0.186 & 0.400 & 0.50 & 2.1 & 1.6 & 0.34 &      & 230 & 10 \\
13 & 0.188 & 0.210 & 0.07 & 3.1 & 1.0 & 0.51 &      & 200 & $-10$ \\
14 & 0.233 &       & 0.04 & 3.5 & 3.6 & 0.96 &      & 320 & 0 \\
19 & 0.211 &       & 0.14 & 3.7 & 2.1 & 0.89 &      & 300 & $-20$ \\
\noalign{\smallskip}
\hline
\end{tabular}

\centering
\caption[]{\label{hd_fit} Parameters obtained from the fit of the observed 
H$\delta$ emission lines for their CD formation region.}
\tabcolsep 3.0pt
\begin{tabular}{rcc|ccc|cc|rr}
\hline\hline
\noalign{\smallskip}
     & \multicolumn{2}{c|}{$S^o_{\lambda}/F_c$} & & && \multicolumn{2}{c|}{$\overline{H}/R_o$} & $V_{\rm rot}$ & $V_{\rm exp}$ \\
Star & model & fit & $\tau_{\lambda_o}$ & $R/R_o$ & $\Delta R/R_o$ & model & fit & km/s & km/s \\ 
\noalign{\smallskip} 
\noalign{\smallskip}
 8 & 0.263 &       & 0.04 & 2.0 & 2.1 & 0.42 &      & 250 & $-130$ \\
 9 & 0.230 & 0.190 & 1.20 & 2.4 & 1.3 & 0.51 &      & 210 & $-40$ \\
11 & 0.224 & 0.500 & 0.20 & 1.9 & 1.7 & 0.30 &      & 230 & 8 \\
13 & 0.230 &       & 0.03 & 2.8 & 1.0 & 0.47 &      & 200 & $-30$ \\
14 & 0.275 &       & 0.04 & 3.2 & 2.5 & 0.64 &      & 320 & $-20$ \\
19 & 0.253 &       & 0.07 & 3.2 & 1.5 & 0.72 &      & 320 & $-20$ \\
\noalign{\smallskip}
\hline
\end{tabular}
\end{table}

\begin{figure}[]
\centerline{\includegraphics[scale=0.7]{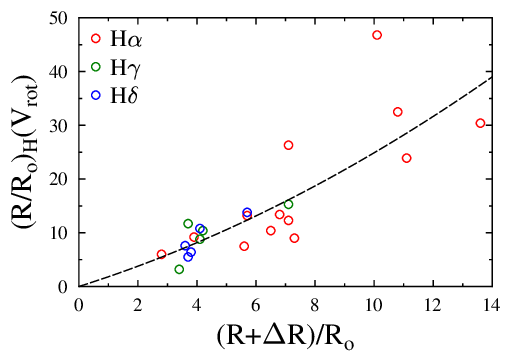}} 
\caption{\label{corr_rp_rd} Correlation of Huang's radii derived with $V_{\rm rot}$ against line formation extends $R+\Delta R$. In the ordinates the values are from Table~\ref{dpdhm}.} 
\end{figure}

\subsection{Fit of the line emission components and determination of disk parameters}\label{flecrp}  

  In our approach the main free parameters are: 1) $\tau_Z$, the opacity of the ring in the $Z-$direction; 2) the radius $R/R_o$ of the ring; and 3) the rotation and expansion velocity parameters $V_{\rm rot}$ and $V_{\rm exp}$, respectively. At each iteration step, the extent $\Delta R$ of the emitting region in the disk is calculated as a function of the chosen value of $R/R_o$.  We initiate the iterative fitting procedure of the line emission profiles by adopting the calculated value of $[\eta/(1+\eta)]^{1/2}B^*/F_{\rm c}$ and of $\overline{H}/R_o$ that corresponds to the chosen value of $R$. In the search for finer adjustments, these quantities can be considered as free parameters. This enables us to test the relevance of the assumptions made on the models of CDs relative to 
$\eta^{1/2}B^*/F_{\rm c}$ and $\overline{H}/R_o$. The inclination angle of the star-disk system is  taken from Table~\ref{corr_par_mode_t} already determined with the remaining astrophysical parameters of stars.\par   
  All fitting parameters were obtained through a trial-and-error procedure, which is not arbitrary because the global intensity of the emission, the central absorption component in emission lines, and the width of the line wings are sensitive to the ratios between the searched quantities. Nevertheless, we cannot discuss their uniqueness -- although it would be  possible perhaps if a robust method of adjustment such as the ``annealing method" were used \citep{metro1953}. The parameters displayed in Tables~\ref{ha_fit}, \ref{hg_fit} and \ref{hd_fit} must then be considered first-order estimates. We note that there are additional uncertainties affecting them which are related to the shape of the emission line profiles proper, mainly when it comes to the extracted ones in H$\gamma$ and H$\delta$ lines. \par
    
\begin{table}[] 
\centering
\caption[]{\label{nh_b2} Mean non-LTE departure coefficients in the line-forming regions, the coefficient $n_1$ for the inner disk density distribution (Eq.~\ref{eq3}), the base density, $\rho_o$, of the CD and the CD mass in the Balmer line formation region, $M_{\rm CD}$.}
\tabcolsep 3.5pt
\begin{tabular}{rcccccc} 
\hline\hline
\noalign{\smallskip}

Star  & $b_2$ & $b_2$ & $b_2$ & $n_1$ &  
$\rho_o\pm\sigma_{\rm \rho_o}$ & 
$M_{\rm CD}$ \\ 
     & $H\alpha$ & $H\gamma$ & $H\delta$ & & g\,cm$^{-3}$ & $M_{\odot}$ \\
\hline\noalign{\smallskip}
   1 & 0.04 & --- & --- &  ---- & $1.5\pm0.2\,10^{-11}$ &  $2.0\,10^{-08}$\\ 
   2 & 0.25 & 1.25 & --- &  4.8 & $6.4\pm1.3\,10^{-11}$ & $2.1\,10^{-08}$\\ 
   3 & 0.01 & --- & --- &  ---- & $1.0\pm0.2\,10^{-10}$ & $4.0\,10^{-07}$\\ 
   4 & 0.36 & --- & --- &  ---- & $6.0\pm0.3\,10^{-11}$ &  $2.4\,10^{-08}$\\ 
   5 & 1.78 & --- & --- &  ---- & $2.0\pm0.1\,10^{-13}$ &  $1.0\,10^{-09}$\\ 
   6 & 0.55 & --- & --- &  ---- & $5.7\pm0.7\,10^{-12}$ & $7.7\,10^{-08}$\\ 
   7 & 0.42 & --- & --- &  ---- & $2.7\pm1.3\,10^{-12}$ &  $5.2\,10^{-09}$\\ 
   8 & 2.05 & 4.26 & 4.31 & 4.8 & $1.5\pm0.4\,10^{-11}$ & $6.5\,10^{-08}$\\ 
   9 & 0.23 & 1.68 & 1.18 & 1.1 & $3.7\pm3.7\,10^{-12}$ & $5.0\,10^{-08}$\\ 
  10 & 0.62 & --- & --- &  ---- & $6.0\pm0.3\,10^{-12}$ & $2.0\,10^{-10}$\\ 
  11 & 0.03 & 1.38 & 1.54 & 7.1 & $1.7\pm0.5\,10^{-10}$ &  $9.5\,10^{-8}$\\ 
  12 & 0.29 & --- & --- &  ---- & $4.0\pm0.3\,10^{-13}$ & $1.6\,10^{-09}$\\ 
  13 & 0.05 & 0.73 & 0.92 & 6.1 & $8.7\pm1.0\,10^{-11}$ &  $9.1\,10^{-08}$\\ 
  14 & 2.53 & 2.53 & 3.11 & 0.5 & $1.2\pm0.5\,10^{-12}$ &  $4.3\,10^{-09}$\\ 
  16 & 0.04 & --- & --- &  ---- & $1.1\pm0.2\,10^{-11}$ &  $1.5\,10^{-08}$\\ 
  17 & 0.07 & --- & --- &  ---- & $1.4\pm0.2\,10^{-11}$ & $1.0\,10^{-08}$\\ 
  19 & 0.53 & 1.12 & 1.55 & 5.9 & $1.9\pm0.4\,10^{-11}$ & $1.5\,10^{-07}$\\  
\noalign{\smallskip}  
\hline
\end{tabular}
\end{table}

\subsection{Extent of the line emission regions}\label{eler}

 The Gaussian fit of the H$\alpha$ emission intensity maps obtained with long baseline interferometry for about 20 Be stars \citep{quirr1997,
tyc2004,tyc2005,tyc2006,tyc2008,koub2010,delaa2011,touh2013,almei2020}, produce an average observational estimate of the extent of the emission region $\langle E/R_o\rangle=7.4\pm3.5$. This value must be compared with our $R+\Delta R$ determinations, which for 17 program Be stars lead to the average $\langle(R+\Delta R)/R_o\rangle=7.4\pm2.8$ that can be considered in good agreement with interferometrical estimates.\par 

\subsection{Density distribution in the CD}\label{ddcd} 
 
       Only a theory for the CD formation and evolution can determine the function $D(R)$ in Eq.~\ref{eq2} as in VDD model that depends on the chosen viscosity. In this work, we simply assume an analytic expression for $D(R)$ to be valid for the studied line formation region. \par 
    In recent interpretations of the photometric, spectroscopic, and polarimetric variations of $\omega$~CMa and 66~Oph \citep{ghore2021,
marr2021}, $D(R)$ is characterized in Eq.~\ref{eq2} by a single global exponents, $n,$ that describe the several disk formation phases. These values range from some $n\lesssim4.0$ at CD formation phases to $n\gtrsim2.0$ at disk dissipation epochs. Nevertheless, much higher values of $n$ may describe the disk formation controlled by periodic mass injection rates, where the injection and dissipation timescales are similar, as shown by \citet{haubois2012}. \par 
    For the H$\alpha$ line formation region, we made the conservative assumption $n_2=3.0$ for all program stars, because in our method we cannot determine it otherwise. Since the formation region of the H$\alpha$ emission component is situated at radii of $R\gtrsim4R_o$, the value of $n_1$ has no incidence on the determination of the hydrogen disk base density 
$\rho_o$ g\,cm$^{-3}$.\par  
    In stars where we could extract reliable emission profiles for the 
H$\gamma$ and H$\delta$ lines, we estimated the exponent $n_1$ that characterizes the inner layers of the CD in a relative way by demanding that the values of $\rho_o$ derived from the optical depths of all three lines be the same and equal to that found with H$\alpha$. In Table~\ref{nh_b2}, we give the values of $n_1$ that were thus derived. These values are, on average, $\langle n_1\rangle=4.3\pm2.5$ -- similar to the estimates by \citet{vieira2017}, where $n$ is nevertheless meant to represent the entire radial density distribution at CD forming phases. We can also note that the average $[n_1+n_2(=3)]/2=3.7\pm1.3$, which is close to the values of $n$ estimated by several authors which used a single $n$ to characterize the entire CD, namely, \citet{silaj2010,arcos2017}. \par 
    The large variety of $n_1$ values estimated in this work may then correspond to different evolution phases of disks, from their formation to their likely steady state, or dissipation. In cases where spectroscopic observations of Be stars exist only for specific dates, a method such as  the one developed in the present work, which is of relative simple use, may still bring useful insights on the nature of CDs. In the present case, the higher values of $n_1$ suggest that in many Be stars the disk formation could respond to periodic mass injection rates, as discussed in \citet{haubois2012}. It is worth noting that further detailed studies of the H$\gamma$ and H$\delta$ emission lines at different mass ejection episodes could reveal the physical properties of regions where  the viscous transfer of angular momentum is probably organized to the rest of the CD regions and responsible for its Keplerian rotation. \par      
  From the radial optical depth $\tau_Z$, the non-LTE departure coefficients averaged over the respective line formation regions, and assuming that the He abundance in number is roughly $N_{\rm He}\simeq N_{\rm H}/4$, we readily determine the CD base density, $\rho_o$, which is given in Table~\ref{nh_b2}. These values are of the same order of magnitude than adopted in other more detailed studies  \citep{silaj2010,catan2013,arcos2017,vieira2017}. \par 
  The integration of Eq.~\ref{eq2} together with the use of the obtained values of $\rho_o$, lead us to an estimate of the total mass, $M_{\rm CD}$, gathered in the formation regions of the studied Balmer line emissions that are listed in Table~\ref{nh_b2}.\par 
  
\subsection{Additional comments on the CDs}  

  Our CD emission model is challenged with the star $n=5$. This object is a binary system, where it is likely that contributions by two distinct regions in the disk would be needed to account for the entire H$\alpha$ line emission profile in a more reliable way. \par
  
  \citet{arcos2017} obtained a correlation at 61\% between Huang's radii and the radii $R_{90}$, which represents the distance in the CD up to which $90$\% of the line emission is formed. We note that the $V\!\sin i$ parameters used to this end were not measured, but obtained as one among several parameters used to fit the H$\alpha$ emission line profiles.  We obtain a similar correlation at 68\% between Huang's radii calculated with the fitting quantity $V_{\rm rot}$ and then distance $R+\Delta R$ (see Fig.~\ref{corr_rp_rd}). However, because in both approaches Huang's radii are calculated with fitted rotational velocities, doubts can be cast on the physical reliability of these correlations. \par   
       
Then, $V_{\rm exp}$ is a fitting parameter needed to account for the line emission asymmetry. It resumes perhaps the effects carried by the known quasi-cyclic V/R variations attributed to the global one-armed disk density oscillations \citep{kato83,okazaki91,okazaki97}.\par    
   Finally, in this work we did not attempt to interpret the correlations implying emission lines and photometric measurements, simply because it would require to include a study of the formation of the continuum energy distribution with approaches like those carried by \citet{mouj1998PhDT,mouj2000,mouj2000b}, which are not in the scope of the present paper.\par  

\section{Discussions}\label{discussions} 
  
\subsection{Comments on the pnrc parameters}\label{cotpp} 

  A rotating star of mass, $M,$ behaves as an object with lower effective mass $M-\Delta M$, where $\Delta M$ is proportional to the pressure-weighted average of the centrifugal to the gravitational force over the entire object \citet{sackmann1970,clement1979}. Consequently, the rotating star produces a lower bolometric luminosity during roughly the first half of the main sequence (MS) evolutionary phase and has longer evolving time scales than its non-rotating counterpart. Rotational mixing carries fresh H-fuel into the convective core, which still prolongs its life in the MS phase \citep{maeder2000}. This mixing also transports helium (and other H-burning products) into the radiative envelope, which lowers its opacity and contributes to the enhancement of the emitted stellar luminosity over that produced by the non-rotating objet with same mass, taking place near to the second half of the MS phase \citep{maeder2009}.\par
\begin{figure}[h]   
\centerline{\includegraphics[scale=1.0]{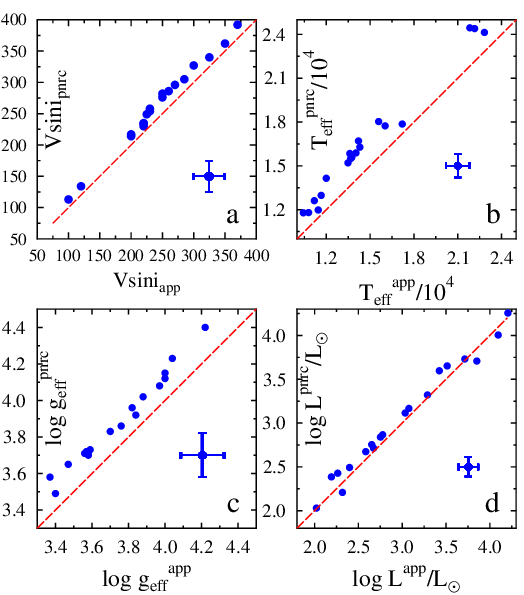}} 
\caption{\label{pnrc-app} Comparison between the  pnrc parametric set $(V\!\sin i,T_{\rm eff},\log g, \log L/L_{\odot})$ obtained for $\Omega/\Omega_{\rm c}=0.95$ with the apparent one. The average uncertainties of the plotted quantities are indicated. The diagonal in red represents the identity relation.}
\end{figure}

  
  The apparent luminosity depends mainly on the geometrical deformation of the star, the gravitational darkening effect (GD) and the inclination angle at which is seen the object. Generally, the GD effect produces an average lower apparent effective temperature and the geometrical deformation of the star leads to an apparent average lower effective gravity. This explains that the pnrc effective temperatures and gravities are larger than the apparent ones. Concerning the bolometric luminosity, the picture is less clear and strongly depends on the combination of the astrophysical parameters, mainly on the rotation rate and on the stellar aspect angle. Roughly, the ratio of the apparent to the pnrc bolometric luminosity is $L^{\rm app}(i,\Omega)/L^{\rm pnrc}\simeq$ $[S(i,\Omega)/S_{\rm pnrc}][\langle T^{\rm app}_{\rm eff}(i,\Omega)\rangle/T^{\rm pnrc}_{\rm eff}]^4$, where ``$S$" is the area of the apparent stellar hemisphere and that of the non-rotating  counterpart, respectively. In this relation we take the effective temperature of the rotating object averaged over the deformed observed stellar hemisphere. See details on the behavior of $L^{\rm app}(i,\Omega)$ and $T^{\rm app}_{\rm eff}(i,\Omega)$ in \citet{zorec_2016}. The conditions $L^{\rm pnrc}\gtrsim L^{\rm app}(i,\Omega)$ or $L^{\rm pnrc} \lesssim L^{\rm app}(i,\Omega)$ are controlled by the competition between $S(i,\Omega)$ and $T^{\rm app}_{\rm eff}(i,\Omega)$. The apparent $V\!\sin i$ parameter is underestimated because the equatorial regions, which have the larger surface linear rotational velocities, reduce their contribution to the rotational line broadening due to the GD effect. A comparison between the  pnrc parametric set $(V\!\sin i,T_{\rm eff},\log g, \log L/L_{\odot})$ obtained for $\Omega/\Omega_{\rm c}=0.95,$ with the apparent one is shown in Fig.~\ref{pnrc-app}.\par
\begin{figure}[h]  
\centerline{\includegraphics[scale=0.85]{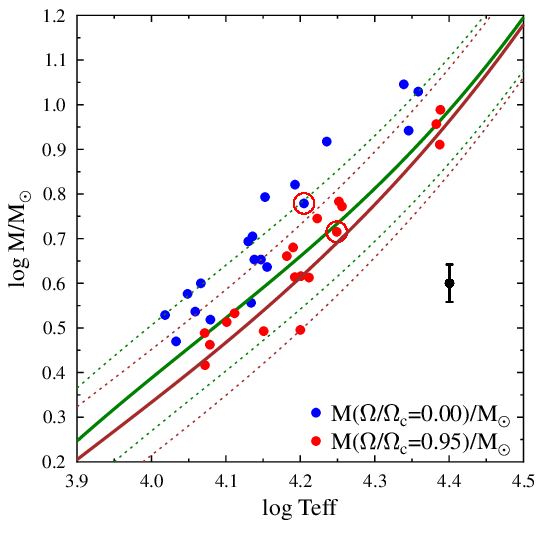}} 
\caption{\label{logm-logt} Comparison of Be stellar masses obtained in this paper with those of non-Be stars against the effective temperature. Blue dots represent the masses obtained from the apparent astrophysical parameters. Red dots represent the pnrc stellar masses determined for $\Omega/\Omega_{\rm c}=0.95$. The green curve and the dotted green lines represent the average empirical relation curve with the error margins obtained by \citet{harmanec1988} ({\color{green}\rule{0.8cm}{0.05cm}}). The red curve and the dotted brown lines represent the average empirical relation with the error margins obtained from LAMOST data \citep{xiong2023}({\color{red}\rule{0.8cm}{0.05cm}}). The typical error bar that characterizes the apparent and pnrc mass determinations is also shown. The points with a red circle identify the star N$^o$~5.}
\end{figure}

  A test on the reliability of the pnrc parameters can be done by comparing the pnrc stellar masses with the masses obtained using detached binary systems. The pnrc stellar masses are meant to represent the actual ones. This comparison is shown in Fig.~\ref{logm-logt}. The blue points are for masses determined using the apparent astrophysical parameters and the red ones are for masses determined with the pnrc parameters. In this figure, the average $(\log M/M_{\odot},\log T_{\rm eff})$ relations are from \citet{harmanec1988} (green curves) and from \citet{xiong2023} (brown curves), with the latter obtained in the frame of the LAMOST medium-resolution spectroscopic survey. We see that the pnrc masses closely lie, within the uncertainty strips, around the mean relation of masses measured in detached binary systems.\par   

  In principle, an O-, B- or A-type star may acquire the required properties to display the Be phenomenon in several ways. It may be due to: formation and pre-main sequence evolution circumstances of single stars; internal angular momentum evolution properties during the MS phase \citep[see references in][]{zorec2023}; a rejuvenating process as a consequence that one of the components has gained mass and angular momentum during mass transfer phases in binary systems \citep{packet1981,
pols1991,demink2013}. Depending on the several circumstances that may characterize the mass transfer phenomenon, the resulting Be stars should be binaries or single stars \citep{shaoli2014,hastings2021}. If they are binaries, their companions can be stripped sub-dwarf O/B (sdO/B) He stars (Be$_{\rm He}$), white dwarfs (Be$_{\rm WD}$, neutron stars (Be$_{\rm NS}$), or black holes (Be$_{\rm BH}$). Single Be stars can be merger objects (Be$_{\rm merger}$), disrupted Be$_{\rm NS}$ or Be$_{\rm BH}$ binaries. A significant number of recent observational data show that Be stars are binaries indeed, where the secondary is probably a stripped sdO/B object \citep[][and references therein]{bodensteiner2020,elBadry2022}. The paradigm of mass transfer may perhaps explain the nature of the star N$^o$~5 studied in this paper and that we briefly discuss in Sect.~\ref{tbsnf}.\par
  From the above-cited set of possible companions of Be stars in binary system, those whose presence was neglected in the present work but may in principle have some effect on the $T_{\rm eff}$ determination include: white dwarfs and the stripped sdO/B companions. However, as white dwarfs have radii of $R_{\rm WD}/R_{\odot}=$ $0.01(M_{\odot}/M_{\rm WD})^{1/3}$, masses of $0.15\lesssim M_{\rm WD}/M_{\rm WD}\lesssim1.4,$ and effective temperatures of $8000\lesssim T^{\rm WD}_{\rm eff}\lesssim 40\,000$ K, they can hardly carry any effect on the estimated $T_{\rm eff}$ of our program Be stars. This may not be the case with hot sub-dwarf companions. 
According to \citet{heber2009}, hot sub-dwarfs have most frequently effective temperatures in the range of $20\,000 \lesssim T_{\rm eff_{\rm sd}}\lesssim40\,000$ K, and radii in the interval $0.15\lesssim R_{\rm sd}/R_{\odot}\lesssim 0.25$. If we neglect their presence, the extreme errors that we will risk to commit on the estimate of the apparent effective temperatures of Be stars, can be estimated using the following extreme approximate bolometric luminosity relations, namely, both stars are seen or the sdO/B eclipses the Be star:  
$(s_{\rm B}+s_{\rm sd})T^4_{\rm Be}=s_{\rm B}T^4_{\rm B}+s_{\rm sd}T^4_{\rm sd}$ and $s_{\rm B}T^4_{\rm Be}=(s_{\rm B}-s_{\rm sd})T^4_{\rm B}+s_{\rm sd}T^4_{\rm sd}$, where $T_{\rm Be}$, $T_{\rm B}$ and $T_{\rm sd}$ are: the effective temperature obtained with GIRFIT of the studied Be star, the actual effective temperature expected for the Be star, and the effective temperature of the sub-dwarf, respectively; the $s_{\rm B}$ and $s_{\rm sd}$ represent the areas of the Be and sub-dwarf respective apparent stellar surfaces. Because $r=s_{\rm sd}/s_{\rm B}\ll1$, we have for both cases $T_{\rm B}\approx$ $T_{\rm Be}\times\{[1-r(T_{\rm sd}/T_{\rm Be})^4]\}$, which applied individually to our program objects enables us to estimate the effective temperature and the radius of the sub-dwarf star that carry on the $T_{\rm eff}$ estimate an error on the order of the uncertainty characterizing our effective temperature determinations. The average effective temperature and radius of sdO/B stars thus calculated are: 
$\langle T_{\rm sd}\rangle=$ $43600\pm13000$ K and
$\langle R_{\rm sd}/R_{\odot}\rangle=$ $0.27\pm0.06$, respectively. According to these results, the sdO/B stars in our sample should belong to the largest and hottest population of sdO/B stars. Such objects would certainly leave spectral signatures in the wavelength region studied in this work by \ion{He}{ii} $\lambda\lambda$ 4100, 4200, 4338, \ion{C}{iv} $\lambda$ 4442 and \ion{N}{iii} $\lambda$ 4379 lines \citep{drilling2013}, which are not obvious to identify in our spectra with certainty. We may then conclude that if these sdO/B are proven to exist, their effect on the $T_{\rm eff}$ determinations will be at most of the order of uncertainties inherent to the spectral resolutions and S/Ns used in this work. However,  star N$^o$~5 decidedly requires a detailed study of the actual nature of the secondary, which (according to the characteristics of eclipses) could be larger than those considered in the above estimates. \par   
   It is worth recalling that the Be phenomenon proper remains linked to the physical characteristics of the O-, B- or A-type star, whatever  the origin of Be stars as either single or merger byproducts. However, when dealing with binary systems, precise determinations of the astrophysical parameters of Be stars still requires that from now on it would be wise to carry out a thorough inquiry on the presence of companions and and on their characteristics, as well as on their perturbation power on the energy distribution of the primary in the studied spectral interval.\par    
 
\subsection{The binary star N$^o$~5}\label{tbsnf}

  The astrophysical parameters of the star N$^o$~5 were determined assuming that the object was a single star, although the light curve clearly indicates that it is a close eclipsing binary system \citep{klagyivik2017}. Neglecting the presence of the companion and the perturbations that it may introduce to the SED of the primary would lead us to infer the improbable inclination angle for the primary $i\sim50^o$ if $\Omega/\Omega_{\rm c}=0.95$. In fact, we could expect that the most probable inclination should be close to  $i\sim90^o$, because we are dealing with an eclipsing binary. However, assuming $i\sim90^o$ and consequently fixing the value of the rotation parameter of the primary at $V\!\sin i=337$ km~s$^{-1}$, the GIRFIT fitting procedure produces bad adjustments of the synthetic spectra with the observed ones, and ends up with the following estimates of the apparent astrophysical parameters: $T_{\rm eff}=14000\pm700$ K and $\log g=3.20\pm0.16$.\par
  Moreover, the H$\alpha$ emission line has likely a shell aspect. Other shell signatures are also found in the unique spectrum at our disposal (\ion{He}{i} 6678 and 7065 \AA). Unfortunately, the scarce data we have on this star do not help in determining whether the shell features are produced by a circumstellar disk or whether they are signatures attributed to the secondary component.\par
  
  Due to the combined effect of gravitational and centrifugal forces, the circumstellar disk of the Be component will tend to be on or around its equatorial plane and probably near or over the binary orbital plan. However, 
we cannot give in to the temptation of speculating on a possible complex circumbinary envelope deploying strongly outside the binary plane, produced by a mass transfer episode between the components, such as those recently predicted by \citet{smallwood2023}. Such complex circumbinary gaseous structures can lead to misinterpretations of the position (inclination) of the circumstellar disk of the Be star. \par

\subsection{Considering the rotational velocity rates}\label{otrvr}

  Regarding the corrections of effects carried by the rapid rotation on the apparent astrophysical parameters, we note that if we had enough independent apparent parameters as in the case of the calculations presented in \citet{frem2005} or \citet{zorec_2016}, we could produce an estimate of the actual rotation frequency for each star individually. Since in this paper we do not have the required number of independent entry parameters, we ought to parameterize the ratios $\Omega/\Omega_{\rm c}$.\par
  Thus, we considered only the two extreme cases $\Omega/\Omega_{\rm c}=0.0$ and 1.0, for which the average velocity ratios corrected for GD effects are $\langle V/V_{\rm c}\rangle =$ $0.85\pm0.20$ and $\langle V\sin i/V_{\rm c}\rangle =$ $0.91\pm0.20$, respectively. In turn, these ratios correspond to $\langle\Omega/\Omega_{\rm c}\rangle =$ $0.96\pm0.10$ and $\langle\Omega/\Omega_{\rm c}\rangle =$ $0.99\pm0.05$, or to equatorial force ratios
$\langle\eta=F_{\rm centrifugal}/F_{\rm gravity}\rangle = $ $0.64\pm0.33$ and $\langle\eta\rangle =$ $0.77\pm0.29$, respectively. Thus, the force ratios clearly indicate that on average our stars have quite under-critical rotations, even though the average  $\Omega/\Omega_{\rm c}$ ratios seem to suggest otherwise \citep[see][]{zorec_2016,zorec2023}.\par  
  The obtained rotation rates can be translated into ratios of rotational frequencies that are in the interval $\langle\nu_{\rm r}(\Omega/\Omega_{\rm c}\!\!=\!\!1.0)/\nu_{\rm r}(\Omega/\Omega_{\rm c}\!\!=\!\!0.95)\rangle=$ $1.10\pm0.02$. This range can be taken as an indicator of the uncertainties affecting the rotational frequencies due to possible undue choices of the individual $\Omega/\Omega_{\rm c}$ values.\par
 
\section{Concluding remarks}\label{conclusions}

  A set of faint 19 B-type stars of which 18 are Be stars, observed by the satellite CoRoT in the fourth long run (LRA02), was studied by spectroscopy. Using spectra in the $\lambda\lambda\, 4000-4500$ \AA\ obtained during a spectroscopic campaign with the VLT/FLAMES instruments at ESO. When necessary, the SEDs of Be stars were corrected for the veiling effect due to the circumstellar disk emission. All studied stars were assumed to be single stars or their SEDs  had not been perturbed in the analyzed spectral range by the presence of possible stripped sdO/B components produced during the mass transfer phase in binary systems.\par 
 The fit of spectra with models of stellar atmospheres using our GIRFIT package based on the MINUIT minimization algorithm developed by CERN, produced apparent astrophysical parameters of stars, which were then corrected for the effects induced by the rapid rotation according to parameterized ratios $\Omega/\Omega_{\rm c}$. The corrected astrophysical parameters enabled us to estimate true rotation frequencies that will be used in a following paper dealing with the analysis of non-radial pulsations derived from the light curves observed by CoRoT. We have concluded that it is difficult to decide whether the estimated effective temperatures in this work are systematically affected by the presence of sub-dwarf companions, because the expected mis-estimate of the effective temperatures depend on the actual characteristics of the companion sdO/B stars, and that 
they could be on the order of uncertainties related  to the resolution and S/Ns of the spectra analyzed.\par 
  The masses of stars derived with the pnrc astrophysical parameters and models of stellar evolution are in agreement with the masses measured in detached binary systems. Moreover, the obtained astrophysical parameters faithfully reproduce the stellar distances measured by GAIA.\par
  We note however, that the inclination angle and  other astrophysical parameters of the star N$^o$~5 remain uncertain because in this case, the perturbation due to the secondary could be not negligible. Unfortunately, we do not have enough data to study this issue and decide whether some spectral features are genuine spectroscopic shell characteristics or they are signatures due to the presence a binary component.\par  
  Using the pnrc astrophysical parameters lead us to conclude that the studied Be stars are rapid rotators, but the relation of the equatorial centrifugal force to the gravity indicate that they are far from being critical rotators. Critical or near-critical rotation is a necessary ingredient to favor massive mass ejections related with the non-radial pulsations \citep{kee2016}. Nevertheless, the inferred  $\Omega/\Omega_{\rm c}$  ratios are large enough to favor the appearance of more or less extended regions in the stellar envelope and in the sub-photospheric regions that become unstable to convection \citep{clement1979,maeder2008,cantiello2009}. Since convection favors the setting out of differential rotation with a concomitant production of internal  magnetic fields, magnetic dynamos, and cyclic activities \citep{zor2011,zorec2023,keszthelyi2023}, we may ask whether instabilities related with these magnetic fields could be the cause of significant upheavals in the outer stellar structure, where the correlated non-radial pulsations and huge sporadic mass ejections could both be phenomena triggered by these disturbances. In the framework of this phenomenology, the study of non-radial pulsations in Be stars may acquire a new major area of interest centered on the study of the internal structure of rapidly rotating stars.\par 
  Spectra were obtained also in the region of the H$\alpha$ line to confirm the Be nature of objects in the chosen set of stars. For some objects, emission components could also be extracted in the H$\gamma$ and H$\delta$ lines. Using a simplified representation of the line emitting regions in a flat CD, we inferred indications on the physical properties of disks surrounding the observed Be stars. A non-LTE approach was employed to calculate the source function of lines and their opacity based on first principles that control the level populations of atoms. Entrusting the fit obtained of the observed H$\alpha$, H$\gamma,$ and H$\delta$ emission lines profiles with model CD disks oriented according to the inclination angles derived from the astrophysical parameters, we have discussed the density distribution in the CDs. We thus found that the exponent of the density distribution in the radial direction of the circumstellar disk is a function of the distance, $R$.\par 
  Although the VDD models also predict exponents of the radial distribution of the density in the CDs that depend on the distance, our results point out on the necessity to carry out controls of the predicted exponents using detailed analysis of spectral lines that preferentially form in regions close to the central object. These regions are fed with material by sporadic huge mass ejections and variable stellar winds that provide the material to build up the CD. The spectral lines formed in these regions may reveal the actual physical properties of layers which make the transition from the star to the CD and where is organized the transport of the angular momentum that in fine determines the Keplerian rotation of disks in Be stars.\par  
  The obtained extents of the H$\alpha$ line emission formation regions are in agreement with similar quantities estimated elsewhere by interferometry in other Be stars. From the derived line opacity, we inferred the CD base densities that closely approach those estimated with much detailed models developed recently by other authors. Assuming that the snapshot image brought by our spectra of the line emission profiles can be attributed to a kind of stationary structure of the CD, the disk base density enabled us to obtain a rough estimate of the mass contained in the region of the line formation regions. These estimates are also in good agreement with other published estimates in the literature, where they were derived considering that the CD are subject to a dynamical evolution as well as that a quasi stationary picture actually results from several mass injection phenomena and to their dissipation. \par   
  
\begin{acknowledgements}   
 This work has made use of data from the European Space Agency (ESA) mission {\it Gaia} ({https://www.cosmos.esa.int/gaia}), processed by the {\it Gaia} Data Processing and Analysis Consortium (DPAC,
{https://www.cosmos.esa.int/web/gaia/dpac/consortium}). Funding for the DPAC has been provided by national institutions, in particular the institutions participating in the {\it Gaia} Multilateral Agreement. This research has made use of the service of the ESO Science Archive Facility.  We are thankful to the referee for his/her attentive reading of the manuscript and for his/her many suggestions and corrections that improved the presentation of our results. We warmly thank Agnes Monod-Gayraud for the language editing of the paper.\par  
\end{acknowledgements} 

\bibliographystyle{../VZAMS/aa} 
\bibliography{46018}  

\onecolumn  

\begin{appendix}

\section{Table containing the information on FLAMES/GIRAFFE spectra and conditions of their exposure time.}\label{lotsowfg}

\setlength{\tabcolsep}{5pt}
\begin{longtable}{|c|ccccccccccc|}
\caption{\label{tabr}Log of the stars observed with FLAMES/GIRAFFE} \\

\hline 
\noalign{\smallskip}
CoRoT-ID  & N$^o$ & RA &  DEC  & Run & Grat & F & EXP  & DATE-OBS & AM &  S/N & Moon \\
          &.   &.   &.      &     &.     &.  & sec. &.         &.   &.    &       \\
\hline
\noalign{\smallskip}

\noalign{\smallskip} 
\endfirsthead
\caption{Continued.} \\
\noalign{\smallskip} 
\hline
\noalign{\smallskip}
CoRoT-ID  & N$^o$ & RA &  DEC  & Run & Grat & F & EXP  & DATE-OBS & AM &  S/N & Moon \\
          &.   &.   &.      &     &.     &.  & sec. &.         &.   &.    &       \\
\hline
\noalign{\smallskip}
\noalign{\smallskip}
\endhead
\noalign{\smallskip}
\endfoot
\noalign{\smallskip}
\hline
\multirow{3}{*}{103000272} & \multirow{3}{*}{1} & \multirow{3}{*}{06\
50\ 39.09} & \multirow{3}{*}{-03\ 18 \ 14.37} & \multirow{3}{*}{LRA2-b} & 
\multirow{1}{*}{LR2} & 1 & 3685 & 2010-11-23T06:11:00 & 1.13 & 170 & 98\% \\
\cline{6-12}
& & & & & \multirow{2}{*}{LR6} & 1 & 780 & 2010-11-23T07:25:16 & 1.07 & 140 & 98\% \\
& & & & & & 2 & 780 & 2010-11-23T07:39:09 & 1.07 & - & 98\% \\
\hline
\multirow{3}{*}{103032255} & \multirow{3}{*}{2} & \multirow{3}{*}{06\ 
51\ 23.79} & \multirow{3}{*}{-03\ 14 \ 46.51} & \multirow{3}{*}{LRA2-c} &
\multirow{1}{*}{LR2} & 1 & 3685 & 2010-12-24T04:18:40 & 1.12 & 170 & 90\% \\
\cline{6-12}
& & & & & \multirow{2}{*}{LR6} & 1 & 780 & 2010-12-24T05:32:01 & 1.07 & 160 & 90\% \\
& & & & & & 2 & 780 & 2010-12-24T05:45:54 & 1.08 & - & 90\% \\
\hline
\multirow{3}{*}{110655185} & \multirow{3}{*}{3} & \multirow{3}{*}{06\
49\ 06.10} & \multirow{3}{*}{-05\ 00 \ 21.86} & \multirow{3}{*}{LRA2-m} & 
\multirow{1}{*}{LR2} & 1 & 3685 & 2011-01-03T02:18:40 & 1.29 & 65 & B/H \\
\cline{6-12}
& & & & & \multirow{2}{*}{LR6} & 1 & 780 & 2010-12-26T04:20:50 & 1.09 & - & 72\% \\
& & & & & & 2 & 780 & 2010-12-26T04:34:43 & 1.07 & - & 72\% \\
\hline
\multirow{3}{*}{110655384} & \multirow{3}{*}{4} & \multirow{3}{*}{06\
49\ 06.80} & \multirow{3}{*}{-05\ 20 \ 41.77} & \multirow{3}{*}{LRA2-m} & 
\multirow{1}{*}{LR2} & 1 & 3685 & 2011-01-03T02:18:40 & 1.29 & 90 & B/H  \\
\cline{6-12}
& & & & & \multirow{2}{*}{LR6} & 1 & 780 & 2010-12-26T04:20:50 & 1.09 & 80 & 72\% \\
& & & & & & 2 & 780 & 2010-12-26T04:34:43 & 1.07 & -  & 72\% \\
\hline
\multirow{3}{*}{110655437} & \multirow{3}{*}{5} & \multirow{3}{*}{06\
49\ 06.98} & \multirow{3}{*}{-05\ 02 \ 27.30} & \multirow{3}{*}{LRA2-m} & 
\multirow{1}{*}{LR2} & 1 & 3685 & 2011-01-03T02:18:40 & 1.29 & 65 & B/H \\
\cline{6-12}
& & & & & \multirow{2}{*}{LR6} & 1 & 780 & 2010-12-26T04:20:50 & 1.09 & 75 & 72\% \\
& & & & & & 2 & 780 & 2010-12-26T04:34:43 & 1.07 & - & 72\% \\
\hline
\multirow{3}{*}{110662847} & \multirow{3}{*}{6} & \multirow{3}{*}{06\
50\ 10.73} & \multirow{3}{*}{-03\ 20 \ 55.79} & \multirow{3}{*}{LRA2-b} & 
\multirow{1}{*}{LR2} & 1 & 3685 & 2010-11-23T06:11:00 & 1.13 & 85 & 98\% \\
\cline{6-12}
& & & & & \multirow{2}{*}{LR6} & 1 & 780 & 2010-11-23T07:25:16 & 1.07 & 90 & 98\% \\
& & & & & & 2 & 780 & 2010-11-23T07:39:09 & 1.07 & - & 98\% \\
\hline
\multirow{3}{*}{110663174} & \multirow{3}{*}{7} & \multirow{3}{*}{06\
50\ 11.94} & \multirow{3}{*}{-04\ 40 \ 53.22} & \multirow{3}{*}{LRA2-k} & 
\multirow{1}{*}{LR2} & 1 & 3685 & 2011-01-20T04:25:44 & 1.09 & 62 & 100\% \\
\cline{6-12}
& & & & & \multirow{2}{*}{LR6} & 1 & 780 & 2011-01-22T02:50:53 & 1.08 & 450 & 92\% \\
& & & & & & 2 & 780 & 2011-01-22T03:04:47 & 1.07 & - & 92\% \\
\hline
\multirow{3}{*}{110663880} & \multirow{3}{*}{8} & \multirow{3}{*}{06\
50\ 14.36} & \multirow{3}{*}{-03\ 23 \ 59.50} & \multirow{3}{*}{LRA2-b} & 
\multirow{1}{*}{LR2} & 1 & 3685 & 2010-11-23T06:11:00 & 1.13 & 170 & 98\% \\
\cline{6-12}
& & & & & \multirow{2}{*}{LR6} & 1 & 780 & 2010-11-23T07:25:16 & 1.07 & 140 & 98\% \\
& & & & & & 2 & 789 & 2010-11-23T07:39:09 & 1.07 & - & 98\% \\
\hline
\multirow{3}{*}{110672515} & \multirow{3}{*}{9} & \multirow{3}{*}{06\
51\ 14.24} & \multirow{3}{*}{-05\ 23 \ 51.02} & \multirow{3}{*}{LRA2-q} & 
\multirow{1}{*}{LR2} & 1 & 3685 & 2010-11-27T06:15:00 & 1.09 & 59 & 69\% \\
\cline{6-12}
& & & & & \multirow{2}{*}{LR6} & 1 & 780 & 2010-11-27T07:28:53 & 1.06 & 60 & 69\% \\
& & & & & & 2 & 780 & 2010-11-27T07:42:46 & 1.07 & - & 69\% \\
\hline
\multirow{3}{*}{110681176} & \multirow{3}{*}{10} & \multirow{3}{*}{06\
52\ 08.32} & \multirow{3}{*}{-03\ 37 \ 52.33} & \multirow{3}{*}{LRA2-f} & 
\multirow{1}{*}{LR2} & 1 & 3685 & 2011-01-17T01:36:53 & 1.27 & 300 & 91\% \\
\cline{6-12}
& & & & & \multirow{2}{*}{LR6} & 1 & 780 & 2011-01-17T02:48:54 & 1.11 & 250 & 91\% \\
& & & & & & 2 & 780 & 2011-01-17T03:02:48 & 1.09 & - & 91\% \\
\hline
\multirow{4}{*}{110688151} & \multirow{3}{*}{11} & \multirow{4}{*}{06\
53\ 00.94} & \multirow{4}{*}{-04\ 40 \ 18.17} & \multirow{4}{*}{LRA2-l} & 
\multirow{1}{*}{LR2} & 1 & 3685 & 2010-12-24T06:16:48 & 1.09 & 350 & 90\% \\
\cline{6-12}
& & & & & \multirow{3}{*}{LR6} & 1 & 780 & 2010-12-24T07:28:55 & 1.24 & 250 & 90\% \\ 
& & & & & & 2 & 780 & 2010-12-25T06:54:23 & 1.16 & - & 82\% \\ 
& & & & & & 3 & 780 & 2010-12-25T07:08:17 & 1.19 & - & 82\% \\ 
\hline
\multirow{3}{*}{110747131} & \multirow{3}{*}{12} & \multirow{3}{*}{06\
50\ 27.04} & \multirow{3}{*}{-05\ 26 \ 36.86} & \multirow{3}{*}{LRA2-q} & 
\multirow{1}{*}{LR2} & 1 & 3685 & 2010-11-27T06:15:00 & 1.09 & 91 & 69\% \\
\cline{6-12}
& & & & & \multirow{2}{*}{LR6} & 1 & 780 & 2010-11-27T07:28:53 & 1.06 & 60 & 69\% \\
& & & & & & 2 & 780 & 2010-11-27T07:42:46 & 1.07 & - & 69\% \\
\hline
\multirow{3}{*}{110751872} & \multirow{3}{*}{13} & \multirow{3}{*}{06\
51\ 17.51} & \multirow{3}{*}{-4\ 36 \ 59.97} & \multirow{3}{*}{LRA2-k} & 
\multirow{1}{*}{LR2} & 1 & 3685 & 2011-01-20T04:25:44 & 1.09 & 450 & 100\% \\
\cline{6-12}
& & & & & \multirow{2}{*}{LR6} & 1 & 780 & 2011-01-22T02:50:53 & 1.08 & 450 & 92\% \\
& & & & & & 2 & 780 & 2011-01-22T03:04:47 & 1.07 & - & 92\% \\
\hline
\multirow{3}{*}{110751876} & \multirow{3}{*}{14} & \multirow{3}{*}{06\
51\ 17.52} & \multirow{3}{*}{-03\ 19 \ 29.71} & \multirow{3}{*}{LRA2-c} & 
\multirow{1}{*}{LR2} & 1 & 3685 & 2010-12-23T04:18:40 & 1.12 & 450 & 90\% \\
\cline{6-12}
& & & & & \multirow{2}{*}{LR6} & 1 & 780 & 2010-12-24T05:32:01 & 1.07 & 300 & 90\% \\
& & & & & & 2 & 780 & 2010-12-24T05:45:54 & 1.08 & - & 90\% \\
\hline
\multirow{3}{*}{110752156} & \multirow{3}{*}{15} & \multirow{3}{*}{06\
51\ 18.19} & \multirow{3}{*}{-04\ 07 \ 40.62} & \multirow{3}{*}{LRA2-h} & 
\multirow{1}{*}{LR2} & 1 & 3685 & 2011-01-21T02:36:14 & 1.09 & 60 & 98\% \\
\cline{6-12}
& & & & & \multirow{2}{*}{LR6} & 1 & 780 & 2011-01-21T03:58:07 & 1.07 & 50 & 98\% \\
& & & & & & 2 & 780 & 2011-01-21T04:12:00 &  1.08 & - & 98\% \\
\hline
\multirow{3}{*}{110827583} & \multirow{3}{*}{16} & \multirow{3}{*}{06\
48\ 58.53} & \multirow{3}{*}{-00\ 13 \ 06.25} & \multirow{3}{*}{LRA2-m} & 
\multirow{1}{*}{LR2} & 1 & 3685 & 2011-01-03T02:18:40 & 1.29 & 130 & B/H \\
\cline{6-12}
& & & & & \multirow{2}{*}{LR6} & 1 & 780 & 2010-12-26T04:20:50 & 1.09 & 100 & 72\% \\
& & & & & & 2 & 780 & 2010-12-26T04:34:43 & 1.07 & - & 72\% \\
\hline
\multirow{3}{*}{300002611} & \multirow{3}{*}{17} & \multirow{3}{*}{06\
48\ 19.69} & \multirow{3}{*}{-03\ 43 \ 27.41} & \multirow{3}{*}{LRA2-d} & 
\multirow{1}{*}{LR2} & 1 & 3685 & 2011-01-16T06:08:08 & 1.29 & 65 & B/H \\
\cline{6-12}
& & & & & \multirow{2}{*}{LR6} & 1 & 780 & 2010-11-23T08:03:29 & 1.09 & 50 & 98\% \\
& & & & & & 2 & 780 & 2010-11-23T08:17:23 &  1.10 & - & 98\% \\
\hline
\multirow{3}{*}{300002834} & \multirow{3}{*}{18} & \multirow{3}{*}{06\
48\ 25.86} & \multirow{3}{*}{-03\ 26 \ 21.14} & \multirow{3}{*}{LRA2-a} & 
\multirow{1}{*}{LR2} & 1 & 3685 & 2010-12-26T04:57:58 & 1.07 & 70 & 72\% \\
\cline{6-12}
& & & & & \multirow{2}{*}{LR6} & 1 & 780 & 2011-01-10T03:07:30 & 1.12 & 60 & B/H \\
& & & & & & 2 & 780 & 2011-01-10T03:21:24 &  1.10 & - & B/H \\
\hline
\multirow{3}{*}{300003290} & \multirow{3}{*}{19} & \multirow{3}{*}{06\
48\ 36.90} & \multirow{3}{*}{-03\ 53 \ 45.79} & \multirow{3}{*}{LRA2-d} & 
\multirow{1}{*}{LR2} & 1 & 3685 & 2011-01-16T06:08:08 & 1.29 & 60 & B/H \\
\cline{6-12}
& & & & & \multirow{2}{*}{LR6} & 1 & 780 & 2010-11-23T08:03:29 & 1.09 & 50 & 98\% \\
& & & & & & 2 & 780 & 2010-11-23T08:17:23 &  1.10 & - & 98\% \\
\hline
\multicolumn{12}{|l|}{Notes: F = ; Grant = ; Run = ; RA = right ascension' DEC = declination; EXP = exposure time in seconds; AM = air mass; } \\
\multicolumn{12}{|l|}{S/N =
signal-to-noise ratio ; Moon = Moon's illumination ; B/H = moon below Horizon}\\
\hline
\end{longtable} 
 
\vskip -10truecm

\twocolumn

\section{Comparison between an observed and the corresponding veiling corrected spectrum}\label{cooavcs}

\begin{figure}[h]
\centerline{\includegraphics[scale=0.85]{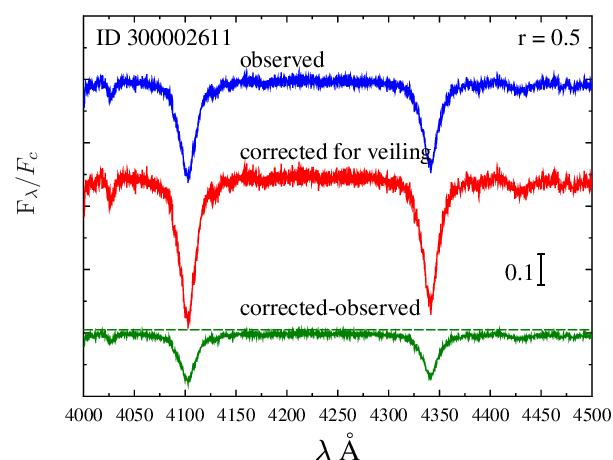}} 
\caption{\label{fig_fit17} Comparison between normalized observed and veiling corrected spectra of the program star N$^o$ 17 (ID 300002611) in the wavelength interval $\lambda\lambda\,4000-4500$~\AA.}
\end{figure}

 According to \citet{sema2013}, the expression to correct the spectra for the veiling effect is 
  
\begin{eqnarray}
\left. \begin{array}{rcl}
\displaystyle F^*(\lambda)/F^*_{\rm c}(\lambda) &
= & \displaystyle \left\{\left[F^{\rm obs}(\lambda)/F^{\rm obs}_{\rm c}(\lambda)\right][1+r(\lambda)]-r(\lambda)\right\} \\
\displaystyle  r(\lambda) & = & \displaystyle E_{\rm c}(\lambda)/[F^*_{\rm c}(\lambda)A_{\rm c}(\lambda)]
\end{array}
\right\rbrace,
\label{veil3} 
\end{eqnarray}   

\noindent where $F^{\rm obs}(\lambda)$ is the observed line spectrum;  $F_{\rm c}^{\rm obs}(\lambda)$ is the continuum spectrum emitted by the star+disk system;  $F^*(\lambda)$ and $F^*_{\rm c}(\lambda)$ are the line and continuum spectra due to the stellar photosphere, respectively. The term $``1+r"$ is called the ``veiling factor." $E_{\rm c}(\lambda)$ and $A_{\rm c}(\lambda)$ respectively represent the amount of continuum emission and absorption due to the CD that can be estimated using the empirical correlations in \citep{ball1995}. The veiling factor in Eq.~(\ref{veil3}) is a function of $\lambda$ that varies by $\Delta r/r\gtrsim0.04$ from $\lambda=4000$ \AA\ to $\lambda=4500$ \AA\ as it can be inferred from relations given in \citet{mouj1999,mouj2000} and using the required parameters calculated in \citet{zorec_2016}. The term  $r(\lambda)$ roughly behaves as   $r(\lambda)=a\times(\lambda/\lambda_{\rm H\gamma})^3+b$. \citet{frem2006} showed that by neglecting the veiling effect in Be stars displaying strong line emissions, the determination of astrophysical parameters $(T_{\rm eff},\log g)$ may suffer from substantial uncertainties. The veiling factors $r$ at $\lambda=4500$ \AA\ determined using the empirical correlations in \citet{ball1995} and the emission intensities $I_{\rm H\gamma}$ in the H$\gamma$ line are given in Table~\ref{w_int_1}. The  emission intensity in the H$\gamma$ line is defined as $I_{\rm H\gamma} = W^{\rm em}_{\rm H\gamma}\times\left\{[F^{\rm c}_{\rm H\gamma}(T_{\rm eff}, \log g)]/[F^{\rm c}_{\rm H\gamma}(22500,4.0)]\right\}$, where $W^{\rm em}_{\rm H\gamma}$ represents the equivalent width of the emission component of the $H\gamma$ line and $F^{\rm c}_{\rm H\gamma}$ is the flux of the continuum spectrum at $H\gamma$. In Table~\ref{w_int_1}, stars N$^o$ 1, 6, 12, and 16 have very low emission and accordingly it is $r\simeq0$. An example of the effect introduced on spectra by the correction by the veiling effect is shown in Fig.~\ref{fig_fit17}.\par  

 As an example of spectra before and after correction made for the veiling effect are shown in Fig.~\ref{fig_fit17}. They correspond to the Be star ID 300002611 (N$^o$ 17). The correction is made over the entire spectral range  of $\lambda\lambda 4000-4500\ \AA$. The spectra in Fig.~\ref{fig_fit17} are shown to be  arbitrarily shifted in the ordinates. In this figure is also shown the difference between the observed and veiling corrected spectra to appreciate the change in the equivalent widths of spectral lines and the almost negligible variation introduced on the energy distribution in the treated wavelength interval. \par

\begin{table}[]
\centering
\caption[]{\label{w_int_1} Equivalent width and normalized flux of the emission component in the H$\gamma$ line, and the veiling factor $1+r$.}
\tabcolsep 3.0pt
\begin{tabular}{l|cccccccccc} 
\hline\hline
\noalign{\smallskip}
Star & 1 & 2 & 3 & 4 & 5 & 6 & 7 & 8 & 9 & 10 \\
$W^{\rm em}_{\rm H\gamma}$ & 0.00 & 0.31 & 0.74 & 0.38 & 0.51 & 0.00 & 0.77 & 1.49 & 1.06 & 0.06 \\
$I_{\rm H\gamma}$ & 0.00 & 0.12 & 0.22 & 0.14 & 0.23 & 0.00 & 0.32 & 1.83 & 0.47 & 0.01 \\
$1+r$ & 1.00 & 1.04 & 1.14 & 1.05 & 1.08 & 1.00 & 1.14 & 1.36 & 1.23 & 1.00 \\
\noalign{\smallskip}\hline
\noalign{\smallskip}
Star & 11 & 12 & 13 & 14 & 15 & 16 & 17 & 18 & 19 & \\
$W^{\rm em}_{\rm H\gamma}$ & 1.24 & 0.00 & 0.37 & 0.96 & 0.20 & 0.00 & 1.89 & 0.31 & 1.36 &  \\
$I_{\rm H\gamma}$ & 0.53 & 0.00 & 0.15 & 0.96 & 0.18 & 0.00 & 0.59 & 0.12 & 1.73 &  \\
$1+r$ & 1.28 & 1.00 & 1.05 & 1.20 & 1.02 & 1.00 & 1.50 & 1.04 & 1.32 & \\
\hline
\multicolumn{11}{l}{Equivalent widths $W^{\rm em}_{\rm H\gamma}$ and intensities $I_{\rm H\gamma}$ are
given in \AA} \\
\hline
\end{tabular}
\end{table}

\section{Spectra in the blue region and the obtained best fits}\label{spect_fits}

\begin{figure*}[]
\centerline{\includegraphics[scale=0.9]{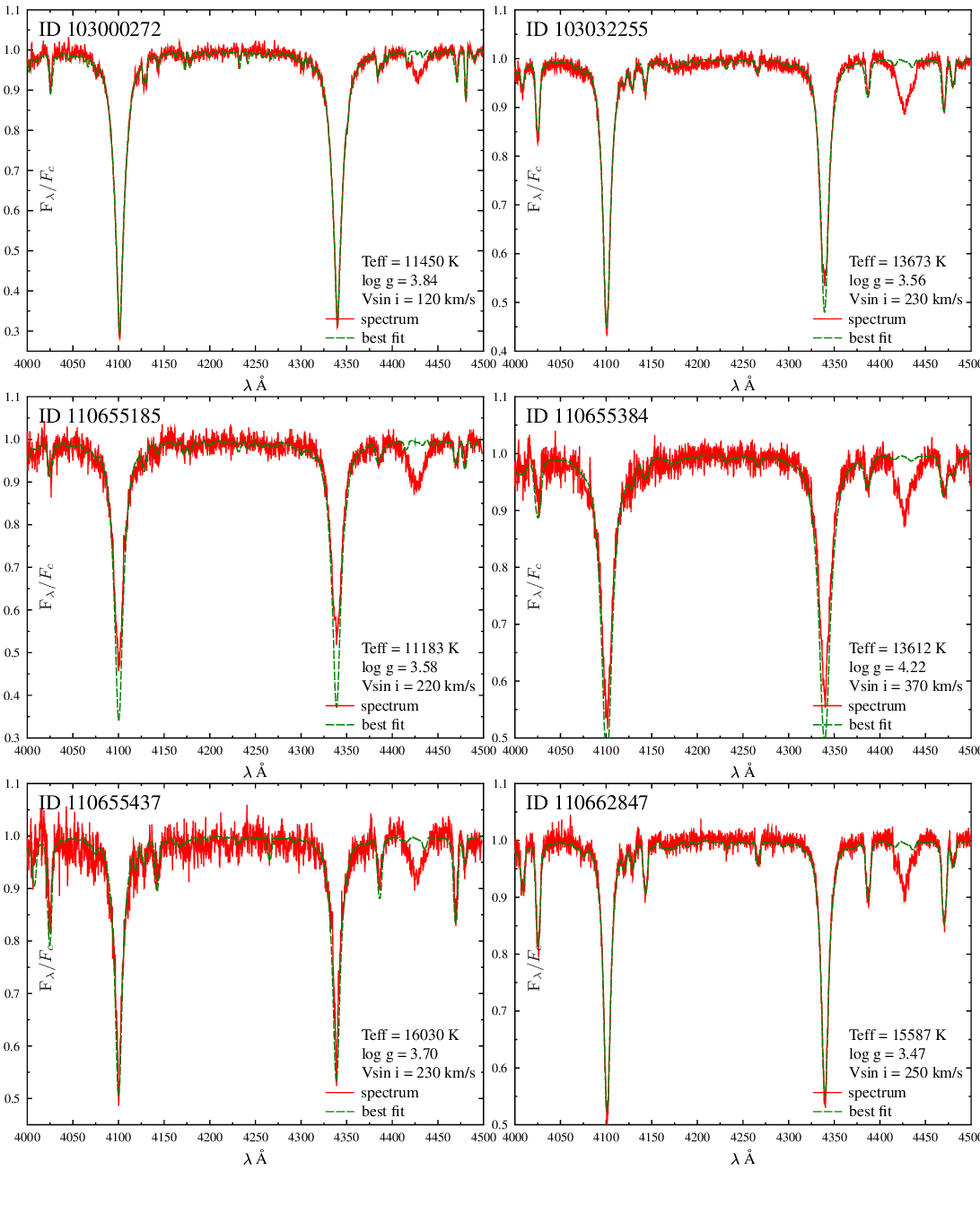}} 
\caption{\label{fig_fit1} Observed spectra corrected for veiling effects of the program stars from N$^o$ 1 to 6 in the wavelength interval $\lambda\lambda\,4000-4500$~\AA.}
\end{figure*}

\begin{figure*}[] 
\centerline{\includegraphics[scale=0.9]{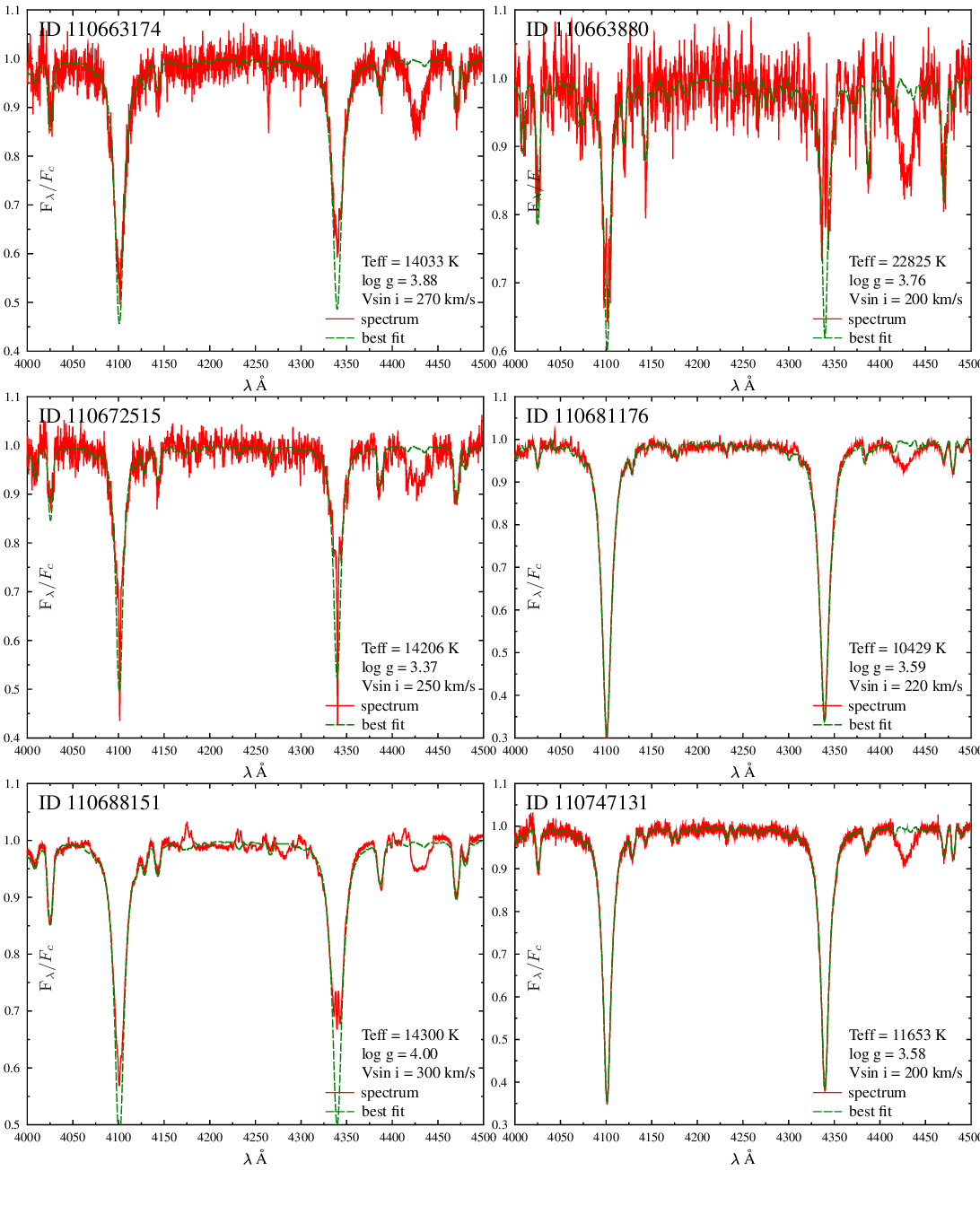}}  
\caption{\label{fig_fit2} Observed spectra corrected for veiling effects of the program stars from N$^o$ 7 to 12 in the wavelength interval $\lambda\lambda\,4000-4500$~\AA.}
\end{figure*} 

\begin{figure*}[]
\centerline{\includegraphics[scale=0.9]{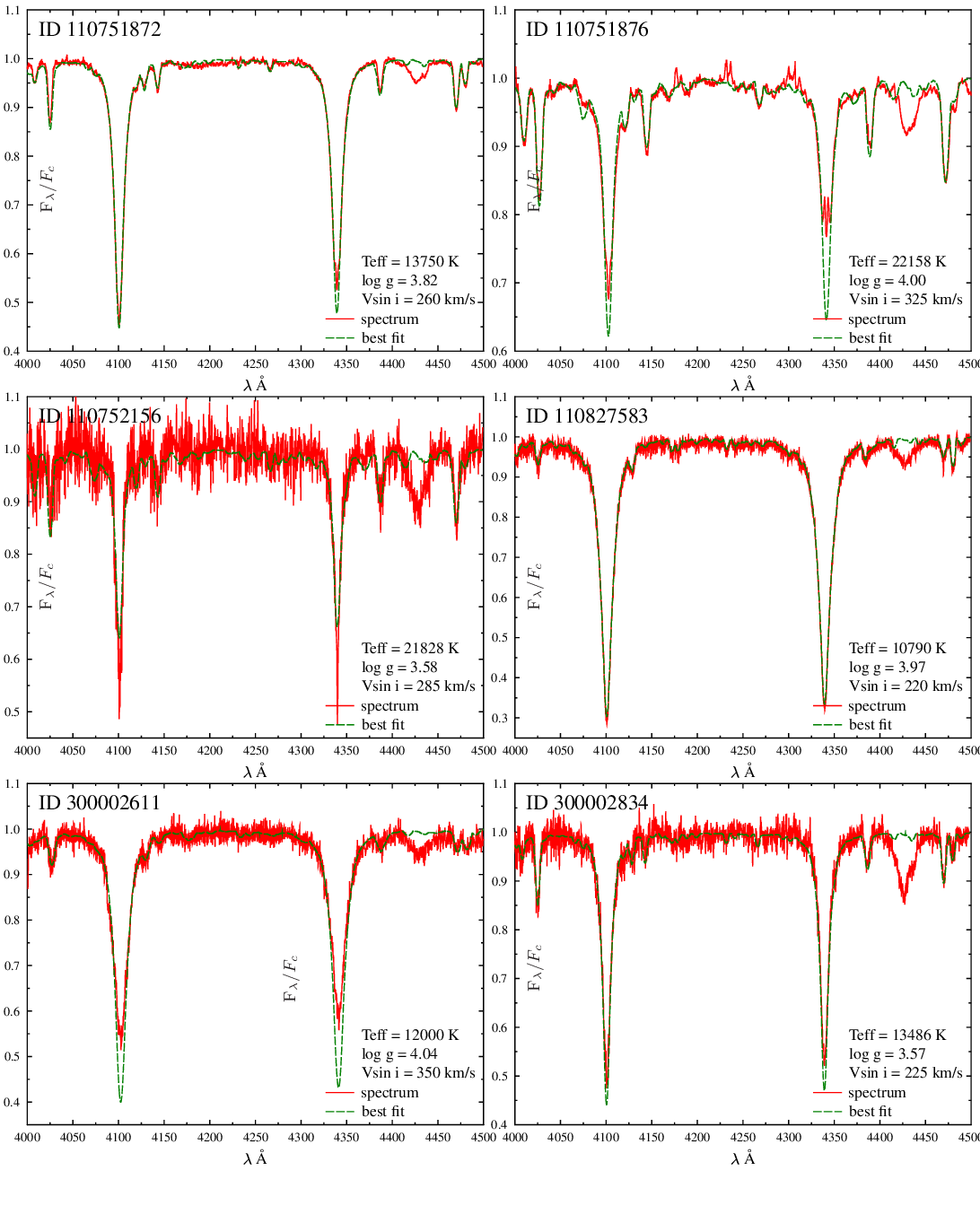}}  
\caption{\label{fig_fit3} Observed spectra corrected for veiling effects of the program stars from N$^o$ 13 to 18 in the wavelength interval $\lambda\lambda\,4000-4500$~\AA.}
\end{figure*} 

\begin{figure*}[]
\centerline{\includegraphics[scale=0.9]{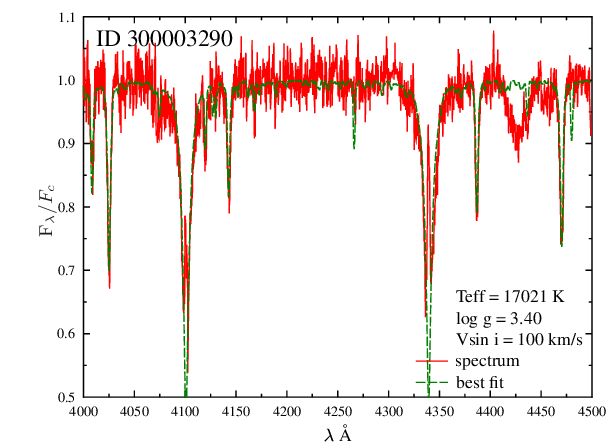}}  
\caption{\label{fig_fit4} Observed spectra corrected for veiling effects of the program star N$^o$ 19 in the wavelength interval $\lambda\lambda\,4000-4500$~\AA.}
\end{figure*}

\section{Astrophysical parameters and rotational frequencies corrected for rotational effects of the observed Be stars assuming the angular velocity ratio $\Omega/\Omega_{\rm c} = 0.95$}\label{aparfcfreotobe}

\begin{table*}[tbp]
\centering
\caption[]{\label{corr_par_prom_t} Averaged astrophysical parameters and rotational frequencies corrected for rotational effects of the observed Be stars assuming the angular velocity ratio $\Omega/\Omega_{\rm c} = 0.95$}
\tabcolsep 5.0pt
{\scriptsize
\begin{tabular}{clccccccccc}
\hline\hline
\noalign{\smallskip}
Star  &  $\langle T_{\rm eff}\rangle\pm\sigma$ & $\langle\log g\rangle\pm\sigma$ & 
$\langle V\!\sin i\rangle\pm\sigma$ & $\langle\log L/L_{\odot}\rangle\pm\sigma$  &
$\langle M/M_{\odot}\rangle\pm\sigma$ & $\langle V_{\rm c}\rangle\pm\sigma$ & 
$\langle i\rangle\pm\sigma$ & $\langle t/t_{\rm MS}\rangle\pm\sigma$ & $\langle{\rm age}\rangle\pm\sigma$ & 
$\langle\nu_{\rm r}\rangle\pm\sigma$ \\ 
      & \ \ \ \ \ K & dex &  km~s$^{-1}$ & dex & & km~s$^{-1}$ & deg. & & yr & c/d \\
\hline
\noalign{\smallskip} 
   1 & 11980 $\pm$  590 &  3.92 $\pm$  0.15 & 134 $\pm$ 13 &  2.209 $\pm$  0.085 &  2.90 $\pm$  0.35 & 368 $\pm$ 25 & 26 $\pm$  9 &  0.366 $\pm$  0.256 &  2.11E+08 $\pm$  8.01E+07 &  1.61 $\pm$  0.25 \\
   2 & 15500 $\pm$  880 &  3.71 $\pm$  0.14 & 254 $\pm$ 25 &  3.167 $\pm$  0.108 &  4.79 $\pm$  0.47 & 366 $\pm$ 21 & 58 $\pm$ 18 &  0.703 $\pm$  0.211 &  9.36E+07 $\pm$  2.74E+07 &  0.89 $\pm$  0.15 \\
   3 & 12620 $\pm$  720 &  3.72 $\pm$  0.15 & 234 $\pm$ 22 &  2.674 $\pm$  0.110 &  3.26 $\pm$  0.38 & 341 $\pm$ 21 & 57 $\pm$ 16 &  0.479 $\pm$  0.264 &  1.84E+08 $\pm$  7.22E+07 &  0.97 $\pm$  0.17 \\
   4 & 15850 $\pm$  840 &  4.40 $\pm$  0.12 & 392 $\pm$ 36 &  2.428 $\pm$  0.103 &  3.13 $\pm$  0.68 & 513 $\pm$ 28 & 68 $\pm$ 14 &  0.199 $\pm$  0.275 &  3.66E+09 $\pm$  1.15E+09 &  3.12 $\pm$  0.49 \\
   5 & 17740 $\pm$  810 &  3.83 $\pm$  0.12 & 258 $\pm$ 23 &  3.321 $\pm$  0.093 &  5.19 $\pm$  0.20 & 408 $\pm$ 21 & 50 $\pm$ 14 &  0.292 $\pm$  0.225 &  4.68E+07 $\pm$  2.35E+07 &  1.10 $\pm$  0.17 \\
   6 & 18030 $\pm$  880 &  3.65 $\pm$  0.12 & 282 $\pm$ 24 &  3.652 $\pm$  0.106 &  5.92 $\pm$  0.81 & 378 $\pm$ 19 & 67 $\pm$ 14 &  0.533 $\pm$  0.313 &  4.43E+07 $\pm$  1.82E+07 &  0.71 $\pm$  0.12 \\
\hline
\end{tabular}
}
\end{table*}

\begin{table*}[tbp]
\centering
\caption[]{\label{corr_par_mode_t} Modes of astrophysical parameters and rotational frequencies corrected for rotational effects of the observed Be stars assuming the angular velocity ratio $\Omega/\Omega_{\rm c} = 0.95$}
\tabcolsep 5.0pt
{\scriptsize
\begin{tabular}{clccccccccc}
\hline\hline
\noalign{\smallskip}
Star  &  $\langle T_{\rm eff}\rangle\pm\sigma$ & $\langle\log g\rangle\pm\sigma$ & 
$\langle V\!\sin i\rangle\pm\sigma$ & $\langle\log L/L_{\odot}\rangle\pm\sigma$  &
$\langle M/M_{\odot}\rangle\pm\sigma$ & $\langle V_{\rm c}\rangle\pm\sigma$ & 
$\langle i\rangle\pm\sigma$ & $\langle t/t_{\rm MS}\rangle\pm\sigma$ & $\langle{\rm age}\rangle\pm\sigma$ & 
$\langle\nu_{\rm r}\rangle\pm\sigma$ \\ 
      & \ \ \ \ \ K & dex &  km~s$^{-1}$ & dex & & km~s$^{-1}$ & deg. & & yr & c/d \\
\hline
\noalign{\smallskip}
   1 & 11980 $\pm$   70 &  3.98 $\pm$  0.02 & 134 $\pm$  1 &  2.208 $\pm$  0.010 &  3.12 $\pm$  0.24 & 378 $\pm$  3 & 24 $\pm$  0 &  0.581 $\pm$  0.230 &  2.51E+08 $\pm$  8.17E+06 &  1.65 $\pm$  0.03 \\
   2 & 15500 $\pm$  100 &  3.76 $\pm$  0.02 & 254 $\pm$  2 &  3.167 $\pm$  0.013 &  5.14 $\pm$  0.06 & 366 $\pm$  2 & 55 $\pm$  1 &  0.869 $\pm$  0.018 &  9.81E+07 $\pm$  2.43E+06 &  0.88 $\pm$  0.02 \\
   3 & 12620 $\pm$   80 &  3.78 $\pm$  0.02 & 234 $\pm$  2 &  2.717 $\pm$  0.013 &  3.10 $\pm$  0.55 & 349 $\pm$  2 & 53 $\pm$  1 &  0.431 $\pm$  0.109 &  1.72E+08 $\pm$  1.40E+07 &  0.95 $\pm$  0.02 \\
   4 & 15850 $\pm$  100 &  4.45 $\pm$  0.01 & 392 $\pm$  4 &  2.428 $\pm$  0.012 &  3.12 $\pm$  0.20 & 525 $\pm$  3 & 62 $\pm$  1 &  0.034 $\pm$  0.132 &  6.28E+09 $\pm$  2.58E+07 &  3.11 $\pm$  0.05 \\
   5 & 17740 $\pm$   90 &  3.88 $\pm$  0.01 & 258 $\pm$  2 &  3.321 $\pm$  0.011 &  5.29 $\pm$  0.21 & 416 $\pm$  2 & 47 $\pm$  0 &  0.282 $\pm$  0.093 &  4.31E+07 $\pm$  4.56E+06 &  1.12 $\pm$  0.02 \\
   6 & 18020 $\pm$  100 &  3.70 $\pm$  0.01 & 282 $\pm$  2 &  3.651 $\pm$  0.012 &  5.46 $\pm$  0.17 & 386 $\pm$  2 & 61 $\pm$  1 &  0.288 $\pm$  0.085 &  3.36E+07 $\pm$  5.72E+06 &  0.73 $\pm$  0.01 \\
\hline
\end{tabular}  
}
\end{table*} 

\end{appendix}

\onecolumn

{\scriptsize
\setlength{\tabcolsep}{5pt}
\begin{longtable}{clccccccccc}
\caption{\label{corr_par_prom} Averaged astrophysical  parameters and rotational frequencies corrected for rotational effects of the observed Be stars assuming different ratios $\Omega/\Omega_{\rm c}$
of angular velocities.} \\
\hline\hline
\noalign{\smallskip}
Star  &  $\langle T_{\rm eff}\rangle\pm\sigma$ & $\langle\log g\rangle\pm\sigma$ & 
$\langle V\!\sin i\rangle\pm\sigma$ & $\langle\log L/L_{\odot}\rangle\pm\sigma$  &
$\langle M/M_{\odot}\rangle\pm\sigma$ & $\langle V_{\rm c}\rangle\pm\sigma$ & 
$\langle i\rangle\pm\sigma$ & $\langle t/t_{\rm MS}\rangle\pm\sigma$ & $\langle{\rm age}\rangle\pm\sigma$ & 
$\langle\nu_{\rm r}\rangle\pm\sigma$ \\ 
      & \ \ \ \ \ K & dex &  km~s$^{-1}$ & dex & & km~s$^{-1}$ & deg. & & yr & c/d \\
\hline
\noalign{\smallskip} 
\endfirsthead
\caption{Continued.} \\
\noalign{\smallskip} 
\hline
\noalign{\smallskip}
Star  &  $\langle T_{\rm eff}\rangle\pm\sigma$ & $\langle\log g\rangle\pm\sigma$ & 
$\langle V\!\sin i\rangle\pm\sigma$ & $\langle\log L/L_{\odot}\rangle\pm\sigma$ 
& $\langle M/M_{\odot}\rangle\pm\sigma$ & $\langle V_{\rm c}\rangle\pm\sigma$ & 
$\langle i\rangle\pm\sigma$ & $\langle t/t_{\rm MS}\rangle\pm\sigma$ & $\langle{\rm age}\rangle\pm\sigma$ & $\langle\nu_{\rm r}\rangle\pm\sigma$ \\ 
\noalign{\smallskip}
\hline
\noalign{\smallskip}
\endhead
\noalign{\smallskip}
\endfoot
\noalign{\smallskip}
\multicolumn{11}{l}{$\Omega/\Omega_{\rm c} = 0.70$}\\
\noalign{\smallskip}
   1 & 11850 $\pm$  580 &  3.89 $\pm$  0.15 & 122 $\pm$ 11 &  2.319 $\pm$  0.085 &  3.06 $\pm$  0.09 & 362 $\pm$ 24 & 41 $\pm$  12 &  0.503 $\pm$  0.232 &  2.23E+08 $\pm$  7.49E+07 &  1.00 $\pm$  0.15 \\
   2 & 14660 $\pm$  730 &  3.67 $\pm$  0.16 & 229 $\pm$ 24 &  3.165 $\pm$  0.086 &  4.71 $\pm$  0.45 & 352 $\pm$ 24 & 89 $\pm$  12 &  0.808 $\pm$  0.141 &  1.09E+08 $\pm$  2.21E+07 &  0.71 $\pm$  0.12 \\
   3 & 11970 $\pm$  630 &  3.68 $\pm$  0.16 & 220 $\pm$ 24 &  2.680 $\pm$  0.090 &  3.20 $\pm$  0.12 & 330 $\pm$ 23 & 89 $\pm$ 12 &  0.578 $\pm$  0.231 &  2.14E+08 $\pm$  6.59E+07 &  0.79 $\pm$  0.14 \\
   4 & 14610 $\pm$  700 &  4.33 $\pm$  0.13 & 370 $\pm$ 39 &  2.362 $\pm$  0.076 &  3.94 $\pm$  1.68 & 482 $\pm$ 31 & 89 $\pm$  9 &  0.001 $\pm$  0.006 &  4.29E+07 $\pm$  2.75E+07 &  2.86 $\pm$  0.46 \\
   5 & 17170 $\pm$  770 &  3.81 $\pm$  0.13 & 230 $\pm$ 20 &  3.376 $\pm$  0.077 &  5.26 $\pm$  0.19 & 399 $\pm$ 24 & 87 $\pm$  8 &  0.426 $\pm$  0.251 &  5.17E+07 $\pm$  2.21E+07 &  0.76 $\pm$  0.11 \\
   6 & 16710 $\pm$  810 &  3.58 $\pm$  0.13 & 250 $\pm$ 24 &  3.598 $\pm$  0.081 &  5.50 $\pm$  0.16 & 356 $\pm$ 21 & 89 $\pm$  11 &  0.638 $\pm$  0.214 &  6.05E+07 $\pm$  1.97E+07 &  0.60 $\pm$  0.10 \\
   7 & 15050 $\pm$  770 &  3.98 $\pm$  0.16 & 269 $\pm$ 29 &  2.841 $\pm$  0.088 &  4.11 $\pm$  0.10 & 412 $\pm$ 31 & 89 $\pm$  12 &  0.313 $\pm$  0.260 &  8.07E+07 $\pm$  4.07E+07 &  1.27 $\pm$  0.22 \\
\hline
\end{longtable} }

\onecolumn
{\scriptsize
\setlength{\tabcolsep}{5pt}
\begin{longtable}{clccccccccc}
\caption{\label{corr_par_mode} Modes of astrophysical  parameters and rotational frequencies corrected for rotational effects of the observed Be stars assuming different ratios $\Omega/\Omega_{\rm c}$
of angular velocities.} \\
\hline\hline
\noalign{\smallskip}
Star  &  $ \langle T_{\rm eff}\rangle\pm\sigma$ & $\langle\log g\rangle\pm\sigma$ & 
$\langle V\!\sin i\rangle\pm\sigma$ & $\langle\log L/L_{\odot}\rangle\pm\sigma$  &
$\langle M/M_{\odot}\rangle\pm\sigma$ & $\langle V_{\rm c}\rangle\pm\sigma$ & 
$\langle i\rangle\pm\sigma$ & $\langle t/t_{\rm MS}\rangle\pm\sigma$ & $\langle{\rm age}\rangle\pm\sigma$ & 
$\langle\nu_{\rm r}\rangle\pm\sigma$ \\ 
      & \ \ \ \ \ K & dex &  km~s$^{-1}$ & dex & & km~s$^{-1}$ & deg. & & yr & c/d \\
\hline
\noalign{\smallskip} 
\endfirsthead
\caption{Continued.} \\
\noalign{\smallskip} 
\hline
\noalign{\smallskip}
Star  &  $\langle T_{\rm eff}\rangle\pm\sigma$ & $\langle\log g\rangle\pm\sigma$ & 
$\langle V\!\sin i\rangle\pm\sigma$ & $\langle\log L/L_{\odot}\rangle\pm\sigma$ 
& $\langle M/M_{\odot}\rangle\pm\sigma$ & $\langle V_{\rm c}\rangle\pm\sigma$ & 
$\langle i\rangle\pm\sigma$ & $\langle t/t_{\rm MS}\rangle\pm\sigma$ & $\langle{\rm age}\rangle\pm\sigma$ & $\langle\nu_{\rm r}\rangle\pm\sigma$ \\ 
\noalign{\smallskip}
\hline
\noalign{\smallskip}
\endhead
\noalign{\smallskip}
\endfoot
\noalign{\smallskip}
\multicolumn{11}{l}{$\Omega/\Omega_{\rm c} = 0.70$}\\
\noalign{\smallskip}
   1 & 11850 $\pm$   70 &  3.95 $\pm$  0.02 & 122 $\pm$  1 &  2.318 $\pm$  0.010 &  3.05 $\pm$  0.06 & 371 $\pm$  2 & 38 $\pm$  0 &  0.548 $\pm$  0.047 &  2.36E+08 $\pm$  8.94E+06 &  1.03 $\pm$  0.02 \\
   2 & 14660 $\pm$   80 &  3.73 $\pm$  0.02 & 229 $\pm$  2 &  3.198 $\pm$  0.010 &  5.09 $\pm$  0.04 & 361 $\pm$  2 & 89 $\pm$  0 &  0.951 $\pm$  0.011 &  1.04E+08 $\pm$  2.30E+06 &  0.70 $\pm$  0.01 \\
   3 & 11970 $\pm$   70 &  3.75 $\pm$  0.02 & 220 $\pm$  2 &  2.678 $\pm$  0.011 &  3.17 $\pm$  0.08 & 330 $\pm$  2 & 89 $\pm$  0 &  0.566 $\pm$  0.067 &  2.19E+08 $\pm$  1.08E+07 &  0.79 $\pm$  0.02 \\
   4 & 14610 $\pm$   80 &  4.38 $\pm$  0.02 & 370 $\pm$  4 &  2.362 $\pm$  0.009 &  5.38 $\pm$  0.34 & 495 $\pm$  3 & 89 $\pm$  0 &  0.000 $\pm$  0.000 &  4.29E+07 $\pm$  2.75E+07 &  2.86 $\pm$  0.05 \\
   5 & 17170 $\pm$   90 &  3.86 $\pm$  0.02 & 230 $\pm$  2 &  3.406 $\pm$  0.009 &  5.36 $\pm$  0.02 & 408 $\pm$  2 & 89 $\pm$  0 &  0.429 $\pm$  0.031 &  4.82E+07 $\pm$  4.13E+06 &  0.72 $\pm$  0.01 \\
   6 & 16710 $\pm$   90 &  3.64 $\pm$  0.02 & 250 $\pm$  2 &  3.630 $\pm$  0.009 &  5.56 $\pm$  0.10 & 356 $\pm$  2 & 89 $\pm$  0 &  0.664 $\pm$  0.043 &  6.36E+07 $\pm$  2.46E+06 &  0.58 $\pm$  0.01 \\
   7 & 15050 $\pm$   90 &  4.05 $\pm$  0.02 & 269 $\pm$  3 &  2.841 $\pm$  0.010 &  4.09 $\pm$  0.07 & 411 $\pm$  3 & 89 $\pm$  0 &  0.230 $\pm$  0.075 &  6.32E+07 $\pm$  8.87E+06 &  1.27 $\pm$  0.02 \\
\hline
\end{longtable} }      

\end{document}